\documentclass[prb,twocolumn,superscriptaddress,showpacs]{revtex4}

\usepackage{amsmath}
\usepackage{amssymb}
\usepackage{epsfig}
\usepackage{graphicx}
\usepackage{wasysym}
\usepackage{bbm}
\usepackage{multirow}
\usepackage{lipsum}

\usepackage{hyperref}

\usepackage{color}

\usepackage{tikz}

\usetikzlibrary{arrows,shapes,positioning,snakes}

\def\beq{\begin{equation}}
\def\eeq{\end{equation}}

\newcommand{\tabincell}[2]{\begin{tabular}{@{}#1@{}}#2\end{tabular}}
\newcommand{\vS}{\boldsymbol{S}}

\newcommand{\vsigma}{\boldsymbol{\sigma}}
\newcommand{\nd}{{\vphantom{\dagger}}}
\newcommand{\bea}{\begin{eqnarray}}
\newcommand{\eea}{\end{eqnarray}}

\newcommand{\dif}{\mathrm{d}}

\tikzstyle arrowstyle=[scale=1]
\tikzstyle directed=[postaction={decorate,decoration={markings,
    mark=at position .65 with {\arrow[arrowstyle]{stealth}}}}]
\tikzstyle reverse directed=[postaction={decorate,decoration={markings,
    mark=at position .65 with {\arrowreversed[arrowstyle]{stealth};}}}]

\begin{document}

\title{Schwinger boson spin liquid states on square lattice}

\author{Xu Yang\footnote{Current address: Department of Physics, Boston College, Chestnut Hill, Massachusetts 02467, USA}}
\affiliation{International Center for Quantum Materials, School of Physics, 
Peking University, Beijing 100871, China}

\author{Fa Wang}
\affiliation{International Center for Quantum Materials, School of Physics, 
Peking University, Beijing 100871, China}
\affiliation{
Collaborative Innovative Center for Quantum Materials, 
Beijing 100871, China}

\date{\today}

\begin{abstract}
We study possible spin liquids on square lattice that respect 
all lattice symmetries and time-reversal symmetry
within the framework of Schwinger boson (mean-field) theory.
Such spin liquids have spin gap and emergent $Z_2$ gauge field excitations.
We classify them by the projective symmetry group method, 
and find six spin liquid states 
that are potentially relevant to the $J_1$-$J_2$ Heisenberg model. 
The properties of these states are studied 
under mean-field approximation.
Interestingly we find a spin liquid state
that can go through continuous phase transitions
to either the N\'eel magnetic order or
magnetic orders of the wavevector at Brillouin zone 
edge center.
We also discuss the connection between our results and the Abrikosov fermion spin liquids.
\end{abstract}


\maketitle


Quantum spin liquid is 
the ground state of a quantum spin system in two and higher spatial dimensions
that does not show any spontaneous symmetry breaking. 
Since this concept was first introduced four decades ago\cite{Anderson-MRB73}, 
a lot of theoretical and experimental progress have been made in search for this 
unconventional phase\cite{review-Balents}.
In particular several promising experimental candidates have been identified\cite{review-Balents,review-Lee}. 
On the theoretical side, 
it is understood that 
quantum spin liquid is more likely to be found in frustrated spin-1/2 models 
with large classical ground state degeneracy,
where strong quantum fluctuations within the degenerate classical ground states 
may prevent long-range symmetry breaking order\cite{book-frustrated}.

The square lattice Heisenberg model with 
nearest-neighbor ($J_1$) and next-nearest-neighbor ($J_2$)
Heisenberg couplings
(abbreviated as $J_1$-$J_2$ model hereafter), 
\begin{equation}
H=J_1\sum\limits_{\langle ij\rangle} \vS_i\cdot \vS_j+J_2\sum\limits_{\langle \langle ij\rangle \rangle} \vS_i \cdot \vS_j,
\end{equation}
is one of the simplest frustrated spin models. 
It has attracted a lot of attention 
for its possible relevance to the copper oxides\cite{Lee-Nagaosa-Wen-RMP06,Kastner-RMP98} and 
iron-based high-temperature superconductors(HTSC) \cite{Hirschfeld+Korshunov-Mazin-RPP11,Dai-Hu-Dagotto-NatPhys12}.
Ground state of this model is the N\'eel order [$\langle \vS_{(x,y)}\rangle \propto (-1)^{x+y}$]
for $J_1\gg |J_2|\geq 0$ which is the magnetic order of undoped cuprates,
or the ``stripe'' collinear order [$\langle \vS_{(x,y)}\rangle \propto $ $(-1)^{x}$ or $(-1)^{y}$]
for $J_2\gg |J_1|\geq 0$ which is the magnetic order of many parent compounds of iron-based HTSCs. 
Spin liquid physics in this model has also been proposed to be relevant 
to the high temperature superconductivity in these materials\cite{Anderson-Science87,Baskaran-arXiv08,WengZY-NJP14}. 
However it should be noted that the appropriate model for iron-based HTSC is possibly spin-1 (unlike spin-1/2 for cuprates) 
and may involved biquadratic spin interactions\cite{Wysocki-NatPhys11}.
In this paper we will mainly focus on the spin-1/2 case.

The classical ground state of the $J_1$-$J_2$ model at $J_2/J_1=1/2$ 
has large degeneracy.
The ground state for quantum spin-$1/2$ model around $J_2/J_1=1/2$
has no magnetic order, although the true nature of this disordered phase
has been under debate for a long time\cite{Chandra-PRB88,Gelfand-PRB89,Dagotto-PRL89,Rokhsar-PRB90,ZhengWH-PRB99,Sorella-PRL00,Lauchli-PRB06}.
Recently a density matrix renormalization group(DMRG) study\cite{JiangHC-PRB12} showed evidence
of a spin liquid phase with spin gap in the region $0.41 < J_2/J_1 < 0.62$.
However a later DMRG result of the same model did not confirm this conclusion\cite{ShengDN-PRL14}. 
We do not intend to answer whether a spin liquid phase
exists in the $J_1$-$J_2$ model. 
The goal of this paper is to classify spin liquid 
states within the Schwinger boson formalism, 
identify candidate states relevant for the 
$J_1$-$J_2$ and related models and study their properties.
It is possible that none of the spin liquid states studied in this paper
can be realized in the Heisenberg models on square lattice. 
Nevertheless they may still be ground state for models 
close to the $J_1$-$J_2$ model, for example models with ring-exchange interactions. 

Spin liquids generically have fractionalized spinon 
and emergent gauge field excitations\cite{Read-Sachdev-PRL91,Wen-PRB91}.
The spinons are spin-1/2 and can be bosonic or fermionic.
In this paper we restrict ourselves to the bosonic spinon 
(Schwinger boson) representation.
The spin operators on every site $i$ are expressed by two bosons as
\begin{equation}
\vS_i=\frac{1}{2}\sum_{\alpha,\beta=\uparrow,\downarrow} b^\dagger_{i\alpha} \vsigma^\nd_{\alpha\beta} b^\nd_{i\beta}.
\label{eq:1}
\end{equation}
This enlarges the onsite Hilbert space, 
a local constraint, 
\begin{equation}
\hat{n}_i=\sum_{\alpha} b_{i\alpha}^\dagger b_{i\alpha}^\nd=2S,
\end{equation}
should be enforced and can be implemented by introducing a U(1) gauge field.
In this formalism, spin liquid will have gapped bosonic spinons 
and boson condensation will produce magnetic orders\cite{Read-Sachdev-PRL91,Sachdev-PRB92}.
In order for this state to be stable in two spatial dimensions(2D), 
the U(1) gauge field should be gapped, 
the simplest scenario is to have a spin-singlet spinon pair condensate which reduces
U(1) to $Z_2$ at low energy by the Higgs mechanism\cite{Sachdev-PRB92}.
Different $Z_2$ spin liquid phases within this formalism may be described by
inequivalent mean-field Hamiltonians, which can be classified by 
the projective symmetry group method\cite{Wen-PRB02,WangF-Vishwanath-PRB06}.
The mean-field states can be converted to physical spin states 
by Gutzwiller projection and can in principle be studied numerically by the 
variational Monte Carlo method\cite{Motrunich-PRB11}.

This paper is outlined as follows.
In Section~\ref{sec:PSG}, we 
report the results of PSG analysis for square lattice
and the mean-field representation of $Z_2$ spin liquids
which are potentially relevant to the $J_1$-$J_2$ model.
We then discuss 
the properties of these $Z_2$ spin liquids under mean-field approximation
in Section~\ref{sec:mean-field-results}. 
Section~\ref{sec:summary} contains further discussion and a summary of results.
Technical and numerical details are put in the appendices.

\section{Projective symmetry group analysis of square lattice Schwinger boson states}
\label{sec:PSG}

The projective symmetry group analysis is based on a specific mean-field theory. 
A mean-field treatment of square lattice Heisenberg model and 
its PSG analysis using fermionic spinons have been studied by Wen\cite{Wen-PRB02}.
In this paper we will study the spin liquid states on square lattice using the Schwinger boson mean-field theory.

With the help of the Schwinger boson representation of the spin operator \eqref{eq:1} 
and SU(2) completeness relation 
$\vsigma_{\alpha\alpha'}\cdot \vsigma_{\beta\beta '}=
2\delta_{\alpha\beta'}\delta_{\alpha'\beta}-\delta_{\alpha\alpha'}\delta_{\beta\beta'}$, 
the Heisenberg interaction can be rewritten as
\begin{equation}
\textbf{S}_i\cdot \textbf{S}_j=
\hat{B}_{ij}^{\dagger}\hat{B}_{ij}-\hat{A}_{ij}^{\dagger}\hat{A}_{ij}
\end{equation}
where the pairing term 
$\hat{A}_{ij}=(1/2)\sum\limits_{\alpha,\alpha '}\epsilon_{\alpha\alpha '}b_{i\alpha}b_{j\alpha '}$ and 
hopping term $\hat{B}_{ij}=(1/2)\sum\limits_{\alpha}b^{\dagger}_{i\alpha}b_{j\alpha}$ 
are both SU(2) invariant bond operators.

After a Hubbard-Stratonovich transformation, 
the quartic terms are decoupled and a mean-field Hamiltonian is obtained:
\begin{equation}\begin{split}
& H_{MF}=\sum\limits_{ij}(-A_{ij}^*\hat{A}_{ij}+B_{ij}^*\hat{B}_{ij}+H.c.) \\
&+\sum\limits_{ij}(|A_{ij}|^2-|B_{ij}|^2)/J_{ij}-\mu_i\sum\limits_i(\hat{n}_i-\kappa),
\end{split}
\end{equation}
where complex numbers $A_{ij}=-A_{ji}$ and $B_{ij}=B_{ji}^*$ are called the mean-field ansatz 
and the real Lagrangian multiplier $\mu_i$ is introduced to enforce 
the average boson number $\langle \hat{n}_i\rangle\equiv\kappa=2S$ condition on every site.
This mean-field treatment can become a controlled approximation 
in the large-$N$ limit\cite{Arovas-Auerbach-PRB88,Sachdev-PRB92,Flint-Coleman-PRB09},
we will however not pursue this direction.

Minimizing the variational energy with respect to 
the ansatz $A_{ij},B_{ij}$ and $\mu_i$ yields the self-consistent equations:
\begin{equation}
\langle \hat{n}_i\rangle=\kappa,\quad
\langle \hat{A}_{ij} \rangle=A_{ij}/J_{ij},\quad
\langle \hat{B}_{ij} \rangle=B_{ij}/J_{ij}.
\end{equation}

In the mean-field level, we assume that 
the Lagrangian multipulier $\mu_i$ is independent of site $i$ 
and the bonds $A_{ij}$ and $B_{ij}$ related by symmetry operations have the same amplitude.

This mean-field theory has an emergent U(1) gauge symmetry, 
namely that a local gauge transformation
\begin{subequations}
\begin{eqnarray}
b_{j\alpha}&\rightarrow  & e^{i\phi(j)}b_{j\alpha},\quad\alpha=\uparrow,\downarrow,\\
A_{ij} &\rightarrow & e^{i[\phi(i)+\phi(j)]}A_{ij},\\
B_{ij} &\rightarrow & e^{i[-\phi(i)+\phi(j)]}B_{ij},
\end{eqnarray}
\end{subequations}
will leave the physical observables unaffected.

The emergent gauge symmetry makes the symmetry of the mean-field ansatz not manifest: 
a mean-field ansatz after symmetry operation might seem different from the original one, 
but if they are connected by a gauge transformation 
the two ansatz still describe the same physical state and hence should be considered identical.

In order to solve this problem, Wen and collaborators suggest that one should use the projective representation of the space group to classify different kinds of mean-field ansatz.

In the projective representation, every symmetry operation $X$ is accompanied by a U(1) gauge transformation $G_{X}$, 
\begin{equation}
G_{X}X(b_{j\alpha})\rightarrow e^{i\phi_{X}[X(j)]}b_{X(j)\alpha}.
\end{equation}

The mean-field ansatz should be invariant under the combined operation $G_X X$ instead of symmetry operation $X$ alone.
The collection of all the combined operations which leave the mean-field ansatz invariant form the projective symmetry group (PSG).

Different PSGs characterize different kinds of spin liquid states all sharing the same symmetry. Under certain gauge, the PSG can be fully determined by the commutative relations of the generators of the symmetry group. 
 
The PSG defined through the algebraic relations of the group generators is called the algebraic PSG since there might be PSGs that cannot be realized by any mean-field ansatz. 

\subsection{PSG classification of the symmetric spin liquid on square lattice}
We set up a rectangular coordinate system and represent the lattice site with two unit vectors $\hat{e}_x$ and $\hat{e}_y$: $\textbf{r}=x\hat{e}_x+y\hat{e}_y$, where $x,y$ are integers.

The space group of the square lattice is generated by translation $T_1$ along $\hat{e}_x$, translation $T_2$ along $\hat{e}_y$, a reflection $\sigma$ around X-axis and the $90^{\circ}$ rotation $C_4$ around the origin $(x,y)=(0,0)$. 

\begin{figure}[h]
\includegraphics[scale=0.7]{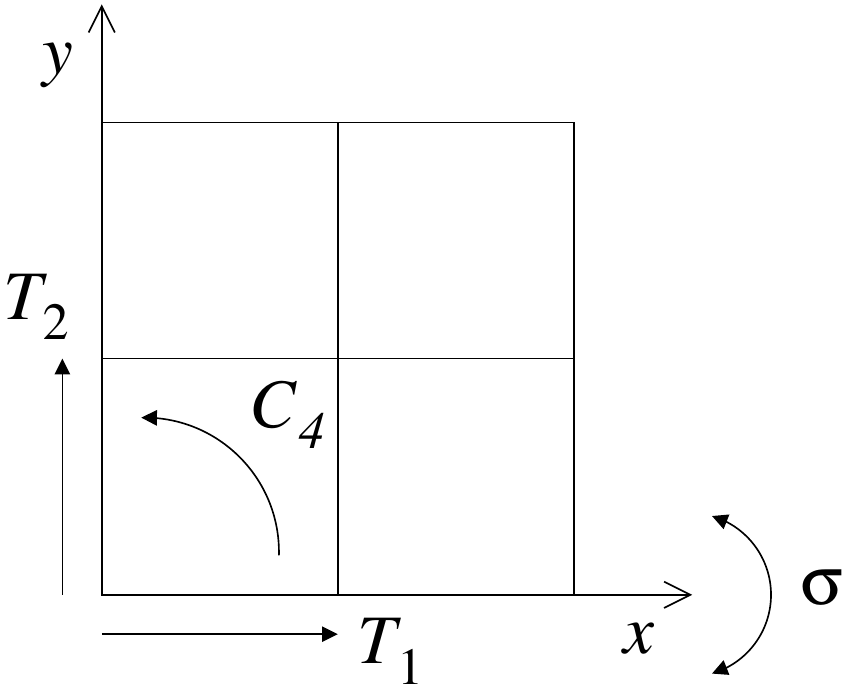}\\
\caption{
The coordinate system and space group generators $T_1$,$T_2$,$C_4$,$\sigma$ of the square lattice.
}
\label{fig1}
\end{figure}

The action of four generators on a lattice site therefore reads 
(see also FIG.~\ref{fig1}):
\begin{subequations}
\begin{eqnarray}
T_1&:&(x,y)\rightarrow(x+1,y),\\
T_2&:&(x,y)\rightarrow(x,y+1),\\
\sigma &:&(x,y)\rightarrow (x,-y),\\
C_4 &:& (x,y)\rightarrow(-y,x).
\end{eqnarray}
\end{subequations}

These generators have the following commutative relations,
which completely define the space group,
\begin{subequations}
\begin{eqnarray}
&& T_1T_2=T_2T_1,\\
&& T_1\sigma=\sigma T_1,\\
&& T_2\sigma=\sigma T_2^{-1},\\
&& \sigma^2=1,\\
&& C_4\sigma=\sigma C_4^{-1},\\
&& C_4^4=1,\\
&& T_1C_4=C_4T_{2}^{-1}.
\end{eqnarray}
\end{subequations}

Besides the space group generators, we can also add the time-reversal operator $\mathcal{T}$ which is commutative with all the space group generators. 

We have solved the algebraic PSG in the Appendix~\ref{sec:algebraicpsg}.
Here we will only list the results,
\begin{subequations}
\begin{eqnarray}
\phi_{T_1}&=& 0,\\ 
\phi_{T_2}&=& p_1\pi x,\\
\phi_{\sigma}&=& p_2\pi x+p_3\pi y +\frac{p_4\pi}{2},\\
\phi_{C_4} &=& p_1\pi xy+(p_2+p_3)\pi y+\frac{p_7\pi}{2},\\
\phi_{\mathcal{T}}&=&p_8(x+y-\frac{1}{2})\pi.
\end{eqnarray}
\end{subequations}
where the number $p_1,p_2,p_3,p_4,p_7,p_8$ are numbers which can be either 0 or 1.
Therefore there are at most 64 kinds of PSGs. 

\subsection{Physical realization of the PSGs on square lattice}

Demanding non-vanishing ansatz on certain bonds will impose further constraints on the algebraic PSGs.
In Appendix~\ref{sec:realization} we have analyzed the constraints imposed on PSGs 
when demanding an arbitrary non-vanishing bond. 
Here for simplicity and with the $J_1$-$J_2$ model in mind, 
we will only report 
the results for nearest-neighbor(NN) and next-nearest-neighbor(NNN) bonds.

Because of the strong antiferromagnetic $J_1$ coupling,
it is reasonable to assume that the nearest-neighbor pairing ansatz $A_{1}$ is nonzero.
Under this condition there are two classes of spin liquid states 
distinguished by the gauge-invariant flux $\Phi$ in the elementary plaquette, 
the zero-flux states with $\Phi=0$ and the $\pi$-flux states with $\Phi=\pi$, 
where $\Phi$ is defined modulo $2\pi$ on all plaquettes as\cite{Tchernyshyov-EPL06}
\begin{equation}
A_{ij}(-A_{jk}^*)A_{kl}(-A_{li}^*)=|A_{1}|^4 e^{i\Phi}.
\end{equation}

In the PSG language, the quantum number of flux $\Phi$ is $p_1$, with zero-flux states corresponding to $p_1=0$ and $\pi$-flux states corresponding to $p_1=1$.

The NN pairing ansatz is invariant under a staggered U(1) gauge transformation, 
it is necessary to gap out this low energy U(1) gauge field 
by either nonzero NN hopping $B_{1}$ or 
nonzero NNN pairing $A_{2}$,
which will reduce the low energy gauge field to $Z_2$\cite{Read-Sachdev-PRL91,WangF-Vishwanath-PRB06}.

The existence of nonzero $A_{1}$ demands that $p_2+p_3=1$ and $p_4=p_2$.
Nonzero $A_{2}$ requires that $p_1=1$ and $p_4+p_7=1$.
Nonzero $B_{1}$ requires that $p_2=0$ and $p_3=1$, and that $B_{1}$ must be pure imaginary. 
Nonzero $B_{2}$ requires that $p_1=0$ and $B_{2}$ is real. 

The simplest zero-flux $Z_2$ states can be constructed if 
we demand the existence of NN pairing $A_{1}$ and NN hopping $B_{1}$,
with possible existence of NNN hopping $B_{2}$.
NNN pairing $A_{2}$ is however forbidden in the zero-flux states.

Therefore the PSG for zero-flux states becomes
\begin{subequations}
\begin{eqnarray}
\phi_{T_1} &=& \phi_{T_2}=0, \\
\phi_{\sigma} &=& \pi y,\\
\phi_{C_4}&=&\pi y+\frac{p_7}{2}\pi,\\
\phi_{T}&=&(x+y-\frac{1}{2})\pi.
\end{eqnarray}
\end{subequations}

We are still left with a $Z_2$ number characterizing two different kinds of spin liquid states. 
We can further distinguish these two states with the help of another gauge-invariant flux defined on all plaquettes: $\Phi_2=\text{Arg}[A_{ij}(-A_{jk}^*)(-B_{kl}^*)(-B_{li}^*)]$. 

Therefore we obtain two zero-flux ($p_1=0$) spin liquid states:
\begin{enumerate}
\item $p_7=0$, $Z_2$[0,0] state with fluxes $\Phi_1=0$ and $\Phi_2=0$ through a plaquette. $A_{1}$ and $B_{2}$ are real, $B_{1}$ is pure imaginary.
The ansatz are given by
\begin{equation}
\begin{split}
& A_{(x,y)(x+1,y)}=-A_{(x,y)(x,y+1)}=A_1,\\
& B_{(x,y)(x+1,y)}=-B_{(x,y)(x,y+1)}=B_1,\\
& B_{(x,y)(x+1,y+1)}=-B_{(x,y)(x-1,y+1)}=B_2.
\end{split}
\end{equation}

\item $p_7=1$, $Z_2$[0,$\pi$] state with fluxes $\Phi_1=0$ and $\Phi_2=\pi$ through a plaquette. $A_{1}$ and $B_{2}$ are real, $B_{1}$ is pure imaginary. The ansatz are given by
\begin{equation}
\begin{split}
& A_{(x,y)(x+1,y)}=A_{(x,y)(x,y+1)}=A_1,\\
& B_{(x,y)(x+1,y)}=-B_{(x,y)(x,y+1)}=B_1,\\
& B_{(x,y)(x+1,y+1)}=-B_{(x,y)(x-1,y+1)}=B_2.
\end{split}
\end{equation}

\end{enumerate}

We can also obtain $\pi$-flux ($p_1=1$) states by demanding that $A_{1}$ and $A_{2}$ are non-vanishing,
with possible nonzero $B_{1}$.
NNN hopping $B_2$ is forbidden in $\pi$-flux states.

Therefore the PSG for $\pi$-flux states becomes
\begin{subequations}
\begin{eqnarray}
\phi_{T_1} &=& 0 ,\\
\phi_{T_2} &=& \pi x,\\
\phi_{\sigma} &=& p_2\pi x+(1-p_2)\pi y+\frac{p_2\pi}{2},\\
\phi_{C_4} &=& \pi xy+\pi y+\frac{(1-p_2)\pi}{2},\\
\phi_{\mathcal{T}}&=& p_8(x+y-\frac{1}{2})\pi.
\end{eqnarray}
\end{subequations}

There are four $\pi$-flux states depending on the values of $p_2$ and $p_8$.
We can further distinguish the four states with gauge-invariant flux 
$\Phi_3=\text{Arg}[A_{ij}(-A_{jk}^*)A_{kl}(-A_{li}^*)]$ 
defined on the triangle 
$i\rightarrow i+\hat{x}\rightarrow i+2\hat{x}\rightarrow i+\hat{x}+\hat{y}\rightarrow i$.

Nonzero $B_{1}$ can only be realized in the $p_2=0$, $p_8=1$ state since the coexistence of $A_{1}$ and $B_{1}$ requires that $p_2=0$ and $B_{1}$ is pure imaginary.

The NNN pairing $A_{2}$ can be real (corresponding to $p_8=0$) or pure imaginary ($p_8=1$), 
which is denoted by the letter $\mathcal{R}$ or $\mathcal{I}$ respectively.

Therefore we obtain four $\pi$-flux states:
\begin{enumerate}
\item $p_2=0$, $p_8=0$: $Z_2[\pi,0]\mathcal{R}$ with fluxes $\Phi_1=\pi$ through the plaquette and $\Phi_3=0$ through the triangle. $A_{1}$ and $A_{2}$ are real.
\begin{equation}
\begin{split}
& (-1)^yA_{(x,y)(x+1,y)}=A_{(x,y)(x,y+1)}=A_1,\\
& (-1)^yA_{(x,y)(x+1,y+1)}=(-1)^{y+1}A_{(x,y)(x-1,y+1)}=A_2.
\end{split}
\end{equation}

\item $p_2=0$, $p_8=1$: $Z_2[\pi,0]\mathcal{I}$ with fluxes $\Phi_1=\pi$ through the plaquette and $\Phi_3=0$ through the triangle. $A_{1}$ is real and $A_{2}$ is pure imaginary. In all four $\pi$-flux states $B_{1}$ can be nonzero only in this one, and must be pure imaginary.
\begin{equation}
\begin{split}
 &(-1)^yA_{(x,y)(x+1,y)}=A_{(x,y)(x,y+1)}=A_1,\\
& (-1)^yA_{(x,y)(x+1,y+1)}=(-1)^{y+1}A_{(x,y)(x-1,y+1)}=A_2,\\
 &B_{(x,y)(x+1,y)}=-B_{(x,y)\rightarrow(x,y+1)}=B_1.
\end{split}
\end{equation}

\item $p_2=1$, $p_8=0$: $Z_2[\pi,\pi]\mathcal{R}$ with fluxes $\Phi_1=\pi$ through the plaquette and $\Phi_3=\pi$ through the triangle. $A_{1}$ and $A_{2}$ are real.
\begin{equation}
\begin{split}
& (-1)^yA_{(x,y)(x+1,y)}=-A_{(x,y)(x,y+1)}=A_1,\\
& (-1)^yA_{(x,y)(x+1,y+1)}=(-1)^{y}A_{(x,y)(x-1,y+1)}=A_2.
\end{split}
\end{equation}

\item $p_2=1$, $p_8=1$: $Z_2[\pi,\pi]\mathcal{I}$ with fluxes $\Phi_1=\pi$ through the plaquette and $\Phi_3=\pi$ through the triangle. $A_{1}$ is real and $A_{2}$ is pure imaginary.
\begin{equation}
\begin{split}
& (-1)^{y}A_{(x,y)(x+1,y)}=-A_{(x,y)(x,y+1)}=A_1,\\
& (-1)^yA_{(x,y)(x+1,y+1)}=(-1)^{y}A_{(x,y)(x-1,y+1)}=A_2.
\end{split}
\end{equation}

\end{enumerate}

\section{Mean-field theory results}
\label{sec:mean-field-results}
In this section we will study the properties of the two zero-flux states and four $\pi$-flux states in the mean-field level. 

We will treat the average boson density $\kappa$ and $\alpha=J_2/J_1$ as variational parameters and obtain the phase diagram of the six states with respect to them. 

Since spinons can only be created in pairs, the physical spin excitation spectrum should be the two-spinon continuum spectrum. We compute the lower boundary of the two-spinon spectrum at every given total momentum $k$ for the six spin liquid states, through which we can distinguish different kinds of spin liquid states.

Another measurable quantity is the static spin structure factor. It can be calculated using the formula in the mean-field level:
\bea\label{ssf}
S(\textbf{k})=\frac{1}{N_{\text{site}}}\sum\limits_{i,j}\langle \textbf{S}_i\cdot \textbf{S}_j     \rangle e^{i\textbf{k}\cdot (\textbf{r}_i-\textbf{r}_j)},
\eea
where $N_{\text{site}}$ is the number of sites.

The two-spinon spin excitation spectrum and static spin structure factor 
can in principle be measured experimentally by (inelastic) neutron scattering,
and numerically by measuring spin-spin correlation functions.

\subsection{Zero-flux state}
\subsubsection{$Z_2[0,0]$ state}
After Fourier transformation $b_{\textbf{r}}=\frac{1}{\sqrt{N_s}}\sum\limits_{\textbf{r}}e^{-i\textbf{k}\cdot \textbf{r}}b_{\textbf{k}}$,
the mean-field Hamiltonian becomes 
\begin{equation}\label{zerozeromfH}
\begin{split}
& H_{MF}=\sum\limits_{\textbf{k}}\Psi^{\dagger}_{\textbf{k}}D_{\textbf{k}}\Psi_{\textbf{k}} \\
&+N_s[\mu+\mu\kappa+\frac{2|A_{1}|^2-2|B_{1}|^2}{J_1}-\frac{2|B_{2}|^2}{J_2}],
\end{split}
\end{equation}
where we have used the Nambu spinor 
$\Psi_{\textbf{k}}=(b_{\textbf{k}\uparrow},b_{-\textbf{k}\downarrow}^{\dagger})^T$ and the $2\times 2$ matrix \begin{equation}
D_{\textbf{k}}=(-2B_{2}f_1-\mu)\textbf{1}-A_{1}f_-\sigma_y-
\text{Im}(B_{1})f_-\sigma_z,
\end{equation}
where $f_1=\text{sin}(k_x)\text{sin}(k_y)$, $f_{\pm}=\text{sin}(k_x)\pm \text{sin}(k_y)$,
$\textbf{1}$ is the $2\times 2$ identity matrix,
$\sigma_{y,z}$ are Pauli matrices.

After a Bogoliubov transformation, the mean-field Hamiltonian can be diagonalized to yield
\begin{equation}
\begin{split}
&H_{MF}=\sum\limits_{\textbf{k}}\omega_{\textbf{k}}(\gamma^{\dagger}_{\textbf{k}\uparrow}\gamma_{\textbf{k}\uparrow}+\gamma^{\dagger}_{\textbf{k}\downarrow}\gamma_{\textbf{k}\downarrow}
+1)\\
&+N_s[\mu+\mu\kappa+\frac{2|A_{1}|^2-2|B_{1}|^2}{J_1}-\frac{2|B_{2}|^2}{J_2}].
\end{split}
\end{equation}

The dispersion relation has two branches $\omega_{\pm}$,
\begin{equation}\label{dispersionforzeroflux}
\omega_{\textbf{k}\pm }=\sqrt{ (2B_{2}f_1+\mu)^2-A_{1}^2f_-^2}\pm \text{Im}B_{1}f_-.\end{equation}

The minima of dispersion are located at $\pm \textbf{Q}$ where $\textbf{Q}=(\pi/2,-\pi/2)$, at which points there is an energy splitting between the two branches which is proportional to $|B_{1}|$.

The self-consistent equations are 
\begin{subequations}
\bea\label{selfconsistentequationforkappa1}
1+\kappa=-\int_{BZ}\frac{1}{2}\left [\frac{\partial \omega_{\textbf{k}+}}{\partial \mu}+\frac{\partial \omega_{\textbf{k}-}}{\partial \mu}\right ]\,\dif^2k,\\
4A_{1}/J_1=-\int_{BZ}\frac{1}{2}\left [\frac{\partial \omega_{\textbf{k}+}}{\partial A_{1}}+\frac{\partial \omega_{\textbf{k}-}}{\partial A_{1}}\right ]\,\dif^2k,\\
4|B_{1}|/J_1=\int_{BZ}\frac{1}{2}\left [\frac{\partial \omega_{\textbf{k}+}}{\partial |B_{1}|}+\frac{\partial \omega_{\textbf{k}-}}{\partial |B_{1}|}\right ]\,\dif^2k,\\
\label{selfconsistentequationforB2}
4B_{2}/J_2=\int_{BZ}\frac{1}{2}\left [\frac{\partial \omega_{\textbf{k}+}}{\partial B_{2}}+\frac{\partial \omega_{\textbf{k}-}}{\partial B_{2}}\right ]\,\dif^2k.
\eea
\end{subequations}

It can be seen from the dispersion relation Eq.~\eqref{dispersionforzeroflux} that the ansatz $B_{1}$ does not enter the self-consistent equations in the mean-field level. A closer examination also shows that $B_{1}$ does not affect the Bogoliubov transformation used to diagonalize the Hamiltonian. But $B_{1}$ does have effect on the low-energy effective field theory and magnetic ordered states as we shall see later.

\begin{figure}[h]
\includegraphics[width=60mm]{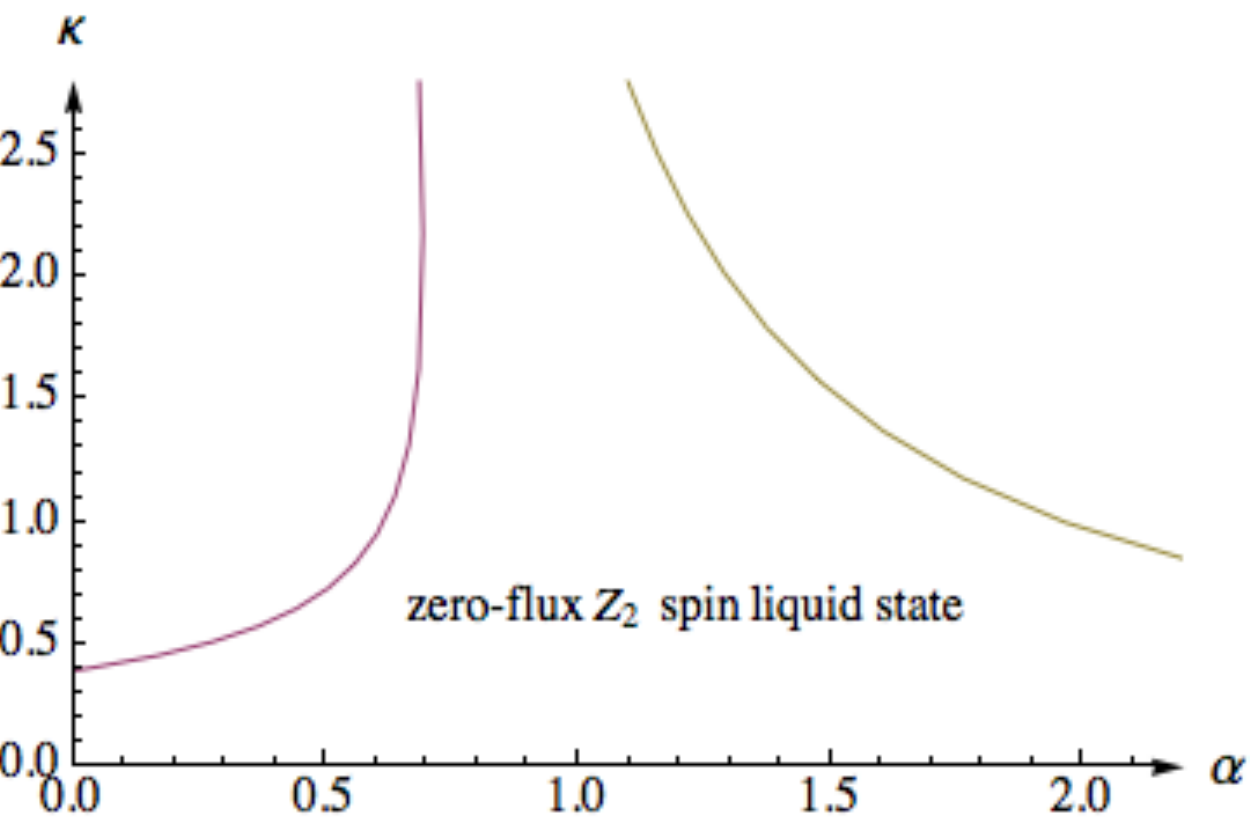}\\
\caption{The mean-field phase diagram for the zero-flux states as a function of ratio $\alpha=J_2/J_1$ and the average boson density $\kappa$. The two kinds of zero-flux states cannot be distinguished by their energetics since the NN hopping ansatz $B_{1}$ does not enter the mean-field equations. The curves of critical $\kappa_c$ are plotted. This state has lower energy than the $\pi$-flux states in the region where $\alpha$ is relatively low compared to unity. Above the critical value of $\kappa$, magnetic order will develop at several $k$ points where the spinon dispersion becomes zero. The magnetic ordered states for the two kinds of zero-flux states are different. The magnetic order obtained from the $Z_2[0,0]$ state after condensation of boson is the canted N\'eel order, and the magnetic order from the $Z_2[0,\pi]$ state is the N\'eel order. 
}\label{fig2}
\end{figure}

\begin{figure}[h]
\includegraphics[width=50mm]{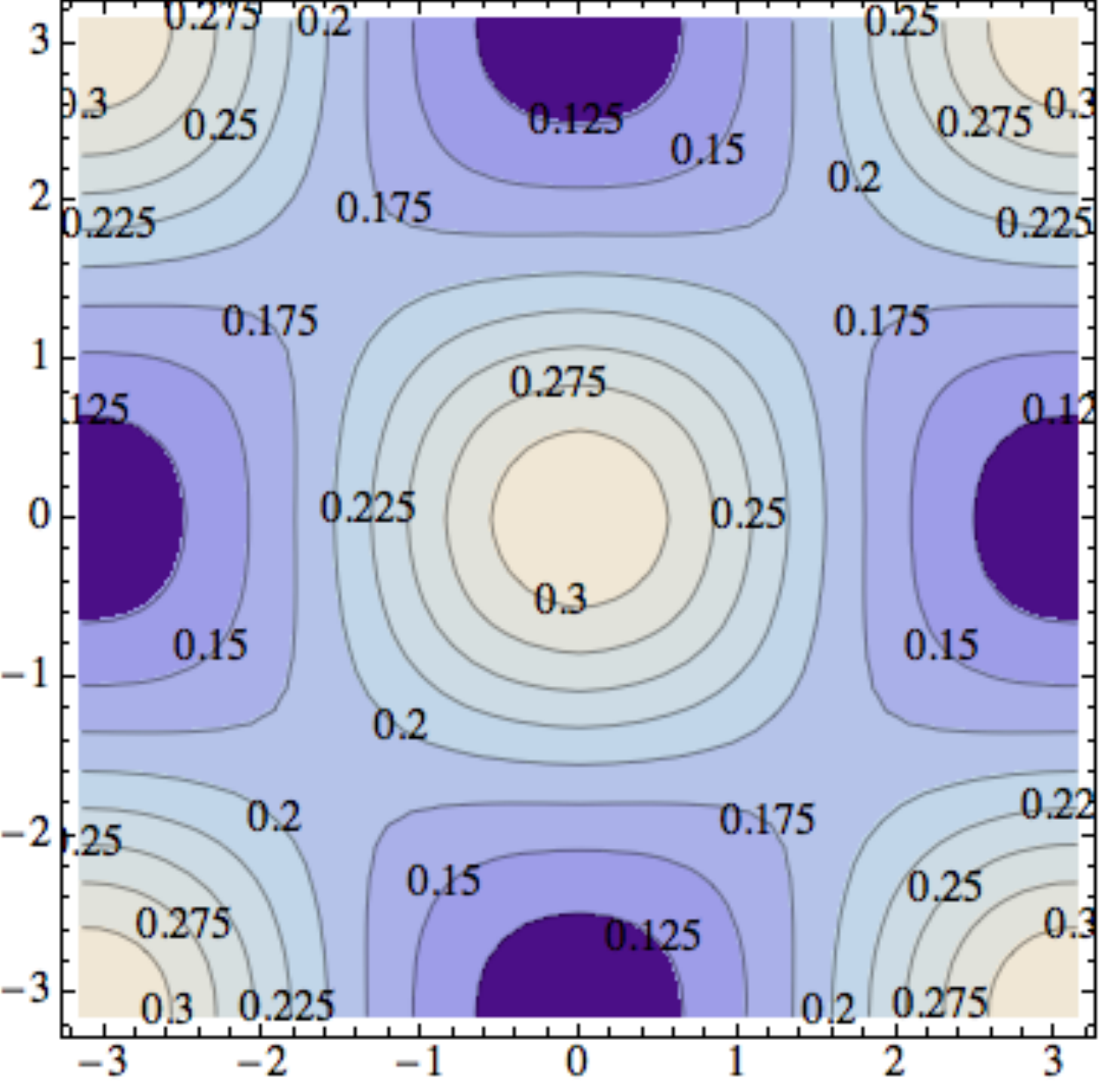}\\
\caption{Static structure factor for the zero-flux state. The spin structure factors for the two zero-flux states are qualitatively similar so here we only show the static structure factor for $Z_2[0,0]$ state at $\kappa=0.3$. Axes are $k_x$ and $k_y$ in dimensionless units and the static structure factor is calculated using Eq.~\eqref{ssf}. The global maxima are located at two wavevectors: $(0,0)$ and $(\pi,\pi)$. When the phase transition from the zero-flux $Z_2$ spin liquid states to the magnetic ordered states happens, the Bragg peak will be located at these two wavevectors, which represents the N\'eel order.
}\label{fig3}
\end{figure}

\subsubsection{$Z_2[0,\pi]$ state}
The mean-field Hamiltonian after Fourier transformation becomes:
\begin{equation}\label{zeropimfH}
\begin{split}
&H_{MF}=\sum\limits_{\textbf{k}}\Psi^{\dagger}_{\textbf{k}}D_{\textbf{k}}\Psi_{\textbf{k}}\\
&+N_s[\mu+\mu\kappa+\frac{2|A_{1}|^2-2|B_{1}|^2}{J_1}-\frac{2|B_{2}|^2}{J_2}],\\
\end{split}
\end{equation}
where
 \begin{equation}
D_{\textbf{k}}=(-2B_{2}f_1-\mu)\textbf{1}-A_{1}f_+\sigma_y-
\text{Im}(B_{1})f_-\sigma_z,
\end{equation}

The spinon dispersion relations are $\omega_{\textbf{k}\pm }=\sqrt{ (2B_{2}f_1+\mu)^2-A_{1}^2f_+^2}\pm \text{Im}(B_{1})f_-$.

The minima of dispersion are located at $\pm \textbf{Q}$ where $\textbf{Q}=(\pi/2,\pi/2)$.

The self-consistent equations are Eq.~\eqref{selfconsistentequationforkappa1}-\eqref{selfconsistentequationforB2}.
The bond $B_{1}$ serves as the Higgs field to break the U(1) gauge symmetry down to $Z_2$. Since $B_{1}$ does not enter the self-consistent equations, the two kinds of zero-flux states can not be distinguished by their energy at least in the mean-field level. 

We have obtained the phase diagram of the zero-flux state in FIG.~\ref{fig2}, treating average boson density $\kappa$ and $\alpha=J_2/J_1$ as variational parameters. The curve of critical boson density $\kappa_c$ is plotted for the zero flux state, above which the energy gap closes and magnetic order is developed.

\subsection{Magnetic order from zero-flux state}
In the Schwinger boson formalism, magnetic order is obtained via the condensation of bosons. When boson density $\kappa$ exceeds critical value $\kappa_c$, spinon dispersion will become zero at several $\textbf{Q}$ points. Bosons will condense at these $\textbf{Q}$ points and develop magnetic orders.

We have numerically computed the static spin structure factor for the two kinds of zero-flux states and find that there is no qualitative distinction between the two cases. Therefore we only show the static structure factor of the $Z_2[0,0]$ state at a relatively low $\kappa=0.3$ in Fig.~\ref{fig3}. We know from Fig.~\ref{fig3} that the global maxima of the static structure factor are located at $(0,0)$ and $(\pi,\pi)$, which indicates that the spin liquid state is adjacent to the N\'eel ordered state when the phase transition from the zero-flux $Z_2$ spin liquid states to the magnetic ordered states happens.

The analytical analysis of magnetic order from $Z_2[0,0]$ and $Z_2[0,\pi]$ states are given in the Appendix~\ref{appendix:magneticorder}. Here we will report only the final results. For convenience of later discussion, we devide the square lattice into two sublattices, sublattice $u$ for $(-1)^{x+y}=1$ and sublattice $v$ for $(-1)^{x+y}=-1$.

The magnetic ordered state obtained from the $Z_2[0,0]$ state is a non-collinear order, given by 
\begin{equation}
\langle \textbf{S}(\textbf{r})\rangle=
\frac{1+(-1)^{x+y}}{2}\vec{n}_{u}+\frac{1-(-1)^{x+y}}{2}\vec{n}_{v},
\end{equation}
where $\vec{n}_{u,v}$ are two vectors of the same length, $|\vec{n}_{u}|=|\vec{n}_{v}|$.
The angle between $\vec{n}_{u}$ and $\vec{n}_{v}$ are in general less than $180^{\circ}$ unless $B_{1}=0$.
This represents the ``canted N\'eel order'', which is the classical ground state for 
nearest-neighbor Heisenberg model under a small uniform magnetic field.

Aside from the spin (magnetic dipole) order parameter $\vec{n}_{u,v}$, this canted N\'eel order also has another vector spin chirality order parameter  
\beq
\vec{C}_{ij}=\nu_{ij}\langle \textbf{S}_i\times \textbf{S}_j\rangle,
\eeq
where $\nu_{ij}=\pm 1$ on the nearest neighbor bond $\langle ij\rangle$, and takes value $+1$($-1$) if site $i$ belongs to sublattice $u$($v$).

The magnetic order obtained from the $Z_2[0,\pi]$ state is the collinear N\'eel order,
$\langle \textbf{S}(\textbf{r})\rangle = (-1)^{x+y}\vec{n}$.
But this state is not the conventional N\'eel order and will also acquire a nonzero ground state expectation value of the vector spin chirality order parameter $\langle \textbf{S}_i \times \textbf{S}_j\rangle$ due to quantum fluctuations.

In the effective field theory, the transition from $Z_2$ spin liquid state to the magnetic ordered state occurs through the condensation of the charge-1 [in terms of the emergent U(1) gauge field] spinon field $z_{\alpha}$. $B_{1}$ serves as a charge-``-2" Higgs field. The physical observables are charge-``0" combinations of the spinon field $z_{\alpha}$. The three possible bilinear combinations are as follows:
\beq
\vec{n}_1=z^{\dagger}\vec{\sigma}z,\ 
\vec{n}_2=\text{Re}[B_1\,z^{T}i\sigma^y\vec{\sigma}z],\ 
\vec{n}_3=\text{Im}[B_1\,z^{T}i\sigma^y\vec{\sigma}z],
\eeq
which are orthogonal to each other.

It is easy to see that $\vec{n}_1$ can be identified with N\'eel order parameter. And since the vector spin chirality order parameter is even under time-reversal transformation, we can identify it with $\text{Re}[B_{1}\,z^{T}i\sigma^y\vec{\sigma}z]$. 
The order parameter is in fact the triad $\vec{n}_{1,2,3}$, 
even if the expectation values of spins form the collinear N\'eel order.
This kind of magnetic order has been obtained before on honeycomb lattice in proximity to certain spin liquid state\cite{Lu-Ran-PRB11}.

\subsection{Critical field theory for transition from zero-flux spin liquid states to magnetic order}
When $\kappa$ exceeds the critical value $\kappa_c$, there will be a phase transition from the $Z_2$ spin liquid state into the magnetic ordered state. 
Close to the phase transition point a continuous field theory can be derived.

The low-energy effective Lagrangian for the transition from $Z_2[0,0]$ state to the canted N\'eel order reads 
\begin{equation}\begin{split}
&\mathcal{L}=\int d^2\textbf{r}\{
\frac{8}{(-\mu+2A_{1}+2B_{2})a^2}\text{Im}(B_{1})\omega^*_{\alpha}\frac{d}{d\tau}\omega_{\alpha}
\\
&+(A_{1}-2B_{2}+\frac{4|B_{1}|^2}{-\mu+2A_{1}+2B_{2}})\partial_{\textbf{r}}\omega^*_{\alpha}\cdot
\partial_{\textbf{r}}\omega_{\alpha}\\
&+(\frac{-2\mu -4A_{1}+4B_{2}}{a^2}-\frac{8 |B_{1}|^2}{(-\mu+2A_{1}+2B_{2})a^2})\omega^*_{\alpha}\omega_{\alpha}\},
\\
\end{split}
\end{equation}
where $\omega$ is related to the spinon field $z$ and is defined in Eq.~\eqref{equ:omega-vs-z}.
Summation over repeated indices is implied hereafter.

The coefficient of the mass term $\omega^*_{\alpha}\omega_{\alpha}$ will change sign upon approaching the critical point, indicating that the critical chemical potential is $\mu\rightarrow 2B_{2}-\sqrt{4A_{1}^2+4|B_{1}|^2}$.
The low-energy effective field theory now flows to a fixed point where space and time scales differently with dynamical critical exponent $z=2$, which is different from previous theories of such phase transitions\cite{Read-Sachdev-PRL91}.
By power counting one can find that this theory approaches the upper critical dimension $d=4$ where the physics is controll by a Gaussian fixed point and the fluctuation around the mean-field theory is negilgible.
The scaling properties of correlation functions can be determined by naive power counting since in the upper critical dimension, the anomalous dimension correction approaches zero in the $4-\epsilon$ analysis when $\epsilon \rightarrow 0$.
For example,
$
\langle \textbf{S}(0)\cdot \textbf{S}(\textbf{r})\rangle\propto |\textbf{r}|^{-\eta}
$,
where $\eta$ is 4 plus correction which is proportional to $\epsilon$ and hence is negligible.

The low-energy effective Lagrangian for the transition from $Z_2[0,\pi]$ state to the N\'eel ordered state is
\begin{equation}
\mathcal{L}=\int d^2\textbf{r}\{  |D_{\tau}z|^2+c^2|D_{\textbf{r}}z|^2+m^2|z|^2
\}.
\end{equation}

The couplings between Higgs and spinon fields have at least two spatial derivatives and one time derivative, which is irrelevant by naive power counting. Anomalous dimension will not change this result.
Upon approaching the critical point, the mass term will disappear and this theory has an enlarged O(4) symmetry. The scaling properties of the field theory at O(4) critical point have been studied both analytically and numerically. The scaling behaviour of the correlation functions is therefore known. As an example, the spin-spin correlation function has a relatively large anomalous dimension, which behaves as
$
\langle \textbf{S}(0)\cdot \textbf{S}(\textbf{r})\rangle\propto |\textbf{r}|^{-\eta}
$,
where $\eta$ is numerically determined as $\eta=1.373(3)$ in contrast to the result for $Z_2[0,0]$ state\cite{Isakov-Senthil-PRB05}.

The transformation rules for spinon fields ($\omega$ or $z$) under space-group symmetry can also be readily deduced from the PSG and are listed in TABLE~\ref{tab:spinon-psg}.
It is easy to verify that the form of Lagrangian is invariant under these transformation rules.

\begin{table}[h]
\begin{tabular}[t]{|c|c|c|}\hline
 $Z_2[0,0]$ state  & $Z_2[0,\pi]$ state &  Symmetry Operation \\
\hline
$\omega\rightarrow -i\omega$ &  $z\rightarrow -i\sigma^y z^*$ & $T_1$  \\
\hline
$\omega\rightarrow i\omega$ &  $z\rightarrow i\sigma^y z^*$ & $T_2$   \\
\hline
$\omega\rightarrow  \omega$ & $z\rightarrow z$ &  $C_4$   \\
\hline
$\omega\rightarrow \omega$ & $z\rightarrow z$ & $\sigma$     \\
\hline
$\omega\rightarrow \omega$ &  $z\rightarrow \sigma^y z^*$ &  $\mathcal{T}$   \\
\hline
\end{tabular}
\caption{Transformation rules of the spinon fields under PSG.}
\label{tab:spinon-psg}
\end{table}

\subsection{$\pi$-flux state}
\subsubsection{$Z_2[\pi,0]\mathcal{R}$ state}
There are two sites in a unit cell of $\pi$-flux ansatz distinguished by $(-1)^y=\pm 1$ which can be labeled by $u$ and $v$ respectively.
The unit cells are labeled by integers $x$ and $\tilde{y}$, 
where the $u$($v$) site in the unit cell at $(x,\tilde{y})$ are at position
$(x,y)=(x,2\tilde{y})$ [$(x,y)=(x,2\tilde{y}+1)$].

After Fourier transformation 
\begin{equation}
b_{(u,v)\textbf{r}\alpha}=\frac{1}{\sqrt{N_s}}\sum\limits_{x,\tilde{y}}e^{-i(k_xx+2k_y \tilde{y})}b_{(u,v)\textbf{k}\alpha},
\end{equation}
where $k_{x,y}\equiv \textbf{k}\cdot \textbf{e}_{x,y}$,
the mean-field Hamiltonian becomes:
\begin{equation}H_{MF}=\sum\limits_{\textbf{k}} \Psi_{\textbf{k}}^{\dag}D_{\textbf{k}}\Psi_{\textbf{k}}+N_s[\mu+\mu\kappa+2 |A_{1}|^2/J_1+2 |A_{2}|^2/J_2],\end{equation}
in which,$$D_{\textbf{k}}=\begin{pmatrix}
\mu\cdot 1 & A_{1}P_1+A_{2}P_2 \\
-A_{1}P_1-A^*_2P_2 & \mu\cdot 1
\end{pmatrix}.$$
And we have also used the Nambu spinor $$\Psi_i=\begin{pmatrix}
b_{u\textbf{k}\uparrow}\\
b_{v\textbf{k}\uparrow}\\
b^{\dag}_{u-\textbf{k}\downarrow}\\
b^{\dag}_{v-\textbf{k}\downarrow}
\end{pmatrix}$$
and two $2\times 2$ antihermitian matrices
\begin{equation}
P_1=\frac{1}{2}\begin{pmatrix}
2i\text{sin}(k_x)&-1+e^{2ik_y} \\
1-e^{-2ik_y}&-2i\text{sin}(k_x)
\end{pmatrix},
\end{equation}
\begin{equation}
P_2=\frac{1}{2}\begin{pmatrix}
0& -\xi^*_{\textbf{k}}  \\
\xi_{\textbf{k}}&0\\
\end{pmatrix},
\end{equation}
where $\xi_{\textbf{k}}=e^{i k_x}-e^{-i k_x}+e^{-i(k_x+2k_y)}-e^{i(k_x-2k_y)}$.\par
The Hamiltonian can be diagonized by a Bogoliubov transformation using a SU(2,2) matrix to yield :
\begin{equation}
\begin{split}
&H_{MF}=\sum\limits_{\textbf{k};a=u,v} \omega_{\textbf{k}}(\gamma_{a\textbf{k}\uparrow}^{\dagger}\gamma_{a\textbf{k}\uparrow}+\gamma_{a\textbf{k}\downarrow}^{\dagger}\gamma_{a\textbf{k}\downarrow}+1)\\
&+N_s[\mu+\mu\kappa+2 |A_{1}|^2/J_1+2 |A_{2}|^2/J_2].\\
\end{split}
\end{equation}

In this state $A_{2}$ is real, the dispersion relation is therefore fourfold degenerated: 
\begin{equation}
\omega_{\textbf{k}}=\sqrt{\mu^2-A_{1}^2f_1-4A_{2}^2f_2},
\end{equation}
where $f_1=\text{sin}^2(k_x)+\text{sin}^2(k_y)$, $f_2=\text{sin}^2(k_x)\text{sin}^2(k_y)$.

The minima of spinon dispersion are located at $(k_x,k_y)=\pm(\pi/2,\pi/2)$.

\begin{figure}[h]
  \includegraphics[width=60mm]{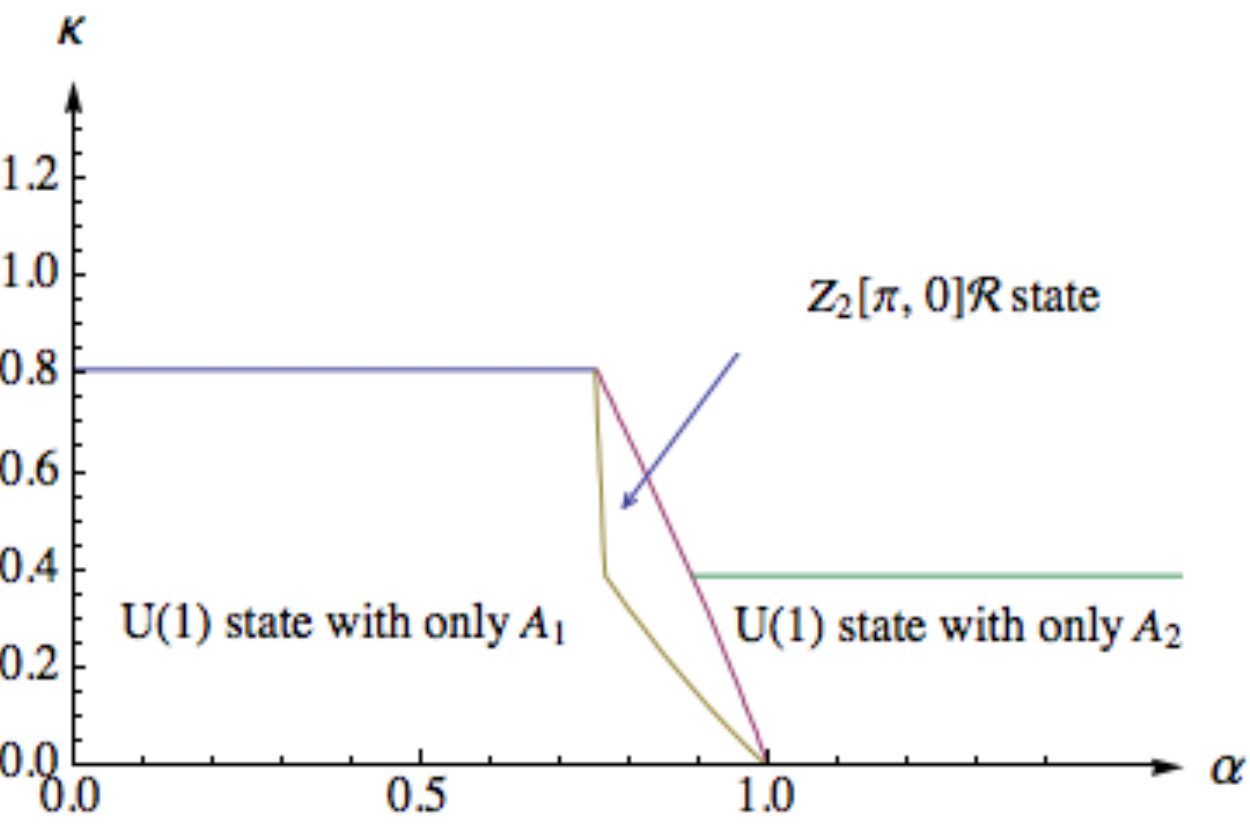}\\
  \caption{The mean-field phase diagram for the $Z_2[\pi,0]\mathcal{R}$ $\pi$-flux state as a function of ratio $\alpha=J_2/J_1$ and the average boson density $\kappa$. This state has a critical value of $\kappa\approx 0.81$, and occupy a finite area in the regime $\alpha=J_2/J_1\approx 0.75\sim 1$. Upon decreasing the value of $\alpha$, the $Z_2$ spin liquid state will become a U(1) spin liquid state with only non-zero nearest-bond $A_{1}$ through a first-order phase transition. The U(1) spin liquid state with only $A_{1}$ has a critical value of $\kappa\approx 0.81$. Upon increasing the value of $\alpha$, the $Z_2$ spin liquid state will become a U(1) spin liquid state with only non-zero $A_{2}$ through a first-order phase transition. The U(1) spin liquid state with only $A_{2}$ can be regarded as two copies of decoupled $\pi$-flux states on the square lattice and has a critical value of $\kappa\approx 0.40$.
}\label{fig4}
\end{figure}

\begin{figure}[h]
\includegraphics[width=50mm]{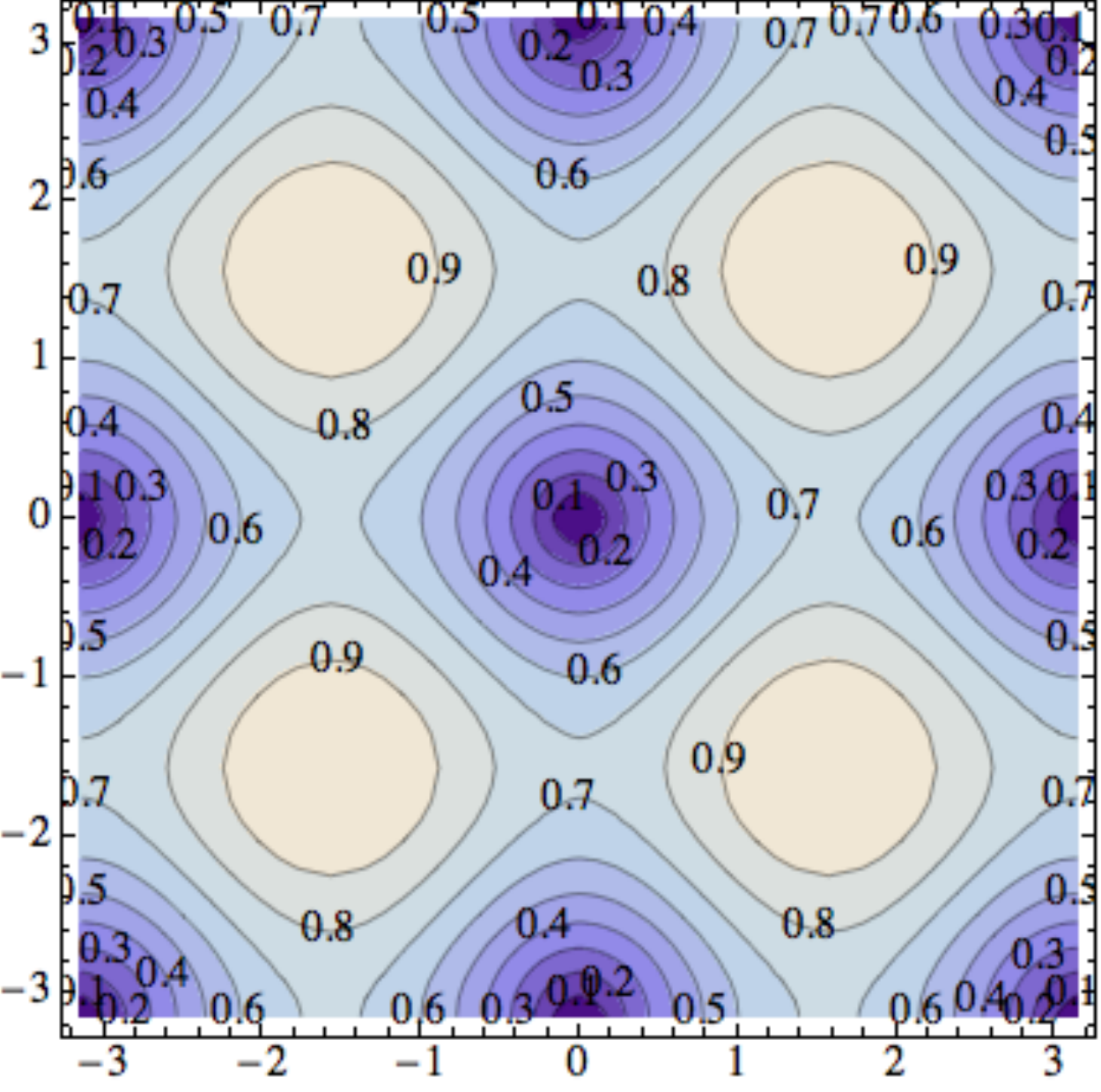}\\
\caption{The lower edge of the two-spinon spectrum for the $Z_2[\pi,0]\mathcal{R}$ state at $\kappa=0.81$.
}\label{fig5}
\end{figure}

The mean-field self-consistent equations are as follows:
\begin{subequations}
\bea\label{selfconsistentequationforkappa}
1+\kappa=-\int_{BZ}\frac{\partial \omega_{\textbf{k}}}{\partial \mu}\,\dif^2k,\\
4A_{1}/J_1=-\int_{BZ}\frac{\partial \omega_{\textbf{k}}}{\partial A_{1}}\,\dif^2k,\\
\label{selfconsistentequationforA2}
4A_{2}/J_2=-\int_{BZ}\frac{\partial \omega_{\textbf{k}}}{\partial A_{2}}\,\dif^2k.
\eea
\end{subequations}

Solving the self-consistent equations we have obtained a mean-field phase diagram for this state as shown in FIG.~\ref{fig4}. This state has $\kappa_c\approx 0.81$ but exist in the regime $\alpha=J_2/J_1\approx 0.75\sim 1$ far away from the physically interesting regime $\alpha\approx 0.5$. Therefore this state is unlikely to be the physical ground state for the square lattice $J_1-J_2$ Heisenberg model found in the numerical methods. The lower edge of the two-spinon spectrum is displayed in Fig.~\ref{fig5} for the $Z_2[\pi,0]\mathcal{R}$ state at $\kappa=0.5$, which could be used as a probe for this state in numerical simulations.

\subsubsection{$Z_2[\pi,0]\mathcal{I}$ state}
This state has nonzero $B_{1}$, so after Fourier transformation, the Hamiltonian becomes
\begin{equation}
H_{MF}=\sum\limits_{\textbf{k}} \Psi_{\textbf{k}}^{\dag}\begin{pmatrix}
\mu\cdot 1+B_{1}P_3 & A_{1}P_1+A_{2}P_2 \\
-A_{1}P_1-A^*_2P_2 & \mu\cdot 1-B_{1}P_3
\end{pmatrix}\Psi_{\textbf{k}}.
\end{equation}

In this state $A_{2}$ is pure imaginary, we have two twofold-degenerate dispersion :
\begin{widetext}
\begin{equation}\label{eq}
\omega_{\textbf{k}\pm}=\sqrt{\mu^2-(A_{1}^2-|B_{1}|^2)f_1-4|A_{2}|^2f_2 \pm 2\sqrt{(|B_{1}|^2+4(A_{1}^2-|B_{1}|^2)|A_{2}|^2f_2)f_1}},
\end{equation}
\end{widetext}
where $f_1=\text{sin}^2(k_x)+\text{sin}^2(k_y)$, $f_2=\text{sin}^2(k_x)\text{sin}^2(k_y)$ and
\begin{equation}
P_3=\frac{1}{2}\begin{pmatrix}
2i \text{sin}(k_x)& -1+e^{2ik_y} \\
1-e^{-2ik_y}&-2i \text{sin}(k_x) \\
\end{pmatrix},
\end{equation}

Minima of the spinon dispersion are at $(k_x,k_y)=\pm(\pi/2,\pi/2)$.

The self-consistent equations are as follows:
\begin{subequations}
\bea\label{selfconsistentequationforkappa2}
1+\kappa=-\int_{BZ}\frac{1}{2}[\frac{\partial \omega_{\textbf{k}+}}{\partial \mu}+\frac{\partial \omega_{\textbf{k}-}}{\partial \mu}]d^2k,\\
4A_{1}/J_1=-\int_{BZ}\frac{1}{2}[\frac{\partial\omega_{\textbf{k}+}}{\partial A_{1}}+\frac{\partial\omega_{\textbf{k}-}}{\partial A_{1}}]d^2k,\\
4|B_{1}|/J_1=+\int_{BZ}\frac{1}{2}[\frac{\partial\omega_{\textbf{k}+}}{\partial |B_{1}|}+\frac{\partial\omega_{\textbf{k}-}}{\partial |B_{1}|}]d^2k,\\
\label{selfconsistentequationforimaginaryA2}
4|A_{2}|/J_2=-\int_{BZ}\frac{1}{2}[\frac{\partial\omega_{\textbf{k}+}}{\partial |A_{2}|}+\frac{\partial\omega_{\textbf{k}-}}{\partial |A_{2}|}]d^2k.
\eea
\end{subequations}

\begin{figure}[h]
  \includegraphics[width=60mm]{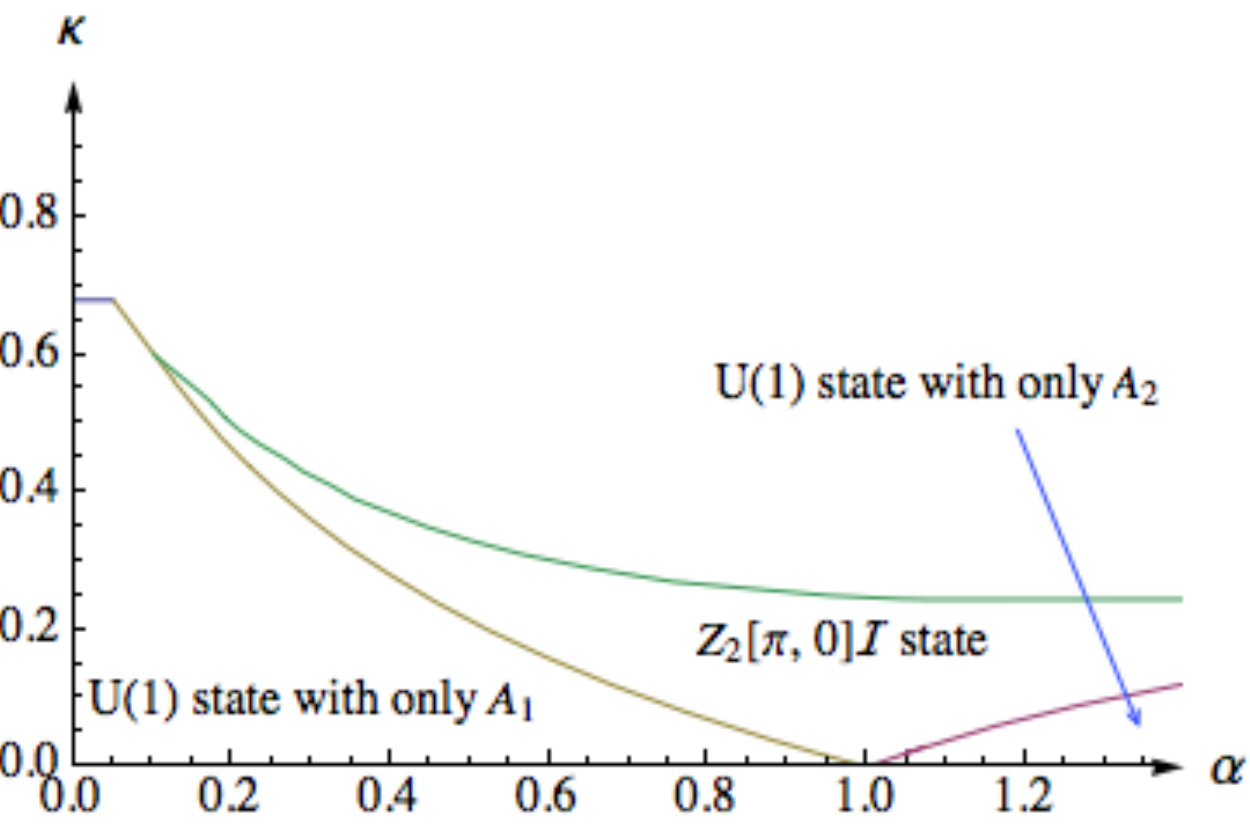}\\
  \caption{The mean-field phase diagram for the $Z_2[\pi,0]\mathcal{I}$ $\pi$-flux state as a function of ratio $\alpha=J_2/J_1$ and the average boson density $\kappa$. This state has a critical value of $\kappa\approx 0.60$, and occupy a finite area in the regime $\alpha \textgreater 0.10$.
}\label{fig6}
\end{figure}

\begin{figure}[h]
  \includegraphics[width=50mm]{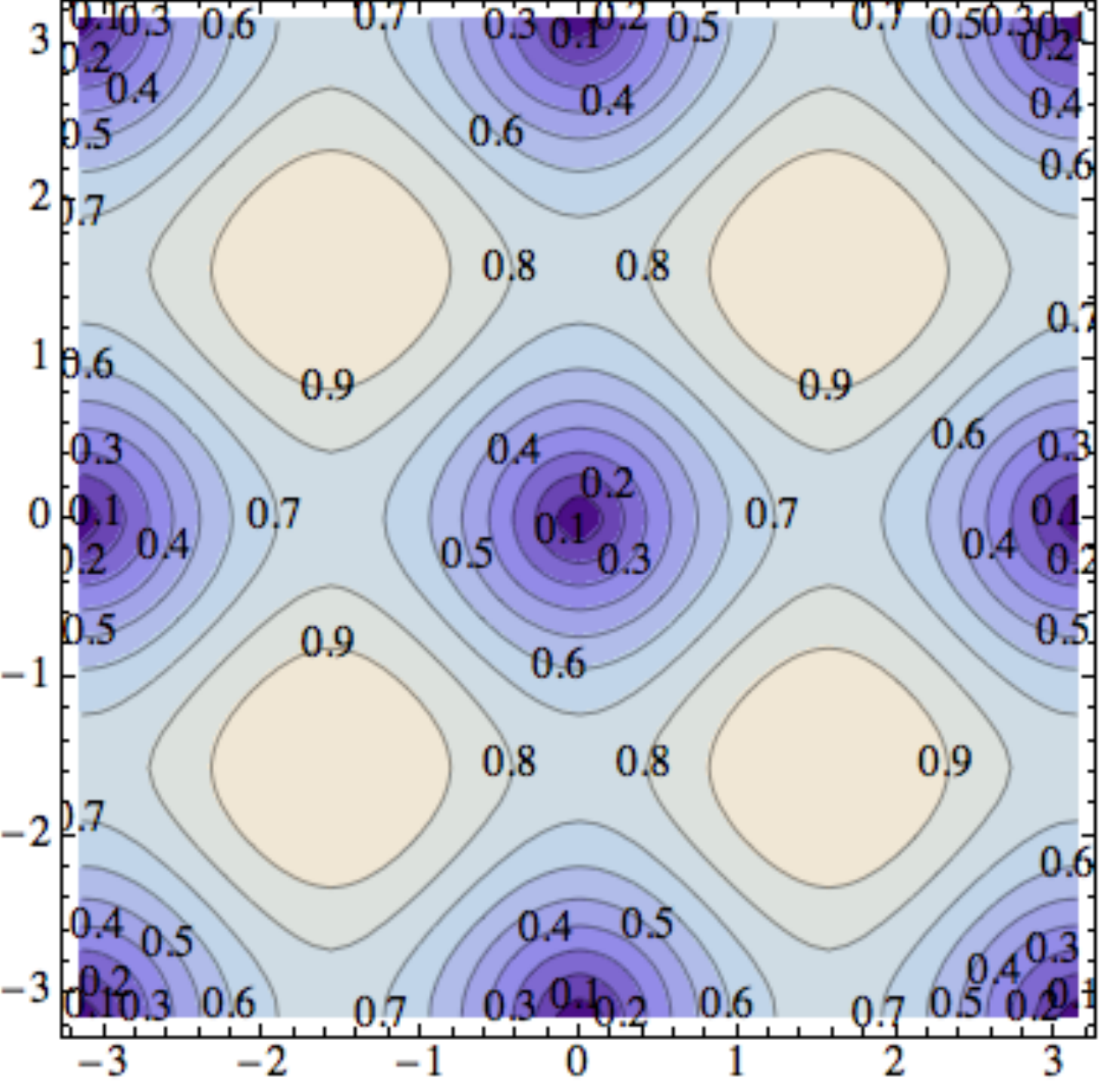}\\
  \caption{The lower edge of the two-spinon spectrum for the $Z_2[\pi,0]\mathcal{I}$ state at $\kappa=0.60$.
}\label{fig7}
\end{figure}

The self-consistent equations are Eq.~\eqref{selfconsistentequationforkappa}-\eqref{selfconsistentequationforA2}. The mean-field phase diagram for the $Z_2[\pi,0]\mathcal{I}$ state is obtained as shown in FIG.~\ref{fig6}. This state has a relatively low $\kappa_c$. The lower edge of the two-spinon spectrum is displayed in Fig.~\ref{fig7} for the $Z_2[\pi,0]\mathcal{I}$ state at $\kappa=0.6$.

\subsubsection{$Z_2[\pi,\pi]\mathcal{R}$ state}
The Hamiltionian after Fourier transformation is now (up to a constant):
\begin{equation}\label{meanfieldHamiltonian2}
H_{MF}=\sum\limits_{\textbf{k}} \Psi_{\textbf{k}}^{\dag}\begin{pmatrix}
\mu\cdot 1 & A_{1}P_1+A_{2}P_2 \\
-A_{1}P_1-A^*_2P_2 & \mu\cdot 1
\end{pmatrix}\Psi_{\textbf{k}}.
\end{equation}

And we have used two $2\times 2$ matrices:
\begin{equation}
P_1=\frac{1}{2}\begin{pmatrix}
2i\text{sin}(k_x)&1-e^{2ik_y} \\
-1+e^{-2ik_y}&-2i\text{sin}(k_x)
\end{pmatrix},
\end{equation}
\begin{equation}
P_2=\frac{1}{2}\begin{pmatrix}
0& -\xi_{\textbf{k}}^* \\
\xi_{\textbf{k}}&0 \\
\end{pmatrix},
\end{equation}
where $\xi_{\textbf{k}}=e^{ik_x}+e^{-ik_x}+e^{-i(k_x+2k_y)}+e^{i(k_x-2k_y)}$.

The dispersion relation is fourfold degenerate:
\begin{equation}
\omega=\sqrt{\mu^2-A_{1}^2f_1-4A_{2}^2f_2},
\end{equation}
where $f_1=\text{sin}^2(k_x)+\text{sin}^2(k_y)$, $f_2=\text{cos}^2(k_x)\text{cos}^2(k_y)$.

We denote $|A_{2}|/A_{1}=\text{tan}(\theta)$. When $\theta\textless \text{arctan}(1/\sqrt{2})$, the minima are located at $\pm(\pi/2,\pi/2)$, and when $\theta\textgreater \text{arctan}(1/\sqrt{2})$, the minima jump to  $(k_x,k_y)=(0,0)$ and $(\pi,0)$.

\begin{figure}[h]
  \includegraphics[width=60mm]{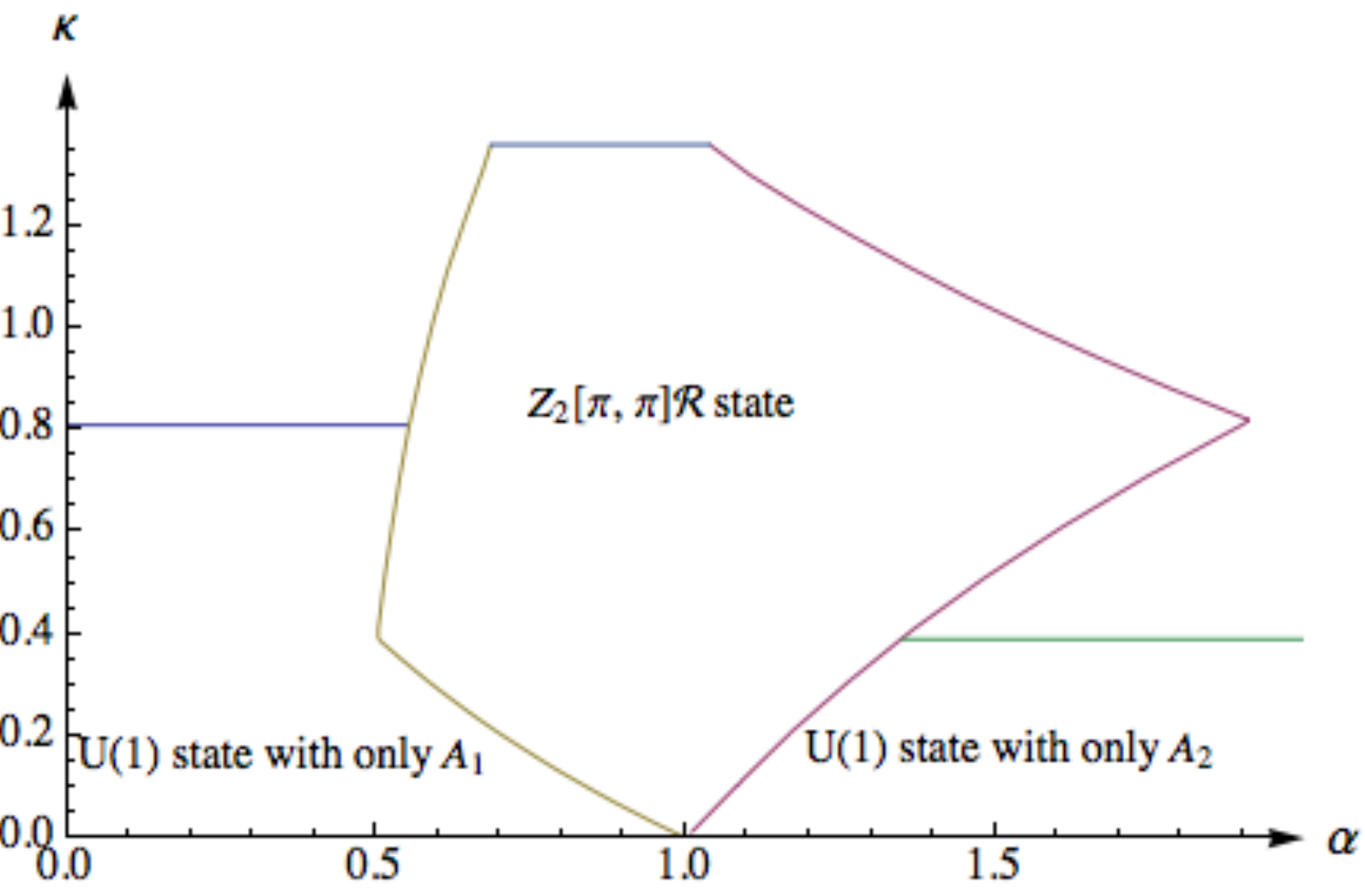}\\
  \caption{The mean-field phase diagram for the $Z_2[\pi,\pi]\mathcal{R}$ $\pi$-flux state as a function of ratio $\alpha=J_2/J_1$ and the average boson density $\kappa$. This state has a critical value of $\kappa\approx 1.36$, and occupy a finite area in the regime $\alpha=J_2/J_1\approx 0.5\sim 1.9$.
}\label{fig8}
\end{figure}

\begin{figure}[h]
  \includegraphics[width=50mm]{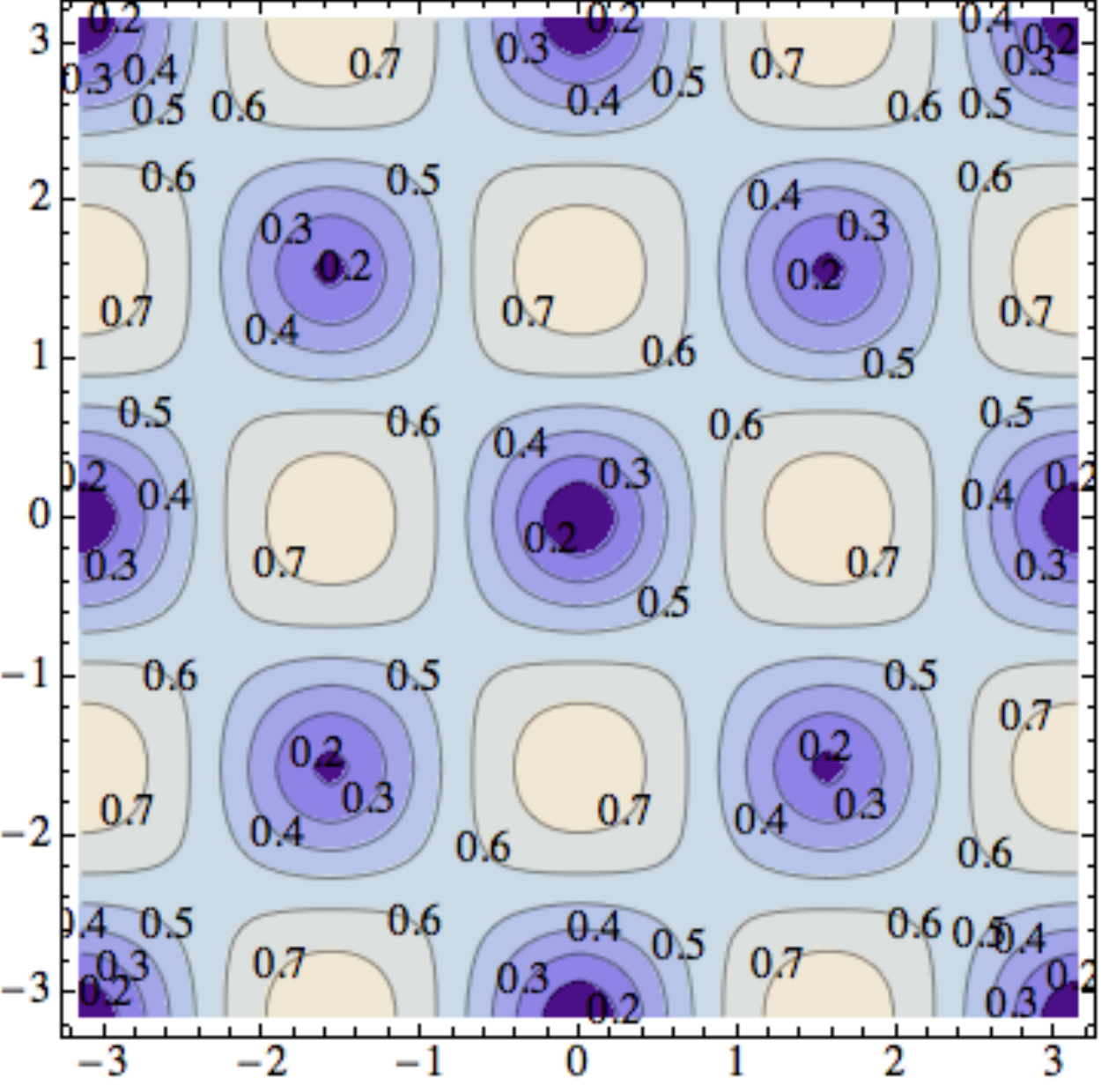}\\
  \caption{The lower edge of the two-spinon spectrum for the $Z_2[\pi,\pi]\mathcal{R}$ state at $\kappa=1.36$.
}\label{fig9}
\end{figure}

The curve of critical $\kappa$ is obtained via the self-consistent equation as shown in FIG.~\ref{fig8}. This state has a relatively high value of $\kappa_c\approx 1.36$ even above the physical value $\kappa=1$. Close to the physically interesting regime where $\kappa=1$ and $\alpha=J_2/J_1\approx 0.5$, this state occupy a finite area of phase space, and hence might be a promising candidate for the numerically found $Z_2$ spin liquid state. But a closer study shows that this state is energetically more unfavorable than the $Z_2[\pi,\pi]\mathcal{I}$ state in the physically interesting regime. The lower edge of the two-spinon spectrum is displayed in Fig.~\ref{fig9} for the $Z_2[\pi,\pi]\mathcal{R}$ state at $\kappa=1.36$.

\subsubsection{$Z_2[\pi,\pi]\mathcal{I}$ state}
The mean-field Hamiltonian is Eq.~\eqref{meanfieldHamiltonian2}.

In this state, $A_{2}$ is imaginary, so the dispersion relations are twofold degenerated:
\begin{equation}
\omega_{\pm}=\sqrt{\mu^2-A_{1}^2f_1-4|A_{2}|^2f_2 \pm 4A_{1}|A_{2}|\sqrt{f_1f_2}}
\end{equation}
where $f_1=\text{sin}^2(k_x)+\text{sin}^2(k_y)$, $f_2=\text{cos}^2(k_x)\text{cos}^2(k_y)$. 

Define $|A_{2}|/A_{1}=\text{tan}(\theta)$. When $\theta\textless \text{arctan}(1/(2\sqrt{2}))$, the minima are at 
$(k_x,k_y)=\pm(\pi/2,\pi/2)$. When $\theta\textgreater \text{arctan}(\frac{1}{2\sqrt{2}})$, the minima will move to incommensurate wave vectors
$(k_x,k_y)=\pm(q,q)$, where $\text{sin}(q)=1/(2\sqrt{2}\text{tan}(\theta))$.

\begin{figure}[h]
  \includegraphics[width=60mm]{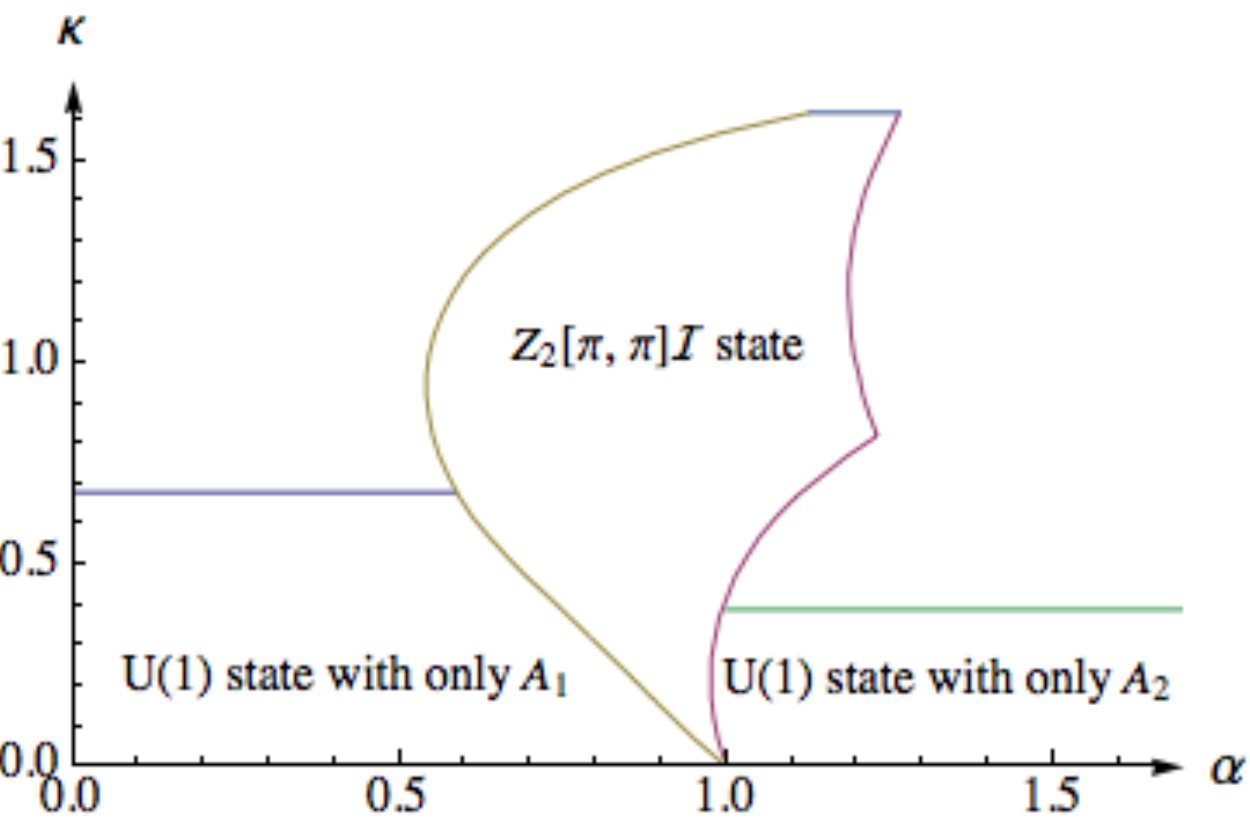}\\
  \caption{The mean-field phase diagram for the $Z_2[\pi,\pi]\mathcal{I}$ $\pi$-flux state as a function of ratio $\alpha=J_2/J_1$ and the average boson density $\kappa$. This state has a critical value of $\kappa\approx 1.62$, and occupy a finite area in the regime $0.55 < \alpha < 1.27$.
}\label{phasediagramforpi4}
\end{figure}

The self-consistent equations are Eq.~\eqref{selfconsistentequationforkappa2}-\eqref{selfconsistentequationforimaginaryA2} and the mean-field phase diagram is obtained as shown in FIG.~\ref{phasediagramforpi4}.

\begin{figure}[h]
  \includegraphics[width=50mm]{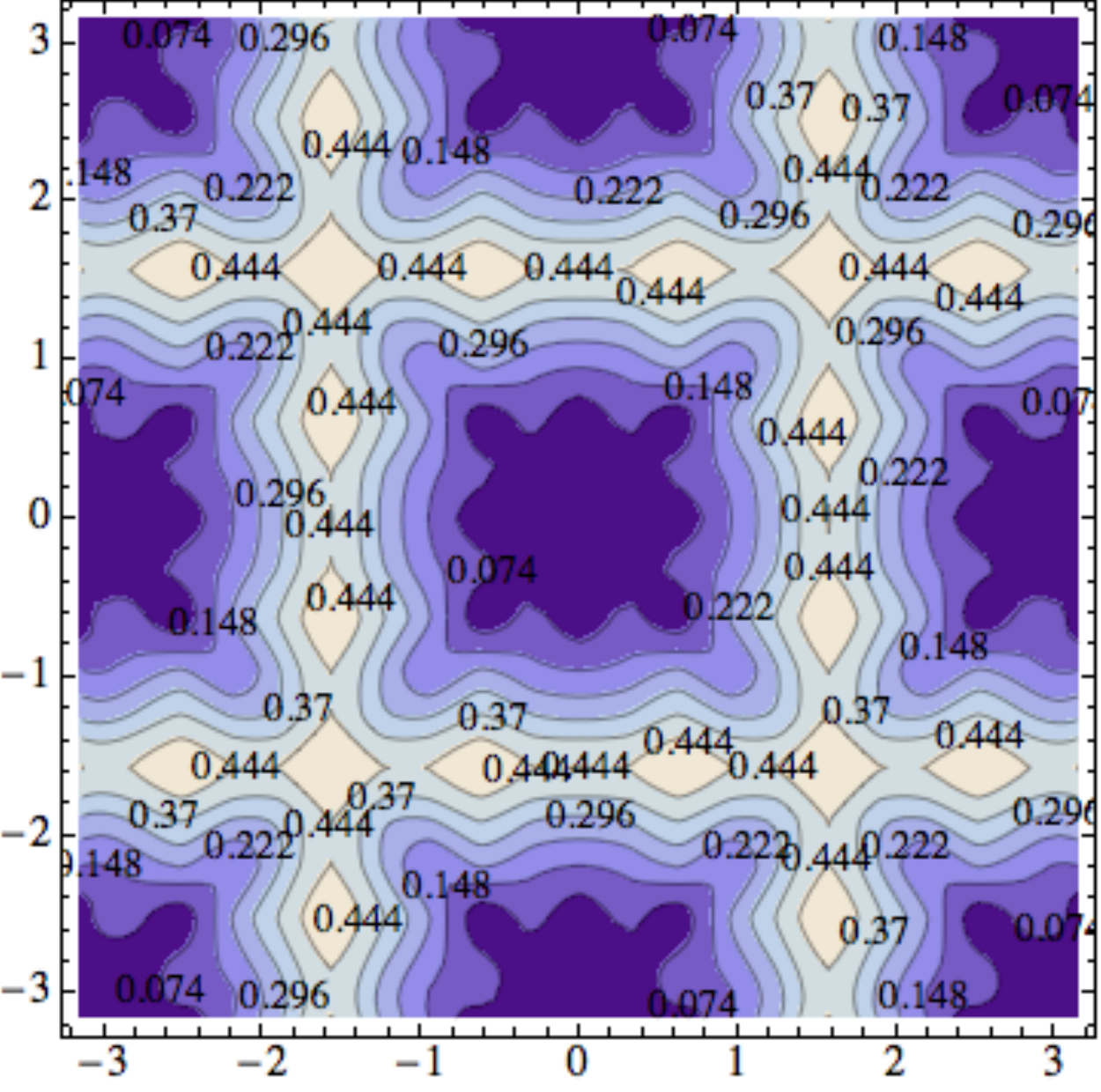}\\
  \caption{The lower edge of the two-spinon spectrum for the $Z_2[\pi,\pi]\mathcal{I}$ state at $\kappa=1.62$.
}\label{fig11}
\end{figure}

This state has a high $\kappa_c\approx 1.62$ even above the physical value $\kappa=2S=1$ and occupy a finite area of phase diagram close to the physically interesting regime $\kappa\approx 1$ and $\alpha\approx 0.5$. Although this state is energetically more unfavorable than the zero-flux states in this regime, we propose that by adding ring-exchange term in the Hamiltonian may favor the $\pi$-flux spin liquid state. The lower edge of the two-spinon spectrum is displayed in Fig.~\ref{fig11} for the $Z_2[\pi,\pi]\mathcal{I}$ state at $\kappa=1.62$.

\begin{figure}[h]
  \includegraphics[width=50mm]{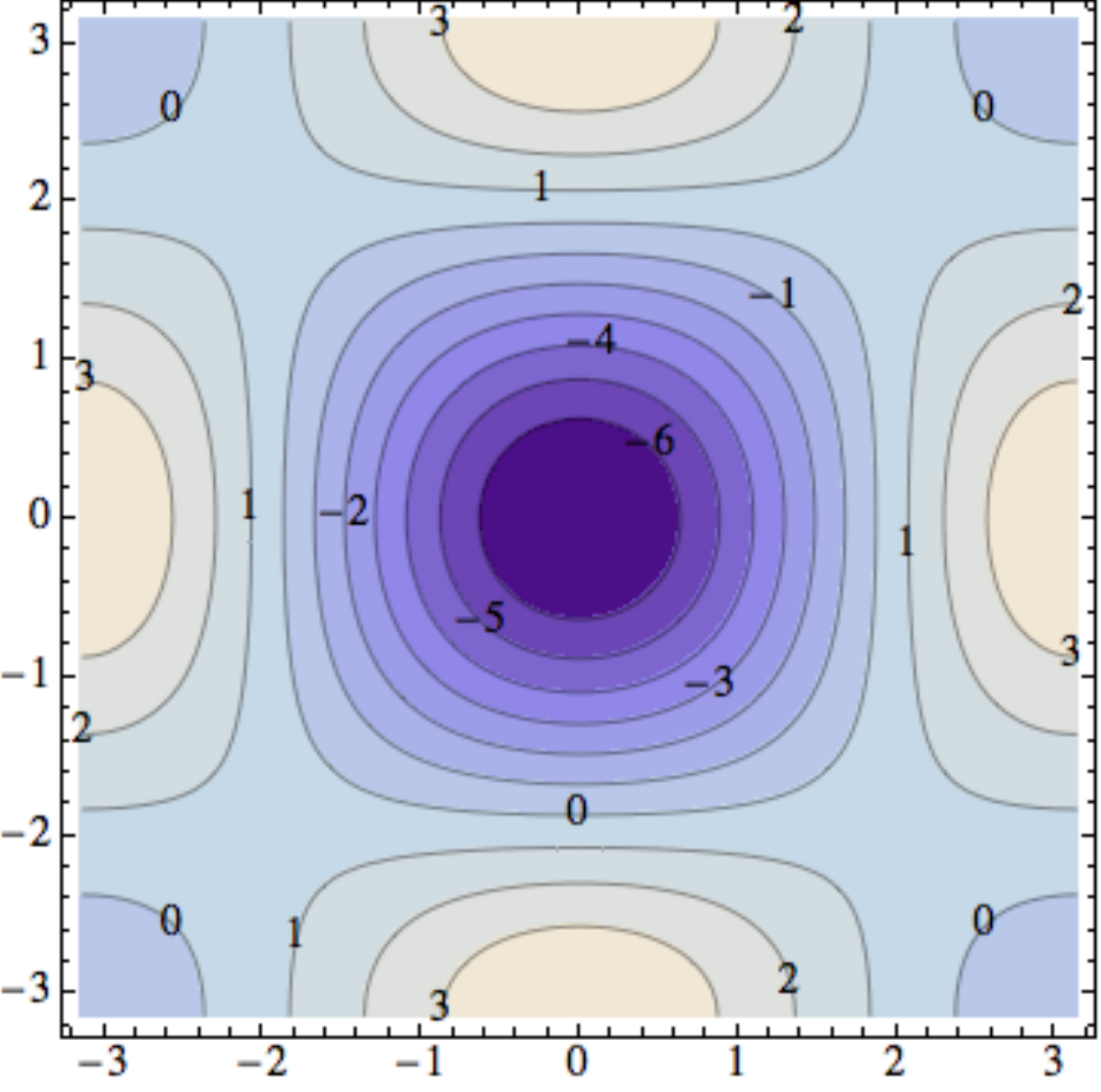}\\
  \caption{Spin structure factor for the $\pi$-flux state. The spin structure factors for the four $\pi$-flux states are qualitatively the same so here we only show the static structure factor for $Z_2[\pi,0]\mathcal{R}$ state at $\kappa=0.2$.  Axes are in dimensionless units $k_x$ and $k_y$ and the static structure factor is calculated using Eq. \ref{ssf}. The global maxima are located at four wave vectors $\pm (\pi,0)$ and $\pm (0,\pi)$. It is expected that when the continuum phase transition from the $\pi$-flux $Z_2$ spin liquid states to the magnetic ordered states happens, the Bragg peak will located at these four wave vectors, which represents a new kind of magnetic order different from the N\'eel order.
}\label{fig12}
\end{figure}

\subsection{Magnetic order from $\pi$-flux states}
We have computed the static spin structure factor for the four kinds of $\pi$-flux states in the mean-field level and find that there is no qualitative distinction between these four cases. Therefore we only show the static structure factor of the $Z_2[\pi,0]\mathcal{R}$ state at a relatively low $\kappa=0.2$ in FIG.~\ref{fig12}. From Fig.~\ref{fig12} that the global maxima of the static structure factor are located at $\pm(\pi,0)$ and $\pm(0,\pi)$, which indicates that the magnetic ordered state adjacent to the $\pi$-flux state is different from the N\'eel order.

We have also analytically obtained the magnetic ordered state starting from the mean-field Hamiltonian using the method of boson condensation. For simplicity, we shall only consider a $\pi$-flux state which only has NN bond $A_{1}$. The mean-field Hamiltonian after Fourier transformation is
\bea
H_{MF}=\sum\limits_{\textbf{k}}\Psi_{\textbf{k}}^{\dagger}\begin{pmatrix}
\mu\cdot 1, & A_{1}P_1\\
-A_{1}P_1, & \mu\cdot 1
\end{pmatrix}
\Psi_{\textbf{k}},
\eea
where 
\bea
P_1=\frac{1}{2}\begin{pmatrix}
2i\text{sin}(k_x), & -1+e^{2ik_y}\\
1-e^{-2ik_y},& -2i\text{sin}(k_x)
\end{pmatrix}.
\eea

The critical spinon vector occurs at two inequivalent $\textbf{k}$ points $\pm \textbf{Q}$, where  $\textbf{Q}=(\pi/2,\pi/2)$. 

Under the constraint that the density of condensate on every lattice site is uniform, we can work out the pattern of spinon condensation.

We then find a set of four-sublattice ordered states consistent with a subset of the classical ground state for $J_2/J_1=1/2$ Heisenberg model. 

The magnetic order is $\langle S(\textbf{r})\rangle=\textbf{m}_1(-1)^x+\textbf{m}_2(-1)^y+\textbf{m}_3(-1)^{x+y}$,
where $\textbf{m}_{1,2,3}$ are three orthogonal vectors and
\begin{equation}
m_1^2+m_2^2+m_3^2=m^2.
\end{equation}

Technical details of the calculation can be found in the Appendix~\ref{app:magneticorder-Z2pi}.

\section{Duality between Schwinger boson spin liquid states and Abrikosov fermion spin liquid states}
In the parton constructions of the Heisenberg model, the physical spin operator can be decomposed into spin-1/2 partons that can be either bosonic (Schwinger bosons) or fermionic (Abrikosov fermions). Whether the two seemingly distinct approaches are equivalent or not remains a long-standing puzzle. In this section we closely follow the work of Hermele \emph{et al.}\cite{Essin-Hermele-PRB13} and Lu \emph{et al.}\cite{LuYM-arXiv1403} and deduce the correspondence between the PSGs of the Schwinger boson representation and Abrikosov fermion representation.

We know clearly from the PSG classification of spin liquid states that topological order alone is not enough to fully characterize the different phases of the spin liquid states when symmetry is also presented\cite{Essin-Hermele-PRB13}. 
Actually, the interplay of topological order and symmetry will yield a richer structure called the ``symmetry enriched topological order"\cite{Wen-PRB02, WangF-Vishwanath-PRB06, Essin-Hermele-PRB13,  Kou-Levin-Wen-PRB08, Levin-Stern-PRB09, Lu-Vishwanath-arXiv1302, Mesaros-Ran-PRB13}.
In different symmetry enriched topological phases, anyon excitations will not only have fractional charges and fractional statistics, but also carry fractional symmetry quantum numbers. To be more concrete, the symmetry operation acts on anyons projectively and when an anyon returns to its original position after a series of symmetry operation, it may gain a nontrivial phase factor due to the gauge structure of the theory. All the fractionalization of symmetry quantum numbers of a certain type of anyon put together defines a fractionalization class for this type of anyon. In the case of Schwinger boson spin liquid states, the fractional symmetry quantum numbers $p_i$ $(i=1,2,3,4,7,8)$ fully determine the fractionalization class for the bosonic spinon. Once the fractionalization classes for each type of anyon excitations of a topological ordered state is determined, we can specify a symmetry class for this kind of symmetry enriched topological phase.

The fusion rules between different types of anyons have certain constraints on the fractionalization classes in the same symmetry class. Therefore we could use the compatibility condition to determine the correspondence between the fermionic PSGs and bosonic PSGs.

In the present case, three kinds of topological excitations, bosonic spinon $b$, fermionic spinon $f$ and the vison $v$, obey the following fusion rules. 
\begin{equation}
\begin{split}
& b\times f=v,\ b\times v=f,\ f\times v=b,\\
& b\times b=f\times f=v\times v=1.
\end{split}
\end{equation}

The fusion rules have a strong constraints on the symmetry fractionalization between the three kinds of topological excitations. Therefore in principle we could obtain the vison PSG from the knowledge of the fermionic PSG and the bosonic PSG due to the fusion rule $b\times v=f$. The additional phase $e^{i\phi_f}$ picked up by a fermion can be obtained from the phase $e^{i\phi_b}$ of a boson, the phase $e^{i\phi_v}$ of a vison, and possibly a twist factor $e^{i\phi_t}$ from the mutual statistics of bosons and visons, therefore we have 
\begin{equation}
e^{i\phi_f}=e^{i\phi_v}\cdot e^{i\phi_b}\cdot e^{i\phi_t}.
\end{equation}

\begin{table}[h]
\centering
\begin{tabular}[t]{cccc} \hline 
Algebraic Identities& bosonic $b_{\alpha}$& fermionic $f_{\alpha}$ &vison $v=b\times f$\\
\hline
$T_2^{-1}T_1T_2T_1^{-1}$ & $(-1)^{p_1}$  &$\eta_{xy}$   & -1 \\
$\sigma^{-1}T_1\sigma T_1^{-1}$ & $(-1)^{p_2}$  &  $\eta_{xpy}$  &   -1 \\
$\sigma^{-1}T_2\sigma T_2$   & $ (-1)^{p_3}  $  &$\eta_{xpx}$   &1\\
$\sigma^2$  &   $(-1)^{p_4}$  &  $\eta_{\sigma}$   &1\\
$\sigma^{-1}C_4\sigma C_4$ &$(-1)^{p_7}$&$ \eta_{\sigma C_4}$&1\\
$C_4^{4}$  &1  &  $\eta_{C_4}$   &   -1\\
$C_4^{-1}T_1C_4T_2$  &  $1$  &  $-1$  &-1\\
$C_4^{-1}T_2C_4T_1^{-1}$  &$(-1)^{p_2+p_3} $ &$-\eta_{xpx}\eta_{xpy} $  &1  \\
$C_4^{-1}\mathcal{T}^{-1}C_4\mathcal{T}$  &  $(-1)^{p_7}$  &$\eta_{C_4\mathcal{T}}$  &1  \\
$\sigma^{-1}\mathcal{T}^{-1}\sigma\mathcal{T}$   &$(-1)^{p_4}$ &$\eta_{\sigma\mathcal{T}}$ &1\\
$T_1^{-1}\mathcal{T}^{-1}T_1\mathcal{T}$& $(-1)^{p_8}$   &$\eta_t$   &1\\
$T_2^{-1}\mathcal{T}^{-1}T_2\mathcal{T}$  & $ (-1)^{p_8}$   &  $\eta_t$   &1 \\
$\mathcal{T}^2$   &   $    -1 $           &$-1 $           &1\\
\hline
\end{tabular}
\caption{
Correspondence between bosonic and fermionic PSGs. The bosonic PSGs are labeled by six integers $p_i=0,1$ $(i=1,2,3,4,7,8)$, while fermionic PSGs are labeled by nine integers $(\eta_{xy},\eta_{xpx},\eta_{xpy},\eta_{\sigma},\eta_{C_4},\eta_{\sigma C_4},\eta_{C_4\mathcal{T}},\eta_{\sigma\mathcal{T}},\eta_{t})$ where $\eta=\pm 1$. The vison PSG are fully determined as shown in Appendix~\ref{app:visonPSG}. The identity $C_4^{-1}T_1C_4T_2$ is not gauge invariant as we can always tune the relative phase between the gauge transformation $G_{T_1}$ and $G_{T_2}$, therefore its PSGs can always be fixed as shown above. The vison PSGs can be obtained from the product of fermionic PSGs and bosonic PSGs except the two cases $C_4^4$ and $\sigma^2$ where an extra factor of $(-1)$ should be taken into consideration as discussed in the main text. 
}
\label{tab:correspondenceofPSG}
\end{table}

The Abrikosov fermion construction of the Heisenberg model and its PSG study is summarized in detail in Wen's paper\cite{Wen-PRB02}. Although the solutions of the fermionic PSG are under certain gauge, the universal data is encoded in the fractionalization classes as in the case of Schwinger boson, and therefore we obtain the fractionalization classes of the Abrikosov fermion as listed in TABLE~ \ref{tab:correspondenceofPSG}.

In the mean-field level, visons can be considered as point-like excitation located on the center of a plaquette. Due to the $Z_2$ nature, two visons will annihilate each other, therefore we shall use $v$ to denote vison creation or annihilation operator. The dynamics of the vison can thus be described by a Ising gauge theory \cite{Huh-Punk-Sachdev-PRB11,Senthil-Fisher-PRB00,Sachdev-Vojta00,Jalabert-Sachdev-PRB91,Xu-Balents-PRB11}
and its symmetry fractionalization can be totally determined\cite{Huh-Punk-Sachdev-PRB11,Xu-Balents-PRB11}. The vison PSGs are determined as listed in TABLE~\ref{tab:correspondenceofPSG}. The detailed calculation is summarized in Appendix~\ref{app:visonPSG}.

\begin{table*}[tb]
\begin{tabular}[t]{|c|c||c|c|c|c|c|c|c|}  \hline 
\multicolumn{2}{|c||}{Schwinger boson}&\multicolumn{7}{c|}{Abrikosov fermion} \\ \hline
$(p_1,p_2,p_3,p_4,p_7,p_8)$ &Label &$(\eta_{xy},\eta_{xpx},\eta_{xpy},\eta_{t})$&  $g_{P_x}$ & $g_{P_y}$ &$g_{P_{xy}}$ &  $g_{\mathcal{T}}$   &Label&Perturbatively gapped? \\  \hline

\multirow{2}{*}{(0,0,1,0,0,1)}   & \multirow{2}{*}{ $Z_2[0,0]$} &\multirow{2}{*}{(-1,-1,-1,-1)}&$i\tau^1$   & $i\tau^1$ &$i\tau^1$&$i\tau^3$ &$Z_2B\tau^1_-\tau^1_-\tau^1\tau^3_-$& Yes \\   \cline{4-9}

 &&&$i\tau^2$  & $i\tau^2$ &$i\tau^1$&$i\tau^3$ &$Z_2B\tau^2_-\tau^2_-\tau^1\tau^3_-$&Yes\\ \cline{1-9}
(0,0,1,0,1,1) & $Z_2[0,\pi]$   &(-1,-1,-1,-1)&$i\tau^1$   & $i\tau^1$ &$\tau^0$&$i\tau^3$ &$Z_2B\tau^1_-\tau^1_-\tau^0\tau^3_- $&Yes \\ \cline{1-9}
(1,0,1,0,1,0)& $Z_2[\pi,0]\mathcal{R}$&(1,-1,-1,1)  &$i\tau^1$   & $i\tau^1$ &$\tau^0$&$i\tau^3$&$Z_2A\tau^1_-\tau^1_-\tau^0\tau^3_+$&Yes\\   \cline{1-9}
(1,0,1,0,1,1)&$Z_2[\pi,0]\mathcal{I}$ &(1,-1,-1,-1)  &$i\tau^1$   & $i\tau^1$ &$\tau^0$&$i\tau^3$ &$Z_2A\tau^1_-\tau^1_-\tau^0\tau^3_- $&Yes \\    \cline{1-9}
\multirow{2}{*}{(1,1,0,1,0,0)}   & \multirow{2}{*}{ $Z_2[\pi,\pi]\mathcal{R}$} &\multirow{2}{*}{(1,1,1,1)}&$\tau^0$   & $\tau^0$ &$\tau^0$&$\tau^0$ &$Z_2A\tau^0_+\tau^0_+\tau^0\tau^0_+$& --- \\   \cline{4-9}
 &&&$\tau^0$  & $\tau^0$ &$\tau^0$&$i\tau^3$ &$Z_2A\tau^0_+\tau^0_+\tau^0\tau^3_+$&Yes\\ \cline{1-9}
 \multirow{2}{*}{(1,1,0,1,0,1)}   & \multirow{2}{*}{ $Z_2[\pi,\pi]\mathcal{I}$} &\multirow{2}{*}{(1,1,1,-1)}&$\tau^0$   & $\tau^0$ &$\tau^0$&$\tau^0$ &$Z_2A\tau^0_+\tau^0_+\tau^0\tau^0_-$& No \\   \cline{4-9}
 &&&$\tau^0$  & $\tau^0$ &$\tau^0$&$i\tau^3$ &$Z_2A\tau^0_+\tau^0_+\tau^0\tau^3_-$&Yes\\ \cline{1-9}
\end{tabular}
\caption{
Correspondence between PSG solutions of bosonic and fermionic $Z_2$ spin liquids. The Schwinger boson $Z_2[0,0]$ state has two fermionic counterparts, which are $Z_2B\tau^1_-\tau^1_-\tau^1\tau^3_-$ state and $Z_2B\tau^2_-\tau^2_-\tau^1\tau^3_-$ state. 
The Schwinger boson $Z_2[0,\pi]$ state, $Z_2[\pi,0]\mathcal{R}$ state and $Z_2[\pi,0]\mathcal{I}$ state correspond to $Z_2B\tau^1_-\tau^1_-\tau^0\tau^3_- $ state, $Z_2A\tau^1_-\tau^1_-\tau^0\tau^3_+$ state and $Z_2A\tau^1_-\tau^1_-\tau^0\tau^3_- $ state respectively. The Schwinger boson $Z_2[\pi,\pi]\mathcal{R}$ state corresponds to $Z_2A\tau^0_+\tau^0_+\tau^0\tau^0_+$ and $Z_2A\tau^0_+\tau^0_+\tau^0\tau^3_+$ state, of which the $Z_2A\tau^0_+\tau^0_+\tau^0\tau^0_+$ state does not have a mean-field realization~\cite{Wen-PRB02}. The Schwinger boson $Z_2[\pi,\pi]\mathcal{I}$ state corresponds to $Z_2A\tau^0_+\tau^0_+\tau^0\tau^0_-$ and $Z_2A\tau^0_+\tau^0_+\tau^0\tau^3_-$ state. Among the eight PSGs that have mean-field realization, only the $Z_2A\tau^0_+\tau^0_+\tau^0\tau^0_-$ is a gapless state whose gaplessness is protected by its PSG~\cite{Wen-Zee-PRB02}. The remaining seven PSGs can all be realized by gapped spin liquid states.
}\label{table:correspondenceofansatz}
\end{table*}

In most cases, the twist factor from the mutual statistics of bosons and visons is trivial, but there are three exceptions where the twist factor are nontrivial ($e^{i\phi_t}=-1$), discussed in Ref.~\onlinecite{Essin-Hermele-PRB13,LuYM-arXiv1403,Qi-Fu-PRB15}. In the following we sketch their arguments for these nontrivial twist phase factors.

The first nontrivial case is $C_4^4$, namely fourfold rotation acted by four times\cite{Essin-Hermele-PRB13,LuYM-arXiv1403}. Consider the bound state of a vison and a bosonic spinon, which fuse into a fermionic spinon. When the bound state is rotated around a loop under the operation of $C_4^4$, the vison will encycle the boson once and therefore pick up a Berry phase of $\pi$\cite{Essin-Hermele-PRB13,LuYM-arXiv1403,Huh-Punk-Sachdev-PRB11}. 

The second case is $\sigma^2$, namely mirror reflection acted by two times, as discussed by Lu \emph{et al.}\cite{LuYM-arXiv1403} and Qi \emph{et al.}\cite{Qi-Fu-PRB15}. Consider a system consisting of two fermions $f_1$ and $f_2$ that can be related by a mirror reflection $\sigma$. Then we have 
$
\sigma f_1\sigma^{-1}=e^{i\phi_1}f_2,\ \sigma f_2\sigma^{-1}=e^{i\phi_2}f_1.
$
Therefore acting $\sigma$ twice on $f_1$ will yield a factor of $e^{i\phi_f}=e^{i\phi_1}\cdot e^{i\phi_2}$. Alternatively, acting $\sigma$ once on the pair $f_1$ and $f_2$ will yield a factor of $-e^{i\phi_1}\cdot e^{i\phi_2}$, in which the minus sign originates from the exchange the two fermions under $\sigma$.

The same argument works for the boson pair and vison pair, but since there is no additional statistical sign as in the case of fermion, we notice that the extra phase factor after acting $\sigma$ twice on the bosonic spinon or vison is the same as the phase factor after acting $\sigma$ once on a pair of bosonic spinons or visons. If we denote the phase acquired by acting $\sigma$ twice on $b$$(v)$ as $e^{i\phi_b}$$(e^{i\phi_v})$, then the above statement is equivalent to $-e^{i\phi_1}\cdot e^{i\phi_2}=e^{i\phi_b}\cdot e^{i\phi_v}$ (since we can simply treat $f$ as the bound state of $b$ and $v$).

Therefore we have the nontrivial fusion rule for the PSG of $\sigma^2$:
$
e^{i\phi_f}=-e^{i\phi_v}\cdot e^{i\phi_b}.
$

The third nontrivial fusion rule comes from $\sigma^{-1}\mathcal{T}\sigma\mathcal{T}$. 

Let's consider a system composed of a pair of fermionic spinons $f_1$ and $f_2$ located on the X-axis, so that they are symmetric under reflection $\sigma$. And we adopt another assumption that the two fermions can be connected by translation $T_x$, which ensures the two fermions share the same symmetry quantum numbers. 

In this case, the system is an eigenstate of the operator $\sigma$, we therefore have $\sigma f_1\sigma^{-1}=e^{i\phi_1}f_1, \sigma f_2\sigma^{-1}=e^{i\phi_1}f_2$. And acting $\sigma$ twice on a fermion will yield a factor of $e^{i\phi_f}=e^{2i\phi_1}$. 

Consider a pair of bosonic spinons and visons located on the X-axis, and we denote the eigenvalue of $b$($v$) under $\sigma$ as $e^{i\phi_2}$($e^{i\phi_3}$). Acting $\sigma$ twice on a boson(vison) will yield a factor of $e^{i\phi_b}=e^{2i\phi_2}$($e^{i\phi_v}=e^{2i\phi_3}$). 

If we treat the fermion as the bound state of a boson and a vison, then from the above discussion we have $e^{2i\phi_1}=-e^{2i\phi_2}\cdot e^{2i\phi_3}$. By splitting $-1$ equally into the two fermion sectors, we have $e^{i\phi_1}=(\pm i)e^{i\phi_2}\cdot e^{i\phi_3}$, which means each fermionic spinon gets an extra $\pm i$ phase in additional to the phase acquired by the boson and vison under reflection $\sigma$. But since the time-reversal transformation will take a number to its complex conjugate, the extra phase will be $(\pm i)^*(\pm i)=1$ under the symmetry operation $(\sigma \mathcal{T})^2$, which indicates that $(\sigma\mathcal{T})^2$ has a trivial fusion rule. 

Consider the following algebraic identity
\bea
(\mathcal{T}\sigma)^2=(\sigma^{-1}\mathcal{T}^{-1}\sigma\mathcal{T})\cdot (\mathcal{T})^2\cdot (\sigma)^2,
\eea
since the fusion rule for $(\mathcal{T}\sigma)^2$ and $\mathcal{T}^2$ are both trivial, while the fusion rule for $\sigma^2$ is nontrivial, we know that the fusion rule for $\sigma^{-1}\mathcal{T}^{-1}\sigma\mathcal{T}$ is nontrivial. 

Notice that the fourfold rotational symmetry $C_4$ is a combination of two reflection symmetry $C_4=P_{xy}\sigma$, where $P_{xy}$ is defined as the reflection along $y=x$. Therefore from the algebraic identity 
\bea
C_4^{-1}\mathcal{T}^{-1}C_4\mathcal{T}=(P_{xy}^{-1}\mathcal{T}^{-1}P_{xy}\mathcal{T})\cdot (\sigma^{-1}\mathcal{T}^{-1}\sigma\mathcal{T}),
\eea
we know that the fusion rule for $C_4^{-1}\mathcal{T}^{-1}C_4\mathcal{T}$ is trivial.

The correspondence of fermionic PSGs and bosonic PSGs are listed in TABLE~\ref{tab:correspondenceofPSG}. From the fusion rule, we find $Z_2$ spin liquids that can be both described by fermionic and bosonic partons if the following conditions are satisfied:
\begin{equation}\begin{split}
&\eta_{xy}=(-1)^{p_1+1},\ \eta_{xpx}=(-1)^{p_3},\ \eta_{xpy}=(-1)^{p_2+1},\\
&\eta_t=(-1)^{p_8},\eta_{\sigma \mathcal{T}}=\eta_{\sigma}=(-1)^{p_4+1},\ \eta_{\sigma C_4}=\eta_{C_4\mathcal{T}}=(-1)^{p_7},\\
&\eta_{C_4}=1,\\
\end{split}
\end{equation}
where the $\eta$s are $\pm 1$ which label different fractionalization classes of fermionic spinons
(see Appendix~\ref{app:fermionicPSG} for their definitions).

All the six Schwinger boson $Z_2$ spin liquid states we have discussed have fermionic counterparts. The Schwinger boson $Z_2[0,0]$ state has two fermionic counterparts, which are $Z_2B\tau^1_-\tau^1_-\tau^1\tau^3_-$ state and $Z_2B\tau^2_-\tau^2_-\tau^1\tau^3_-$ state. 
The Schwinger boson $Z_2[0,\pi]$ state, $Z_2[\pi,0]\mathcal{R}$ state and $Z_2[\pi,0]\mathcal{I}$ state correspond to $Z_2B\tau^1_-\tau^1_-\tau^0\tau^3_- $ state, $Z_2A\tau^1_-\tau^1_-\tau^0\tau^3_+$ state and $Z_2A\tau^1_-\tau^1_-\tau^0\tau^3_- $ state respectively. The Schwinger boson $Z_2[\pi,\pi]\mathcal{R}$ state corresponds to $Z_2A\tau^0_+\tau^0_+\tau^0\tau^0_+$ and $Z_2A\tau^0_+\tau^0_+\tau^0\tau^3_+$ state. The Schwinger boson $Z_2[\pi,\pi]\mathcal{I}$ state corresponds to $Z_2A\tau^0_+\tau^0_+\tau^0\tau^0_-$ and $Z_2A\tau^0_+\tau^0_+\tau^0\tau^3_-$ state. See TABLE~\ref{table:correspondenceofansatz}. Their possible realization on square lattice is discussed in Appendix~\ref{appendix:G}. Among these nine states, the $Z_2A\tau^0_+\tau^0_+\tau^0\tau^0_+$ state cannot be realized by mean-field ansatz, and the $Z_2A\tau^0_+\tau^0_+\tau^0\tau^0_-$ state is a gapless spin liquid state whose gaplessness is protected by its PSG\cite{Wen-Zee-PRB02}, and the remaining PSGs can all be realized by gapped spin liquid states.

\section{Discussions and Summary}
\label{sec:summary}

Here we discuss the relation between our results and related theoretical and experimental works.
The PSG classification of Schwinger boson states has been used in 
Tao Li {\it et al}\cite{LiTao-PRB12}
and Yi-Zhuang You {\it et al}\cite{WengZY-NJP14}.
In particular Tao Li and coworkers studied the energetics of several kinds of projected Schwinger boson wave functions on square lattice for $J_1$-$J_2$ Heisenberg model\cite{LiTao-PRB12}.
However they did not provide a complete projective symmetry group analysis 
and presented numerical results for zero-flux states only.
Yi-Zhuang You and coworkers tried to explain the behavior of local moments in
iron-based superconductors by Schwinger boson spin liquid states\cite{YouYZ-NJP14}.
They achieved a complete solution of algebraic PSG in appendix, 
but did not study all the mean-field states.
As far as we know our paper is the first complete account on 
the PSG classification and mean-field realizations of 
Schwinger boson spin liquid states on square lattice,
as well as the magnetic orders connected to Schwinger boson $Z_2$ spin liquids.

We do not attempt to relate our results directly to any experimental or numerical evidences of spin liquids.
It may be tempting to try to relate these $Z_2$ spin liquids with 
underdoped cuprate superconductors.
This possibility has been thoroughly discussed in the review by
Lee and coworkers\cite{Lee-Nagaosa-Wen-RMP06}, 
albeit using a different slave-particle formalism 
with fermionic spinons.
If the relation between Schwinger boson 
and Abrikosov fermion spin liquids can be firmly established, 
our results may also be relevant to cuprates.
More recently an inelastic neutron scattering experiment observed
evidence of spinons coexisting with N\'eel order\cite{Piazza-NatPhys15}.
This type of states is beyond our current study of \emph{symmetric} spin liquids
without any symmetry breaking.
Classification and numerical studies of this type of ``AFM$^*$'' states
will be an interesting future direction.

Finally we would like to say a few more words about the relation between
Schwinger boson and Abrikosov fermion spin liquid states.

Several groups\cite{Read-Chakraborty-PRB99,Yunoki-Sorella-PRB06,YangFan-PRL12,Qi-Fu-PRB15} have established
the exact mapping 
between these two formulations on extremely short-range(nearest-neighbor) resonating valence bond(RVB) states,
which are limiting cases of Gutzwiller projected wave functions of
mean-field Schwinger boson or Abrikosov fermon states.
This exact mapping however is not very useful on square lattice.
There are two nearest-neighbor RVB states on square lattice 
preserving all lattice symmetry.
They correspond to the
zero-flux Schwinger boson($\pi$-flux Abrikosov fermion) states
and the
$\pi$-flux Schwinger boson(zero-flux Abrikosov fermion) states
respectively.
However the nearest-neighbor RVB states on square lattice 
are not $Z_2$ spin liquids\cite{Wen-PRB02},
and they cannot distinguish different mean-field ansatz with
the same flux in elementary plaquette.
To represent $Z_2$ spin liquids and distinguish all PSG classes, 
valence bonds beyond nearest-neighbors will be required.
We have checked on small lattices that the exact mapping for the nearest-neighbor RVB states\cite{Read-Chakraborty-PRB99,Yunoki-Sorella-PRB06,YangFan-PRL12,Qi-Fu-PRB15}
cannot be established with long range valence bonds.

In summary, we have studied Schwinger boson spin liquid states on square lattice
by the projective symmetry group analysis and Schwinger boson mean-field theory.
Six symmetric $Z_2$ spin liquid states, two zero-flux states and four $\pi$-flux states, 
have been identified to be possibly relevant to the $J_1$-$J_2$ model.
The zero-flux states can go through continuous phase transitions into
canted or collinear N\'eel order. 
However these magnetic orders always show nonzero 
vector spin chirality $\langle \mathbf{S}_i\times \mathbf{S}_j\rangle$,
even if the spin expectation values are collinear.
The $\pi$-flux states are in proximity to certain 4-sublattice magnetic orders including the N\'eel order,
and can be favored energetically over zero-flux states with 
ring-exchange interactions\cite{WangF-Vishwanath-PRB06}.
At mean-field level, two of the $\pi$-flux states, the $Z_2[\pi,\pi]\mathcal{R}$ and $Z_2[\pi,\pi]\mathcal{I}$ states, 
may be stable against magnetic order for spin-1/2 and around $J_2/J_1\sim 1/2$.
We have computed the bottom of two-spinon excitation continuum and 
static spin structure factors for these spin liquid states,
which can be used in numerics and experiments as indirect evidences of these spin liquids.
Our results can be used in further theoretical studies of 
spin liquids on square lattice, for example in variational Monte Carlo calculations of 
projected bosonic spin liquid wave functions.

\acknowledgements
The authors thank Ashvin Vishwanath and Yuan-Ming Lu for helpful discussions,
and thank Shubhayu Chatterjee and Yuan-Ming Lu for pointing out two mistakes in previous version of this paper.
FW acknowledges support from 
National Key Basic Research Program of China (Grant No. 2014CB920902) and
National Natural Science Foundation of China(Grant No. 11374018).

\appendix

\section{Solution of the algebraic PSG}\label{sec:algebraicpsg}
In the following we will solve the algebraic PSGs by using the algebraic constraints on the generators of the square lattice space group.

For reasons discussed in the introduction, the low energy gauge group after Higgs condensation is $Z_2$, which is called the invariant gauge group (IGG) of the PSG.

The two elements of the IGG are identity operator and the generator 
\begin{equation}
b_{j\alpha}\rightarrow -b_{j\alpha}.
\end{equation}

For furture convenience, we introduce two difference operators $\Delta_x$ and $\Delta_y$ defined as $\Delta_x f(x,y)=f(x,y)-f(x-1,y)$ and $\Delta_yf(x,y)=f(x,y)-f(x,y-1)$.

The PSG is solved under certain gauge, therefore it is important to consider how PSG elements change if we do a gauge transformation to the ansatz.

When a gauge transformation $G:\ b_{j\alpha}\rightarrow e^{i\phi(j)}b_{j\alpha}$
is applied to the ansatz, the ansatz will be invariant under $GG_XXG^{-1}=GG_XXG^{-1}X^{-1}X$, therefore $G_X$ is now converted to $GG_XXG^{-1}X^{-1}$ and the phase functions changes according to:
\begin{equation}
\phi_X(\textbf{r})\rightarrow\phi_G(\textbf{r})+\phi_{X}(\textbf{r})-\phi_G[X^{-1}(\textbf{r})].
\end{equation}

With this recipe we can use the gauge freedom to do the gauge fixing procedure on a spanning tree (an open boundary condition is assumed), thus the gauge is fixed to be 
\begin{equation}
\phi_{T_1}(x,y)=0,\quad\phi_{T_2}(0,y)=0.
\end{equation}

The commutative relation $T_1^{-1}T_2T_1T_2^{-1}=\textbf{1}$, when translated into the PSG language, is
\begin{equation}
T_1^{-1}G_{T_1}^{-1}G_{T_2}T_2G_{T_1}T_1T_2^{-1}G_{T_2}^{-1}\in IGG,
\end{equation}
therefore $\Delta_x\phi_{T_2}(x,y)=p_1\pi$ (in this and next section, all the equations are true modulo $2\pi$).
Here the number $p_1$ is either 0 or 1 due to the $Z_2$ gauge structure. Integers $p_{i},i=2,3,\cdots 9$ appeared later are also $Z_2$ integers.

After gauge fixing procedure, we are still left with three gauge freedom which will not change $\phi_{T_1}$, $\phi_{T_2}$ up to IGG elements but may affect other PSG elements.

\begin{eqnarray}
G_1:&&\phi_1(x,y)= \text{const.},\\
G_2:&&\phi_2(x,y)=\pi x,\\
G_3:&&\phi_3(x,y)=\pi y.
\end{eqnarray}

We are left with commutative relations, 
$\sigma^{-1}T_1\sigma T_1^{-1}=\sigma^{-1}T_2\sigma T_2=\sigma^{2}=\sigma^{-1}C_4\sigma C_4=C_4^4=C_4^{-1}T_1C_4T_2=\textbf{1},$ 
and four commutative relations concerning time reversal symmetry, 
$\mathcal{T}^{-1}T_1\mathcal{T}T_1^{-1}=\mathcal{T}^{-1}T_2\mathcal{T}T_2^{-1}=\mathcal{T}^{-1}\sigma\mathcal{T}\sigma^{-1}=
\mathcal{T}^{-1}C_4\mathcal{T}C_4^{-1}=\textbf{1}$.

Using the condition 
$\sigma^{-1}T_1\sigma T_1^{-1}=\textbf{1}$, and $\sigma^{-1}T_2\sigma T_2=\textbf{1}$, we can obtain:
\begin{eqnarray}
\Delta_x\phi_{\sigma}(x,y) &=& p_2\pi,\\
\Delta_y\phi_{\sigma}(x,y) &=& p_3\pi,
\end{eqnarray}
whose solution is $\phi_{\sigma}(x,y)=p_2\pi x+p_3\pi y+\phi_{\sigma}(0,0)$.

The constraint from $\sigma^2=\textbf{1}$ is $2\phi_{\sigma}(0,0)=p_4\pi$. Due the $Z_2$ gauge structure, an overall phase $\pi$ has no consequence and hence we can fix $\phi_{\sigma}(0,0)$ to be $p_4\pi/2$.

Therefore we have \begin{equation}
\phi_{\sigma}(x,y)=p_2\pi x+p_3\pi y+p_4\pi/2.
\end{equation}

The gauge condition $G_1$, $G_2$ and $G_3$ have no effect on it.

Consider the fourfold rotation $C_4$. 

From $C_4^{-1}T_1C_4T_2=\textbf{1}$ and $C_4^{-1}T_2C_4T_1^{-1}=\textbf{1}$ we have
\begin{eqnarray}
\Delta_x\phi_{C_4}&=&p_1\pi y+p_5\pi,\\
\Delta_y\phi_{C_4}&=&p_1\pi x+p_6\pi,
\end{eqnarray}
the solution of which is $\phi_{C_4}(x,y)=p_1\pi xy+p_5\pi x+p_6\pi y +\phi_{C_4}(0,0)$.

From $C_4\sigma C_4\sigma^{-1}=\textbf{1}$ we have 
\begin{equation}
p_2+p_3+p_5+p_6=0,\ 2\phi_{C_4}=p_7\pi.
\end{equation}

The gauge condition $G_2$ can tune $p_5$ to be zero. And the condition $C_4^4=\textbf{1}$ gives no new constraint.

Therefore we have
\begin{equation}
\phi_{C_4}(x,y)=p_1\pi xy +p_6\pi y+p_7\pi/2.
\end{equation}

As for time-reversal symmetry, the condition $\mathcal{T}^{-1}T_1\mathcal{T}T_1^{-1}=\textbf{1}$ and $\mathcal{T}^{-1}T_2\mathcal{T}T_2^{-1}=1$ yield
\begin{eqnarray}
\Delta_x\phi_{\mathcal{T}}&=&p_8\pi,\\
\Delta_y\phi_{\mathcal{T}}&=&p_9\pi.
\end{eqnarray}

Therefore we have
\begin{equation}
\phi_{\mathcal{T}}(x,y)=p_8\pi x+p_9\pi y+\phi_{\mathcal{T}}(0,0)
\end{equation}

The condition $\mathcal{T}^{-1}\sigma\mathcal{T}\sigma^{-1}=\textbf{1}$ produces no constraint.

The condition $\mathcal{T}^{-1}C_4\mathcal{T}C_4^{-1}=\textbf{1}$ requires that $p_8=p_9$.

Note that the gauge condition $G_1(b_{j\alpha})=e^{i\phi_1}b_{j\alpha}$ acts nontrivially in respect of the time reversal symmetry since
time reversal operator will change a number to its complex conjugate.
After gauge transformation $G_1$, the phase function changes as
\begin{equation}
\phi_{\mathcal{T}}(\textbf{r})\rightarrow \phi_{\mathcal{T}}(\textbf{r})+2\phi_1
\end{equation}

Therefore we may use the gauge condition $G_1$ to fix $\phi_{\mathcal{T}}(0,0)$ to be $-p_8\pi/2$, so we finally obtain
\begin{equation}
\phi_{\mathcal{T}}(\textbf{r})=p_8(x+y-1/2)\pi.
\end{equation}

Finally the algebraic solutions of PSG are:
\begin{eqnarray}
\phi_{T_1}&=&0,\\
\phi_{T_2}&=&p_1\pi x,\\
\phi_{\sigma}&=&p_2\pi x+p_3\pi y +\frac{p_4\pi}{2},\\
\phi_{C_4}&=&p_1\pi xy+(p_2+p_3)\pi y+\frac{p_7\pi}{2},\\
\phi_{\mathcal{T}}&=&p_8(x+y-1/2)\pi.
\end{eqnarray}

There are two choices for each $p_i(i=1,2,3,4,7,8)$, so total amounts of PSG solutions are $2^6=64$.

\section{Physical realizations of PSG on square lattice}\label{sec:realization}
PSG solved by the commutative relations of group generators is called the algebraic PSG since it relies only on the structure of the symmetry group and may not be realized by a mean-field ansatz.

Generally speaking, realization of a particular kind of ansatz on square lattice will impose further constraints on algebraic PSG, therefore it is necessary to analyze the constraints imposed on PSG when demanding that an arbitrary bond $u_{ij}$ ($u_{ij}$ represents $A_{ij}$ or $B_{ij}$) is nonvanishing.

We shall only consider bonds that start from the original point $(x,y)=(0,0)$ since other bonds can be obtained by translation.

\subsection{$u_{(0,0)\rightarrow(x,x)}$}
The bond $A_{(0,0)\rightarrow(x,x)}$ becomes $A_{(x,x)\rightarrow(0,0)}$ under symmetry operation $T_2^xT_1^xC_4^2$. Since $A_{ij}$ is antisymmetric, {\it i.e.} $A_{ij}=-A_{ji}$, we have:
\begin{equation}\label{B1}
[G_{T_2}T_2]^x[G_{T_1}T_1]^x[ G_{C_4} C_4]^2A_{(0,0)\rightarrow (x,x)}=-A_{(0,0)\rightarrow (x,x)},
\end{equation}
thus we have
\begin{equation}\label{B2}
\begin{split}
&2\phi_{C_4}(0,0)+\phi_{C_4} (-x,x)+\phi_{C_4}(-x,-x)\\
&+\sum\limits_{i=1}^x(\phi_{T_2}(0,-x+i)+\phi_{T_2}(x,i))=\pi,\\
\end{split}
\end{equation}
therefore $p_1\pi x^2=\pi$.\par
$A_{(0,0)\rightarrow(x,x)}$ is invariant under the combined operation $C_4\sigma$, so we have:
\begin{equation}\label{B3}
\begin{split}
&\phi_{C_4}(0,0)+\phi_{C_4}(x,x)+\phi_{\sigma}(0,0)+\phi_{\sigma}(x,-x)\\
&=p_1\pi x+(p_4+p_7)\pi=0.
\end{split}
\end{equation}

Therefore when $x$ is odd, the condition \eqref{B2} and \eqref{B3} are satisfied only when $p_1=p_4+p_7=1$, and when $x$ is even, the two conditions cannot be satisfied simultaneously and hence $A_{(0,0)\rightarrow (x,x)}$ vanishes.

$B_{(0,0)\rightarrow(x,x)}$ becomes $B^*_{(0,0)\rightarrow(x,x)}$ under $T_2^xT_1^xC_4^2$, so we have  
\begin{equation}\label{B4}
\begin{split}
&\phi_{C_4}(-x,x)+\phi_{C_4}(-x,-x)-2\phi_{C_4}(0,0)\\
&+\sum\limits_{i=1}^x(\phi_{T_2}(0,-x+i)-\phi_{T_2}(x,i)) \\
&=p_1\pi x^2=\text{Arg}(B^*_{(0,0)\rightarrow(x,x)}/B_{(0,0)\rightarrow(x,x)}).\\
\end{split}
\end{equation}

$B_{(0,0)\rightarrow(x,x)}$ is invariant under $C_4\sigma$, so we have
\begin{equation}\label{B5}
-\phi_{C_4}(0,0)+\phi_{C_4}(x,x)-\phi_{\sigma}(0,0)+\phi_{\sigma}(x,-x)=p_1\pi x^2=0.
\end{equation}

When $x$ is even, the condition \eqref{B4} and \eqref{B5} demand that $B_{(0,0)\rightarrow(x,x)}$ is real. When $x$ is odd, \eqref{B5} demands that $p_1=1$ and \eqref{B4} demands that $B_{(0,0)\rightarrow(x,x)}$ is real.

The results are summarized in Table~\ref{table4}.

\begin{table}[h]
\begin{tabular}[t]{|c|c|c|}\hline
 $ A_{(0,0)\rightarrow (x,x)}$  & $ B_{(0,0)\rightarrow (x,x)}$& $x$ (mod 2) \\
\hline
 vanish &  Im(B)=0 &  0    \\
\hline
 $p_1=p_4+p_7=1$ & $p_1=0$, Im(B)=0 &  1     \\
\hline
\end{tabular}
\caption{
Constraint imposed on PSG by nonvanishing bond $u_{(0,0)\rightarrow (x,x)}$
}\label{table4}
\end{table}

\subsection{$u_{(0,0)\rightarrow(x,0)}$}
The bond $A_{(0,0)\rightarrow(x,0)}$ is invariant under $\sigma$, so 
\begin{equation}\label{B6}
\phi_{\sigma}(0,0)+\phi_{\sigma}(x,0)=p_2\pi x+p_4 \pi=0.
\end{equation}

This bond becomes its inverse under $T_1^xC_4^2$, thus we have
\begin{equation}\label{B7}
\begin{split}
&2\phi_{C_4}(0,0)+\phi_{C_4} (0,x)+\phi_{C_4}(-x,0)\\
&=(p_2+p_3)\pi x=\pi.\\
\end{split}
\end{equation}

When $x$ is even, the condition \eqref{B7} cannot be satisfied. When $x$ is odd, the condition \eqref{B6} demands that $p_2+p_4=0$ and the condition \eqref{B7} demands that $p_2+p_3=1$.

$B_{(0,0)\rightarrow(x,0)}$ is invariant under $\sigma$, so 
\bea\label{B8}
-\phi_{\sigma}(0,0)+\phi_{\sigma}(x,0)=p_2\pi x=0.
\eea

This bond becomes its conjugate under $T_1^xC_4^2$, therefore we have 
\begin{equation}\label{B9}
\begin{split}
&-2\phi_{C_4}(0,0)+\phi_{C_4} (0,x)+\phi_{C_4}(-x,0)\\
&=\text{Arg}(B^*_{(0,0)\rightarrow(x,0)}/B_{(0,0)\rightarrow(x,0)}).\\
\end{split}
\end{equation}

When $x$ is even, we have Arg($B^*/B$)=0. When $x$ is odd, we know from \eqref{B8} that $p_2=0$. And the condition \eqref{B9} demands that Arg($B^*/B$)=$p_3\pi$.

The results are summarized in Table~\ref{table5}.

\begin{table}[h]
\begin{tabular}[t]{|c|c|c|}\hline
 $ A_{(0,0)\rightarrow (x,0)}$  & $ B_{(0,0)\rightarrow (x,0)}$& $x$ (mod 2) \\
\hline
vanish &  Im(B)=0 &  0    \\
\hline
 $p_2+p_3=1,p_4=p_2$ & 
\tabincell{c}{$p_2=p_3=0$, Im(B)=0 \\
or $p_2=0,p_3=1$, Re(B)=0} &  1     \\
\hline
\end{tabular}
\caption{
Constraint imposed on PSG by nonvanishing bond $u_{(0,0)\rightarrow (x,0)}$
}\label{table5}
\end{table}

\subsection{$u_{(0,0)\rightarrow(x,y)}$, $x\neq y$}
There is only one constraint:$$[G_{T_2}T_2]^y[G_{T_1}T_1]^x [G_{C_4} C_4]^2A_{(0,0)\rightarrow (x,y)}=-A_{(0,0)\rightarrow (x,y)},$$
which leads to 
\begin{equation}\label{B10}
\begin{split}
&2\phi_{C_4}(0,0)+\phi_{C_4}(-y,x)+\phi_{C_4}(-x,-y)\\
&+\sum\limits_{i=1}^y(\phi_{T_2}(0,-y+i)+\phi_{T_2}(x,i))\\
&=p_1\pi xy +(p_2+p_3)\pi (x-y)=\pi.\\
\end{split}
\end{equation}

When $x$ and $y$ are both even, the condition \eqref{B10} cannot be satisfied, therefore $A_{(0,0)\rightarrow(x,y)}$ vanishes. When $x+y$ is odd, the condition \eqref{B10} is satisfied when $p_2+p_3=1$. When $x$ and $y$ are both odd, the condition \eqref{B10} demands that $p_1=1$.

As for $B_{(0,0)\rightarrow(x,y)}$, we have
\begin{equation}\label{B11}
\begin{split}
&-2\phi_{C_4}(0,0)+\phi_{C_4}(-y,x)+\phi_{C_4}(-x,-y)\\
&+\sum\limits_{i=1}^y(\phi_{T_2}(0,-y+i)-\phi_{T_2}(x,i))\\
&=p_1\pi xy +(p_2+p_3)\pi (x-y)\\
&=\text{Arg}(B_{(0,0)\rightarrow(x,y)}^*/B_{(0,0)\rightarrow(x,y)}).\\
\end{split}
\end{equation}

When $x$ and $y$ are both even, the condition \eqref{B10} demands that $B_{(0,0)\rightarrow(x,y)}$ is real. When $x+y$ is odd, the condition \ref{B10} demands that Arg($B^*/B$)=$(p_2+p_3)\pi$. When $x$ and $y$ are both odd, the condition \eqref{B10} demands that Arg($B^*/B$)=$p_1\pi$.

The results are summarized in Table~\ref{table6}.

\begin{table}[h]\label{IV}
\begin{tabular}[t]{|c|c|c|c|}\hline
 $ A_{(0,0)\rightarrow (x,y)}$  & $ B_{(0,0)\rightarrow (x,y)}$& $x$ (mod 2)& $y$ (mod 2) \\
\hline
vanish &  Im($B$)=0 &  0 &0   \\ \hline
  \multirow{2}{*}{ $p_2+p_3=1$ }&  \multirow{2}{*}{ 
  \tabincell{c}{$p_2+p_3=0$, Im($B$)=0 \\ or $p_2+p_3=1$, Re($B$)=0}
  } 
  & 0&1     \\ \cline{3-4}
&  &  1 &0   \\ \hline
$p_1=1$ &  \tabincell{c}{$p_1=0$, Im($B$)=0 \\or $p_1=1$ Re($B$)=0} &  1 &1   \\
\hline
\end{tabular}
\caption{
Constraint imposed on PSG by nonvanishing bond $u_{(0,0)\rightarrow (x,y)}$
}\label{table6}
\end{table}

\section{Derivation of the magnetic order from the condensation of Schwinger boson in mean-field level}
\label{appendix:magneticorder}

\subsection{Canted N\'eel order from $Z_2[0,0]$ state}
\label{app:magneticorder-Z200}
For $Z_2[0,0]$ state, the spinon condensation happens at $\pm \textbf{Q}, \textbf{Q}=(\pi/2,-\pi/2)$ when $\mu=2B_{2}-2\sqrt{A_{1}^2+B_{1}^2}$.

The zero energy eigenvector of the mean-field Hamiltonian \eqref{zerozeromfH} at $\textbf{Q}=(\pi/2,-\pi/2)$ point is
\begin{equation}
\Psi_1=\begin{pmatrix}
\frac{-i A_{1}}{\sqrt{B_{1}^2+A_{1}^2}-B_{1}}\\
1
\end{pmatrix}.
\end{equation}

When boson condense, a ground state expectation value at $\textbf{Q}$ is obtained: $\langle \Psi(\textbf{Q}) \rangle=z\Psi_1$.

The zero energy eigenvector at $-\textbf{Q}$ point is
\begin{equation}
\Psi_2=\begin{pmatrix}
1\\
\frac{-i A_{1}}{\sqrt{B_{1}^2+A_{1}^2}-B_{1}}
\end{pmatrix}.
\end{equation}

And the condensation value is $\langle \Psi(-\textbf{Q})\rangle =\omega \Psi_2$.

Therefore we can represent the condensation on lattice site $\textbf{r}$ as:
\begin{equation}
x=\begin{pmatrix}
\langle b_{\textbf{r}\uparrow}\rangle\\
\langle b_{\textbf{r}\downarrow}\rangle
\end{pmatrix}
=\begin{pmatrix}
z & w\\
-w^* & z^*
\end{pmatrix}
\begin{pmatrix}
\frac{-i A_{1}}{\sqrt{B_{1}^2+A_{1}^2}-B_{1}}e^{i\textbf{Q}\cdot \textbf{r}}\\
e^{-i\textbf{Q}\cdot \textbf{r}}
\end{pmatrix}.
\end{equation}

The 2$\times$2 matrix is propotional to a SU(2) matrix and its effect on the order parameter $\langle \textbf{S}(\textbf{r})\rangle$ is just a SO(3) rotation.

The magnetic order obtained from the spinor $x$ therefore can be computed using
\begin{equation}\begin{split}
&\langle \textbf{S}(\textbf{r})\rangle=(1/2)x^{\dag}\sigma x\\
&=\frac{1+(-)^{x+y}}{2}\vec{n}_u+\frac{1-(-)^{x+y}}{2}\vec{n}_v,\\
\end{split}
\end{equation}

If we introduce the notion $s=\frac{-iA_{1}}{\sqrt{B_{1}^2+A_{1}^2}-B_{1}}$, the vector $n_u$ and $n_v$ can be conveniently represented as 
\bea
&\vec{n}_u=(0,2is,-1-s^2),\\
&\vec{n}_v=(0,-2is,-1-s^2).
\eea

And when $s\neq 0$, we have $\vec{n}_u\cdot \vec{n}_v/(|\vec{n}_u|\cdot |\vec{n}_v|)=\frac{2(|s|^2-1)^2}{(|s|^2+1)^2}-1 \textgreater -1 $. Therefore the magnetic order obtained from the $Z_2[0,0]$ spin liquid is a non-collinear order.

\subsection{N\'eel order from $Z_2[0,\pi]$ state}
\label{app:magneticorder-Z20pi}
For $Z_2[0,\pi]$ state, the spinon condensation happens at $\pm \textbf{Q}, \textbf{Q}=(\pi/2,\pi/2)$ when $\mu=-2A_{1}-2B_{2}$.

The zero energy eigenvector at the $\textbf{Q}=(\pi/2,\pi/2)$ point is
\begin{equation}
\Psi_1=\begin{pmatrix}
1\\
i
\end{pmatrix}.
\end{equation}

When boson condense, the condensation value at $\textbf{Q}$ is $\langle \Psi(\textbf{Q}) \rangle=z\Psi_1.$

The zero energy eigenvector at $-\textbf{Q}$ point is
\begin{equation}
\Psi_2=\begin{pmatrix}
1\\
-i
\end{pmatrix}.
\end{equation}

And the condensation value is $\langle \Psi(-\textbf{Q})\rangle =\omega \Psi_2$.

Therefore we can represent the condensation on lattice site $\textbf{r}$ as:
\begin{equation}
x=\begin{pmatrix}
\langle b_{\textbf{r}\uparrow}\rangle\\
\langle b_{\textbf{r}\downarrow}\rangle
\end{pmatrix}
=\begin{pmatrix}
z,&w\\
iw^*,&-iz^*
\end{pmatrix}
\begin{pmatrix}
e^{i\textbf{Q}\cdot \textbf{r}}\\
e^{-i\textbf{Q}\cdot \textbf{r}}
\end{pmatrix},
\end{equation}
where the 2$\times$2 matrix is propotional to a SU(2) matrix and its effect on the order parameter $\langle \textbf{S}(\textbf{r})\rangle$ is an SO(3) rotation.

The magnetic order obtained from the spinor $x$ is therefore
\begin{equation}\begin{split}
&\langle \textbf{S}(\textbf{r})\rangle=(1/2)x^{\dag}\sigma x\\
&=\text{cos}(2\textbf{Q}\cdot \textbf{r})\textbf{m}=(-)^{x+y}\textbf{m},\\
\end{split}
\end{equation}
which is the N\'eel order.

\subsection{Magnetic order from $\pi$-flux states}
\label{app:magneticorder-Z2pi}
For $\pi$ flux state with only nearest bond $A_{1}$, the spinon condensation occurs at $\pm \textbf{Q}, \textbf{Q}=(\pi/2,\pi/2)$ when $\mu=-2A_{1}$.

We devide the system into two sets of sublattice which are distinguished by $(-1)^y$ which are labeled by $u$ and $v$ separately.

At $\textbf{Q}$, the Hamiltonian has two linearly independent eigenvectors:
\begin{equation}
\begin{split}
&\Psi_1=(1,0,\frac{i}{\sqrt{2}},\frac{1}{\sqrt{2}})^T,\\
&\Psi_2=(0,1,-\frac{1}{\sqrt{2}},-\frac{i}{\sqrt{2}})^T.
\end{split}
\end{equation}

The condensate at $\textbf{Q}$ is the linear combination of the two eigenvectors 
\bea
\langle\Psi(\textbf{Q})\rangle=z_1\Psi_1+z_2\Psi_2.
\eea

At $-\textbf{Q}$ point, the two eigenvectors are:

\begin{equation}
\begin{split}
&\Psi_3=
(-\frac{i}{\sqrt{2}},\frac{1}{\sqrt{2}},-1,0)^T,\\
&\Psi_4=(-\frac{1}{\sqrt{2}},\frac{i}{\sqrt{2}},0,-1)^T.
\end{split}
\end{equation}

The condensation at $-\textbf{Q}$ is the linear combination of $\Psi_3$ and $\Psi_4$
\bea
\langle\Psi(-\textbf{Q})\rangle=\omega_{1}\Psi_3+\omega_{2}\Psi_4.
\eea

In real space, we have
\begin{equation}
\begin{pmatrix}
\langle b_{u\textbf{r}\uparrow}\rangle\\
\langle b_{v\textbf{r}\uparrow}\rangle\\
\langle b_{u\textbf{r}\downarrow}^*\rangle\\
\langle b_{v\textbf{r}\downarrow}^*\rangle
\end{pmatrix}
=
e^{i\textbf{Q}\cdot \textbf{r}}[z_1\Psi_1+z_2\Psi_2]+e^{-i\textbf{Q}\cdot \textbf{r}}[\omega_{1}\Psi_3+\omega_{2}\Psi_4].
\end{equation}

Therefore the condensate on lattice $\textbf{r}$ can be represented as:

\begin{equation}\label{C14}
\begin{split}
&x_u\equiv \begin{pmatrix}
\langle b_{u\textbf{r}\uparrow}\rangle\\
\langle b_{u\textbf{r}\downarrow}\rangle
\end{pmatrix}\\
&=\begin{pmatrix}
z_1,&-\frac{i}{\sqrt{2}}w_1-\frac{1}{\sqrt{2}}w_2\\
-w_1^*,&\frac{i}{\sqrt{2}}z_1^*-\frac{1}{\sqrt{2}}z_2^*
\end{pmatrix}\begin{pmatrix}
e^{i\textbf{Q}\cdot \textbf{r}}\\
e^{-i\textbf{Q}\cdot \textbf{r}}
\end{pmatrix},
\end{split}
\end{equation}

\begin{equation}\label{C15}
\begin{split}
&
x_v\equiv \begin{pmatrix}
\langle b_{v\textbf{r}\uparrow}\rangle\\
\langle b_{v\textbf{r}\downarrow}\rangle
\end{pmatrix}\\
&=\begin{pmatrix}
z_1,&-\frac{i}{\sqrt{2}}w_1-\frac{1}{\sqrt{2}}w_2\\
-w_1^*,&\frac{i}{\sqrt{2}}z_1^*-\frac{1}{\sqrt{2}}z_2^*
\end{pmatrix}\begin{pmatrix}
e^{i\textbf{Q}\cdot \textbf{r}}\\
e^{-i\textbf{Q}\cdot \textbf{r}}
\end{pmatrix}.
\end{split}
\end{equation}

In order to have uniform magnitude for the ordered moments (1/2)$x_{u,v}^{\dag}\bm{\sigma} x_{u,v}$, we have to impose the constraint $x_u^{\dag}x_u=x_v^{\dag}x_v=const$. 

From \eqref{C14} and \eqref{C15} we find that there are four distinct spinors in a plaquette which are named as $\mathcal{Z}_1$ if $x$ is even and $y$ is even, $\mathcal{Z}_2$ if $x$ is odd and $y$ is even, $\mathcal{Z}_3$ if $x$ is odd and $y$ is odd and $\mathcal{Z}_4$ if $x$ is even and $y$ is odd. Only two of them are linearly independent.

Therefore we choose two linearly independent spinors as:

\bea
&\psi=\mathcal{Z}_1=\begin{pmatrix}
z_1-\frac{i}{\sqrt{2}}w_1-\frac{1}{\sqrt{2}}w_2\\
-w_1^*-\frac{i}{\sqrt{2}}z_1^*-\frac{1}{\sqrt{2}}z_2^*
\end{pmatrix}\\
&i\chi=\mathcal{Z}_2=i\begin{pmatrix}
z_1+\frac{i}{\sqrt{2}}w_1+\frac{1}{\sqrt{2}}w_2\\
-w_1^*+\frac{i}{\sqrt{2}}z_1^*+\frac{1}{\sqrt{2}}z_2^*
\end{pmatrix}\\
\eea

Therefore $\mathcal{Z}_3$ and $\mathcal{Z}_4$ can be represented as:
\bea
&\mathcal{Z}_3
=\begin{pmatrix}
z_2+\frac{1}{\sqrt{2}}w_1+\frac{i}{\sqrt{2}}w_2\\
-w_2^*+\frac{1}{\sqrt{2}}z_1^*+\frac{i}{\sqrt{2}}z_2^*
\end{pmatrix}
\\
&=-\sqrt{2}i\sigma_y\psi^*+i\chi,\\
&\mathcal{Z}_4=i\begin{pmatrix}
z_2-\frac{1}{\sqrt{2}}w_1-\frac{i}{\sqrt{2}}w_2\\
-w_2^*-\frac{1}{\sqrt{2}}z_1^*-\frac{i}{\sqrt{2}}z_2^*
\end{pmatrix}\\
&=\sqrt{2}i\sigma_y\chi^*+i\psi.
\eea

The constraints that condensate on every site is uniform require that
\beq\label{23}
|\psi|^2=|i\chi|^2=|-\sqrt{2}i\sigma_y\psi^*+i\chi|^2=|\sqrt{2}i\sigma_y\chi^*+i\psi|^2.
\eeq

The constraints can be simplified if we define two spinors
\bea
&\phi_{+}=\sqrt{2+\sqrt{2}}(\psi+\sigma_y \chi^*),\\
&\phi_{-}=\sqrt{2-\sqrt{2}}(\psi-\sigma_y\chi^*).
\eea

Therefore we have:
\bea
\mathcal{Z}_1&=&\frac{1}{2\sqrt{2}}(\sqrt{2-\sqrt{2}}\phi_{+}+\sqrt{2+\sqrt{2}}\phi_{-}),\\
\mathcal{Z}_2&=&\frac{1}{2\sqrt{2}}i\sigma_y(-\sqrt{2-\sqrt{2}}\phi_{+}^*+\sqrt{2+\sqrt{2}}\phi_{-}^*),\\
\mathcal{Z}_3&=&\frac{1}{2\sqrt{2}}i\sigma_y(-\sqrt{2+\sqrt{2}}\phi_{+}^*-\sqrt{2-\sqrt{2}}\phi_{-}^*),\\
\mathcal{Z}_4&=&\frac{1}{2\sqrt{2}}(\sqrt{2+\sqrt{2}}\phi_{+}-\sqrt{2-\sqrt{2}}\phi_{-}).
\eea

And the constraints \label{constraint} now become 
\begin{equation}
\begin{split}
|\phi_{+}|^2=|\phi_{-}|^2,\ \text{Re}(\phi_{+}^{\dagger}\phi_{-})=0.
\end{split}
\end{equation}

Define vector order parameter on the four sites as
$\vec{n}_{i}=\mathcal{Z}_i^{\dagger}\bm{\sigma}\mathcal{Z}_i$, ($i=1,2,3,4$). It is easy to verify that $\sum\limits_{i}\vec{n}_i=0$. Therefore we can parameterize vector order parameter on every site as: 
\bea
\vec{n}=(-)^x\vec{n}_{(\pi,0)}+(-)^y\vec{n}_{(0,\pi)}+(-)^{x+y}\vec{n}_{(\pi,\pi)}.
\eea
The three vectors, $\vec{n}_{(\pi,0)}$, $\vec{n}_{(0,\pi)}$ and $\vec{n}_{(\pi,\pi)}$, corresponds to 
$\textbf{m}_1$, $\textbf{m}_2$ and $\textbf{m}_3$ in main text respectively.

The three order parameters $\vec{n}_{(\pi,0)}$, $\vec{n}_{(0,\pi)}$ and $\vec{n}_{(\pi,\pi)}$ can be represented with two spinors $\phi_{+}$ and $\phi_{-}$
\begin{equation}
\begin{split}
&\vec{n}_{(\pi,0)}=((\vec{n}_1+\vec{n}_4)-(\vec{n}_2+\vec{n}_3))/4\\
&=\frac{1}{4}(\vec{n}_{\phi_{+}}+\vec{n}_{\phi_{-}}),\\
&\vec{n}_{(0,\pi)}=((\vec{n}_1+\vec{n}_2)-(\vec{n}_3+\vec{n}_4))/4\\
&=\frac{\sqrt{2}}{8}(\phi_{+}^{\dagger}\vec{\sigma}\phi_{-}+\phi_{-}^{\dagger}\vec{\sigma}\phi_{+}),\\
&\vec{n}_{(\pi,\pi)}=((\vec{n}_1+\vec{n}_3)-(\vec{n}_2+\vec{n}_4))/4\\
&=\frac{\sqrt{2}}{8}(-\vec{n}_{\phi_+}+\vec{n}_{\phi_-}),
\end{split}
\end{equation}
where we have defined $\vec{n}_{\phi_{\pm}}=\phi_{\pm}^{\dagger}\vec{\sigma}\phi_{\pm}$.

The manifold of the order parameter can be described as follows. Given two spinors $\phi_{+}$ and $\phi_{-}$  with the same magnitude, we can perform SU(2) transformations to them separately without altering the condition $|\phi_{+}|^2=|\phi_{-}|^2$. And then we can tune the relative U(1) phase between the two spinors to meet the condition $\text{Re}(\phi_{+}^{\dagger}\phi_{-})=0$.

In principle, the transformation rule of the spinor fields $\phi_{+}$ and $\phi_{-}$ under space group operation can be worked out through the corresponding transformation rule of bosonic field under PSG.

It is easy to deduce from constraints~\eqref{23} that the intersection angle between magnetic moments on neighboring sites should always be greater than $90^{\circ}$. And note that the N\'eel state is not forbidden by this condition and therefore is one of possible magnetic ordered states that can be obtained from the $\pi$-flux state through boson condensation.

\section{Derivation of the continuum-field theory for the transition from $Z_2$ spin liquid states to N\'eel ordered states}
In this Appendix we derive the continuum-field theory for the two $Z_2$ spin liquid states from the microscopic Hamiltonian in the long wavelength limit. 

We divide sublattice $u$ and $v$ with different parity $(-)^{x+y}$: $u$ sublattice is parity-even and $v$ is parity-odd.

Following Read and Sachdev's prescription\cite{Read-Sachdev-IJMPB91}, we represent the boson operator in terms of the two slow-varying fields 

\begin{equation}\label{C1}
\begin{split}
&b_{u,\textbf{r},\alpha}=\psi_{u,\alpha}(\textbf{r})e^{i Q\cdot \textbf{r}},\\
&b_{v,\textbf{r},\alpha}=-\sigma_{\alpha\beta}^y\psi_{v,\beta}^*(\textbf{r})e^{i Q\cdot \textbf{r}},
\end{split}
\end{equation}
where $Q$ represents the $k$ point where condensation of bosons happens.

The bond operator $\hat{A}_{1}$, $\hat{B}_{1}$, $\hat{B}_{2}$ therefore can be represented as:
Nearest bond $\hat{A}_{1}$:
\begin{equation}\label{C2}
\begin{split}
&\hat{A}_{\textbf{r},\textbf{r}'}=\frac{1}{2}(b_{u,\textbf{r},\uparrow}b_{v,\textbf{r}^{\prime},\downarrow}-b_{u,\textbf{r},\downarrow}b_{v,\textbf{r}^{\prime},\uparrow})\\
&=-\frac{i}{2}e^{iQ\cdot \Delta \textbf{r}}\psi_{u,\alpha}[1+\Delta \textbf{r}\cdot \partial_{\textbf{r}}+\frac{(\Delta \textbf{r}\cdot \partial_{\textbf{r}})^2}{2}+\dots]\psi_{v,\alpha}^*.\\
\end{split}
\end{equation}

Nearest bond $\hat{B}_{1}$:
\begin{equation}\begin{split}
&\hat{B}_{\textbf{r},\textbf{r}'}=\frac{1}{2}b_{u,\textbf{r},\alpha}^{\dagger}b_{v,\textbf{r}^{\prime},\alpha}\\
&=-\frac{1}{2}e^{iQ\cdot\Delta \textbf{r}}\sigma_{\alpha\beta}^y\psi_{u,\alpha}^*[1+\Delta \textbf{r}\cdot\partial_{\textbf{r}}+\frac{(\Delta \textbf{r}\cdot \partial_{\textbf{r}})^2}{2}+\dots]\psi_{v,\beta}^*.\\
\end{split}
\end{equation}

Nex-nearest bond $\hat{B}_{2}$:
\begin{equation}\label{C5}
\begin{split}
&\frac{1}{2}b_{u,\textbf{r},\alpha}^{\dagger}b_{u,\textbf{r}^{\prime},\alpha}\\
&=\frac{1}{2}e^{iQ\cdot \Delta \textbf{r}}\psi_{u,\alpha}^*[1+\Delta \textbf{r}\cdot\partial_{\textbf{r}}+\frac{(\Delta \textbf{r}\cdot \partial_{\textbf{r}})^2}{2}+\dots]\psi_{u,\alpha},\\
\end{split}
\end{equation}
and 
\begin{equation}\begin{split}
&\frac{1}{2}b_{v,\textbf{r},\alpha}^{\dagger}b_{v,\textbf{r}^{\prime},\alpha}\\
&=\frac{1}{2}e^{iQ\cdot \Delta \textbf{r}}\psi_{v,\alpha}^*[1+\Delta \textbf{r}\cdot\partial_{\textbf{r}}+\frac{(\Delta \textbf{r}\cdot \partial_{\textbf{r}})^2}{2}+\dots]\psi_{v,\alpha},\\
\end{split}
\end{equation}
where $\Delta \textbf{r}=\textbf{r}'-\textbf{r}$.

\subsection{$Z_2[0,0]$ state}
The minima are located at $\pm (\pi/2,-\pi/2)$. Therefore we choose $Q=(\pi/2,-\pi/2)$ in \eqref{C1}.

Inserting \eqref{C2}-\eqref{C5} into the mean-field Hamiltonian \eqref{zerozeromfH}, we obtain:

\begin{equation}\begin{split}
&\mathcal{L}=\int \frac{d^2\textbf{r}}{a^2}\{\psi_{u,\alpha}^{*}\frac{d}{d\tau}\psi_{u,\alpha}-\psi_{v,\alpha}^{*}\frac{d}{d\tau}\psi_{v,\alpha}-\mu(\psi_{u,\alpha}^*\psi_{u,\alpha}\\
&+\psi_{v,\alpha}^*\psi_{v,\alpha})
-2A_{1}\psi_{u,\alpha}\psi_{v,\alpha}^*+\frac{a^2}{2}A_{1}\partial_{\textbf{r}}\psi_{u,\alpha}\cdot \partial_{\textbf{r}}\psi_{v,\alpha}^*+c.c.\\
&-2B_{1}^*i\sigma^y_{\alpha\beta}\psi_{u,\alpha}^*\psi_{v,\beta}^*
+\frac{a^2}{2}B_{1}^*i\sigma^y_{\alpha\beta}\partial_{\textbf{r}}\psi_{u,\alpha}^*\cdot\partial_{\textbf{r}}\psi_{v,\beta}^*
+c.c.\\
&+2B_{2}(\psi_{u,\alpha}^*\psi_{u,\alpha}+
\psi_{v,\alpha}^*\psi_{v,\alpha})
-a^2B_{2}(\partial_{\textbf{r}}\psi_{u,\alpha}^*\cdot\partial_{\textbf{r}}\psi_{u,\alpha}\\
&+\partial_{\textbf{r}}\psi_{v,\alpha}^*\cdot\partial_{\textbf{r}}\psi_{v,\alpha})
 \}. \\
\end{split}
\end{equation}

Note that terms with odd spatial derivatives vanish due to the geometry of the square lattice.

Introduce two fields:
\begin{equation}\label{C7}
z_{\alpha}=(\psi_{u,\alpha}+\psi_{v,\alpha})/2,
\pi_{\alpha}=(\psi_{u,\alpha}-\psi_{v,\alpha})/2.
\end{equation}
The Lagrangian now becomes

\begin{equation}\begin{split}
&\mathcal{L}=\int \frac{d^2\textbf{r}}{a^2}\{2z^*_{\alpha}\frac{d}{d\tau}\pi_{\alpha}+2\pi^*_{\alpha}\frac{d}{d\tau}z_{\alpha}\\
&+(-2\mu-4A_{1}+4B_{2})z^*_{\alpha}z_{\alpha}
+(-2\mu+4A_{1}+4B_{2})\pi^*_{\alpha}\pi_{\alpha}\\
&+a^2(A_{1}-2B_{2})\partial_{\textbf{r}}z^*_{\alpha}\cdot\partial_{\textbf{r}}z_{\alpha}\\
&-{a^2}B^*_1i\sigma_{\alpha\beta}^y\partial_{\textbf{r}}z^*_{\alpha}\cdot\partial_{\textbf{r}}\pi^*_{\beta}+c.c.
+4B^*_1z_{\alpha}^*i\sigma_{\alpha\beta}^y\pi_{\beta}^*+c.c.\},
\\ 
\end{split}
\end{equation}
where terms involving $\pi$ field and spatial derivatives are omitted since they will generate terms with fourth or higher power of spatial derivatives of $z$ field after integrating out $\pi$ field.

Fields $\pi_{\alpha}$ have a large mass gap $-2\mu+4A_{1}+4B_{2}$ and can be safely integrated out. The low energy effective Lagrangian after integration is :

\begin{equation}\begin{split}
&\mathcal{L}=\int d^2\textbf{r}\{\frac{2}{(-\mu+2A_{1}+2B_{2})a^2}\partial_{\tau}z^*_{\alpha}\cdot \partial_{\tau}z_{\alpha}
\\
&+(A_{1}-2B_{2}+\frac{4|B_{1}|^2}{-\mu+2A_{1}+2B_{2}})\partial_{\textbf{r}}z^*_{\alpha}\cdot
\partial_{\textbf{r}}z_{\alpha}\\
&+(\frac{-2\mu -4A_{1}+4B_{2}}{a^2}-\frac{8}{(-\mu+2A_{1}+2B_{2})a^2}|B_{1}|^2)z^*_{\alpha}z_{\alpha}\\
&+\frac{4}{(-\mu+2A_{1}+2B_{2})a^2}B_{1}\frac{d}{d\tau}z_{\alpha}i\sigma_{\alpha\beta}^yz_{\beta}+c.c.\}.
\\
\end{split}
\end{equation}

Note that the Higgs term $B_{1}\frac{d}{d\tau}z_{\alpha}i\sigma_{\alpha\beta}^y z_{\beta}+c.c.$ plays the role of the vector spin chirality order parameter for reasons discussed in the text.

The Lagrangian can be further simplified after introducing two fields $\omega_{\alpha},\ \alpha=1,2$  (here we assume $\text{Im}(B_{1})\textgreater 0$),
\begin{equation}\begin{split}
&\omega_{1}=\frac{z_{\uparrow}+z_{\downarrow}+i(z_{\uparrow}^*-z_{\downarrow}^*)}{2},\\
&\omega_{2}=\frac{z_{\uparrow}^*+z_{\downarrow}^*+i(z_{\uparrow}-z_{\downarrow})}{2},\\
\end{split}
\label{equ:omega-vs-z}
\end{equation}
and has the form
\begin{equation}\begin{split}
&\mathcal{L}=\int d^2\textbf{r}\{\frac{2}{(-\mu+2A_{1}+2B_{2})a^2}\partial_{\tau}\omega^*_{\alpha}\cdot \partial_{\tau}\omega_{\alpha}
\\
&+(A_{1}-2B_{2}+\frac{4|B_{1}|^2}{-\mu+2A_{1}+2B_{2}})\partial_{\textbf{r}}\omega^*_{\alpha}\cdot
\partial_{\textbf{r}}\omega_{\alpha}\\
&+(\frac{-2\mu -4A_{1}+4B_{2}}{a^2}-\frac{8}{(-\mu+2A_{1}+2B_{2})a^2}|B_{1}|^2)\omega^*_{\alpha}\omega_{\alpha}\\
&+\frac{8}{(-\mu+2A_{1}+2B_{2})a^2}\text{Im}(B_{1})\omega^*_{\alpha}\frac{d}{d\tau}\omega_{\alpha}\}.
\\
\end{split}
\end{equation}

Therefore the Lagrangian will flow to a new fixed point where space and time scales differently. The dispersion relation is now $\omega \propto k^2$ with dynamical critical component $z=2$. 

The term with quadratic powers of time derivatives is irrelevant and hence can be dropped.

\subsection{$Z_2[0,\pi]$ state}
In this case the minima of the spinon dispersion are located at $\pm Q$ where $Q=(\pi/2,\pi/2)$.

Inserting \eqref{C2}-\eqref{C5} into the mean field Hamiltonian, we therefore obtain:
\begin{equation}\begin{split}
&\mathcal{L}=\int \frac{d^2\textbf{r}}{a^2}\{ \psi_{u,\alpha}^{*}\frac{d}{d\tau}\psi_{u,\alpha}-\psi_{v,\alpha}^{*}\frac{d}{d\tau}\psi_{v,\alpha}-\mu(\psi_{u,\alpha}^*\psi_{u,\alpha}\\
&+\psi_{v,\alpha}^*\psi_{v,\alpha})
-2A_{1}\psi_{u,\alpha}\psi_{v,\alpha}^*+\frac{a^2}{2}A_{1}\partial_{\textbf{r}}\psi_{u,\alpha}\cdot \partial_{\textbf{r}}\psi_{v,\alpha}^*+c.c.\\
&+\frac{a^2}{2}B_{1}^*i\sigma^y_{\alpha\beta}\psi_{u,\alpha}^*[\partial_y^2-\partial_x^2]\psi_{v,\beta}^*
+c.c.\\
&-2B_{2}(\psi_{u,\alpha}^*\psi_{u,\alpha}
+\psi_{v,\alpha}^*\psi_{v,\alpha})
+a^2B_{2}(\partial_{\textbf{r}}\psi_{u,\alpha}^*\cdot\partial_{\textbf{r}}\psi_{u,\alpha}\\
&+\partial_{\textbf{r}}\psi_{v,\alpha}^*\cdot\partial_{\textbf{r}}\psi_{v,\alpha}) \}.
 \\ 
\end{split}
\end{equation}

Introduce two fields $z$ and $\pi$ defined in \eqref{C7}, the Lagrangian becomes:

\begin{equation}\begin{split}
&\mathcal{L}=\int \frac{d^2\textbf{r}}{a^2}\{2z^*_{\alpha}\frac{d}{d\tau}\pi_{\alpha}+2\pi^*_{\alpha}\frac{d}{d\tau}z_{\alpha}
\\&+(-2\mu -4A_{1}-4B_{2})z^*_{\alpha}z_{\alpha}+(-2\mu+4A_{1}-4B_{2})\pi^*_{\alpha}\pi_{\alpha}\\
&+a^2(A_{1}+2B_{2})\partial_{\textbf{r}}z^*_{\alpha}\cdot\partial_{\textbf{r}}z_{\alpha}\\
&-{a^2}B^*_1i\sigma_{\alpha\beta}^y z^*_{\alpha}[\partial_y^2-\partial_x^2]\pi^*_{\beta}+c.c.\}.
\\ 
\end{split}
\end{equation}

After integrating out $\pi$ field, we obtain a low energy effective Lagrangian:

\begin{equation}\begin{split}
&\mathcal{L}=\int d^2\textbf{r}\{\frac{2}{(-\mu+2A_{1}-2B_{2})a^2}\partial_{\tau}z^*_{\alpha}\cdot \partial_{\tau}z_{\alpha}\\
&+(A_{1}+2B_{2})\partial_{\textbf{r}}z^*_{\alpha}\cdot
\partial_{\textbf{r}}z_{\alpha}
+\frac{(-2\mu -4A_{1}-4B_{2})}{a^2}z^*_{\alpha}z_{\alpha} \}.
\\
\end{split}
\end{equation}

Here the Higgs terms consist of $B_{1}$ have cubic powers of time and spatial derivatives or higher and hence is irrelevant by naive power counting.

After rescaling and restore the compact gauge field, we obtain an effective field theory cosists of a massive boson $z_{\alpha}$ coupled to a compact U(1) gauge field,
\begin{equation}
\mathcal{L}=\int d^2\textbf{r}[
|D_{\tau}z|^2+c^2|D_{\textbf{r}}z|^2+m^2|z|^2
].
\end{equation}
 The critical point of this theory is $\mu=-2A_{1}-2B_{2}$ consistent with the mean field solution and the spinon velocity $c$ is proportional to $\sqrt{A_{1}(A_{1}+2B_{2})}$.

\section{Calculation of vison PSG}
\label{app:visonPSG}

In this section we will derive the vison PSG for square lattice \cite{Huh-Punk-Sachdev-PRB11,Senthil-Fisher-PRB00,Sachdev-Vojta00,Jalabert-Sachdev-PRB91,Xu-Balents-PRB11}.

The dynamics of vison can be described by an odd Ising gauge theory in the dimer limit. After a duality transformation, the odd Ising gauge theory is transformed to a transverse field Ising model on the dual lattice.

Following Ref.~\onlinecite{Xu-Balents-PRB11}, the dynamics of vison can be described by a fully frustrated transverse-field Ising theory on the dual lattice.\par
\begin{equation}
H=\sum\limits_{i j}J_{ij}\tau^z_i\tau^z_j-\sum\limits_i K_i\tau_i^x\cdots,
\end{equation}
where the product of bonds around each elementary plaquette is 
\begin{equation}
\prod\limits_{\text{plaquette}}\text{sgn}(J_{ij})=-1.
\end{equation}

This Hamiltonian is invariant under $Z_2$ gauge transformation
\begin{equation}
\tau_i\rightarrow \eta_i\tau_i,\ K_i\rightarrow \eta_i K_i,\ J_{ij}\rightarrow \eta_i\eta_j J_{ij},
\end{equation}
where $\eta_i=\pm 1$.

\begin{figure}
\includegraphics[scale=0.42]{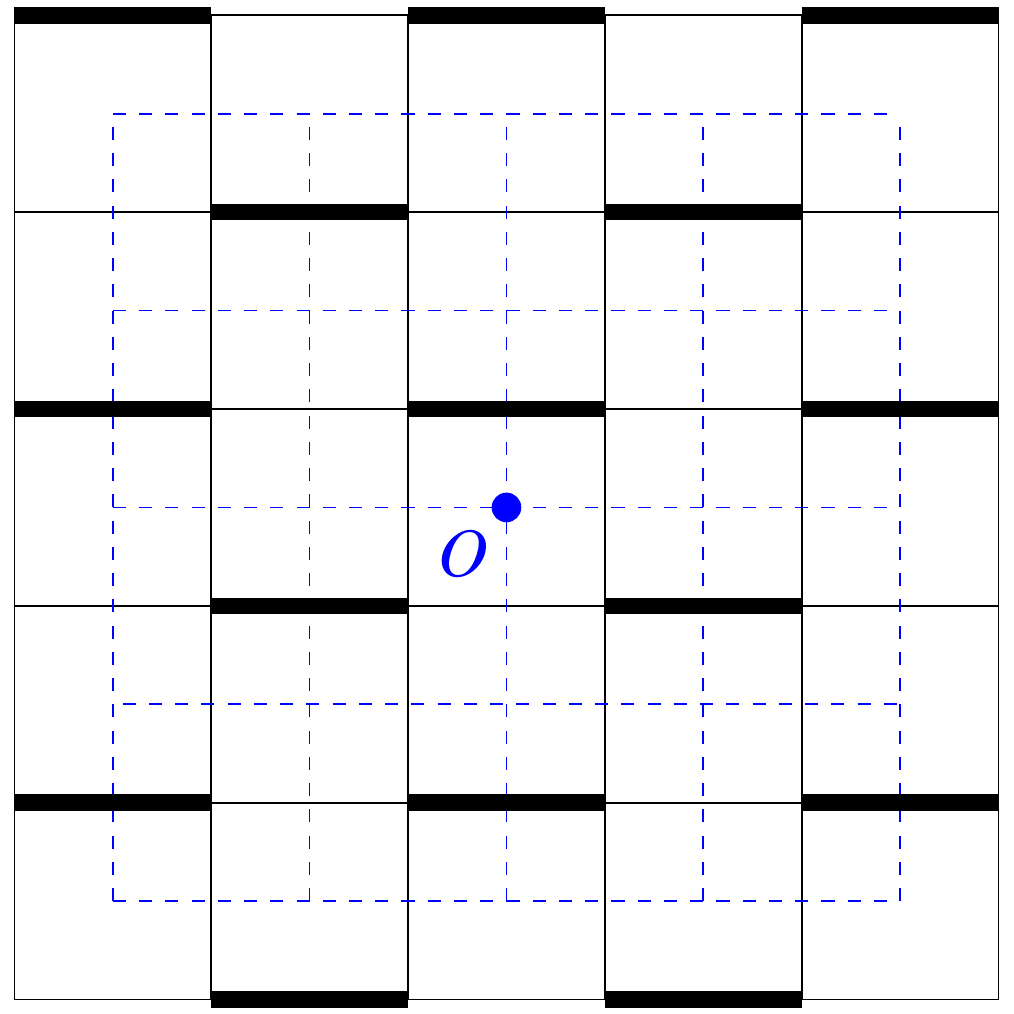}
\caption{\label{fig:gaugechoice}
Gauge choice on the dual lattice (solid black lines). The vison hopping $J_{ij}$ on the thick bonds are negative. The blue dashed line represents the original lattice. Note that all the point group symmetry operations are defined with respect to the original lattice site $O$.
}
\end{figure}

Due to the $Z_2$ gauge structure, we adopt a specific gauge choice (FIG.~\ref{fig:gaugechoice}) for $J_{ij}$ and calculate the vison PSG under this gauge.

\begin{figure}
\includegraphics[scale=0.5]{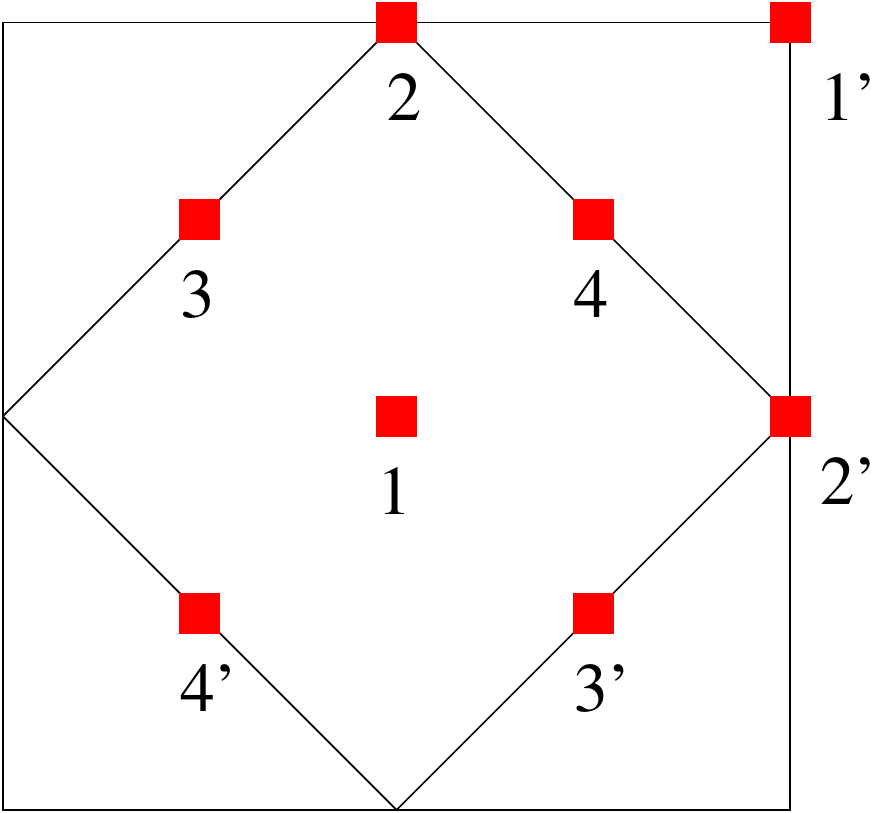}
\caption{
Brillouin zone of the square lattice,
and the eight momenta, labeled as $1,2,3,4,1',2',3',4'$ and indicated by red squares in the figure, for low energy vison modes.
}\label{fig14}
\end{figure}

In the following we shall ignore the kinetic term $K_i\tau_i^x$ and adopt a soft-spin formulation where the vison field $\tau_i^z$'s take real values. Considering nearest and fourth nearest neighbor interaction of the vison fields as considered in Ref.~\onlinecite{Xu-Balents-PRB11}, there are eight (rather than 4 because two sets of sublattice are not equivalent) inequivalent minimal points in the Brillouin zone (see Fig.~\ref{fig14}).
\begin{equation}
\begin{split}
&Q_1=(0,0),Q_2=(0,\pi),Q_3=(\frac{\pi}{2},\frac{\pi}{2}),Q_4=(\frac{\pi}{2},\frac{\pi}{2}) ,\\
&Q_{1'}=(\pi,\pi),Q_{2'}=(\pi,0),Q_{3'}=(\frac{\pi}{2},\frac{-\pi}{2}),Q_{4'}=(-\frac{\pi}{2},-\frac{\pi}{2})\\
\end{split}
\end{equation}

Expand the vison field with slow varying modes $\phi_a\ (a=1,2,3,4,1',2',3',4')$ at these eight momenta:
\begin{equation}
\tau^z=\sum\limits_{a}\phi_ae^{i\vec{Q}_{a}\cdot \vec{r}}.
\end{equation}

Under symmetry operation, the vison fields transform as  
\begin{eqnarray}
T_1:\tau^z_{x,y}&\rightarrow & (-)^{x+1}\tau^z_{x+1,y},\\
T_2:\tau^z_{x,y}&\rightarrow & (-)^{x}\tau^z_{x,y+1},\\
P_x:\tau^z_{x,y}&\rightarrow & (-)^x\tau^z_{x,1-y},\\
P_y:\tau^z_{x,y}&\rightarrow & \tau^z_{1-x,y},\\
P_{xy}:\tau^z_{x,y}&\rightarrow & \frac{(1-i)i^{x+y}+(1+i)i^{-x-y}}{2}\tau^z_{y,x}.
\end{eqnarray}

We choose base functions as $(\phi_1,\phi_2,\phi_3,\phi_4,\phi_{1'},\phi_{2'},\phi_{3'},\phi_{4'})^T$, the vison modes transform under smmetry operation as
\begin{equation}
\phi_n\rightarrow \sum\limits_{m=1}^{8}[O_{\phi}(g)]_{n,m}\phi_m.
\end{equation}\par
Therefore we can directly write out the transformation matrices 
\begin{equation}
O_{\phi}(T_1)=
\begin{pmatrix}
0& 0&0&0&0&-1&0&0  \\
0&0&0&0&-1&0&0&0 \\
0&0&0&i&0&0&0&0 \\
0&0&-i&0&0&0&0&0 \\
0&1&0&0&0&0&0&0\\
1&0&0&0&0&0&0&0\\
0&0&0&0&0&0&0&-i\\
0&0&0&0&0&0&i&0\\
\end{pmatrix},
\end{equation}

\begin{equation}
O_{\phi}(T_2)=
\begin{pmatrix}
0& 0&0&0&0&1&0&0  \\
0&0&0&0&-1&0&0&0 \\
0&0&0&i&0&0&0&0 \\
0&0&i&0&0&0&0&0 \\
0&-1&0&0&0&0&0&0\\
1&0&0&0&0&0&0&0\\
0&0&0&0&0&0&0&-i\\
0&0&0&0&0&0&-i&0\\
\end{pmatrix},\\
\end{equation}

\begin{equation}
O_{\phi}(P_x)=
\begin{pmatrix}
0& 0&0&0&0&1&0&0  \\
0&0&0&0&-1&0&0&0 \\
0&0&0&0&0&0&i&0 \\
0&0&0&0&0&0&0&i \\
0&-1&0&0&0&0&0&0\\
1&0&0&0&0&0&0&0\\
0&0&-i&0&0&0&0&0\\
0&0&0&-i&0&0&0&0\\
\end{pmatrix},\\
\end{equation}

\begin{equation}
O_{\phi}(P_y)=
\begin{pmatrix}
1& 0&0&0&0&0&0&0  \\
0&1&0&0&0&0&0&0 \\
0&0&0&-i&0&0&0&0 \\
0&0&i&0&0&0&0&0 \\
0&0&0&0&-1&0&0&0\\
0&0&0&0&0&-1&0&0\\
0&0&0&0&0&0&0&i\\
0&0&0&0&0&0&-i&0\\
\end{pmatrix},\\
\end{equation}
\\

\begin{equation}
\begin{split}
&O_{\phi}(P_{xy})=\\
&\frac{1}{2}
\begin{pmatrix}
 0 & 0 & 0 & 1-i & 0 & 0 & 0 & 1+i \\
 0 & 0 & 1-i & 0 & 0 & 0 & 1+i & 0 \\
 0 & 1+i & 0 & 0 & 0 & 1-i & 0 & 0 \\
 1+i & 0 & 0 & 0 & 1-i & 0 & 0 & 0 \\
 0 & 0 & 0 & 1+i & 0 & 0 & 0 & 1-i \\
 0 & 0 & 1+i & 0 & 0 & 0 & 1-i & 0 \\
 0 & 1-i & 0 & 0 & 0 & 1+i & 0 & 0 \\
 1-i & 0 & 0 & 0 & 1+i & 0 & 0 & 0 \\
\end{pmatrix}. \\
\end{split}
\end{equation}
\\
\\
\\
\\
\\
\\

In the following we may identify the operation $P_x$ with $\sigma$ and $P_{xy}P_x$ with $C_4$, therefore
\\
\\
\\
\\
\\

\begin{widetext}
\begin{equation}
\begin{split}
&O_{\phi}(C_4)=O_{\phi}(P_{xy})O_{\phi}(P_x)=\frac{1}{2} 
\begin{pmatrix}
 0 & 0 & 0 & 1-i & 0 & 0 & 0 & 1+i \\
 0 & 0 & 1-i & 0 & 0 & 0 & 1+i & 0 \\
 1-i & 0 & 0 & 0 & -1-i & 0 & 0 & 0 \\
 0 & -1+i & 0 & 0 & 0 & 1+i & 0 & 0 \\
 0 & 0 & 0 & -1-i & 0 & 0 & 0 & -1+i \\
 0 & 0 & -1-i & 0 & 0 & 0 & -1+i & 0 \\
 1+i & 0 & 0 & 0 & -1+i & 0 & 0 & 0 \\
 0 & -1-i & 0 & 0 & 0 & 1-i & 0 & 0 \\
\end{pmatrix}.\\
\end{split}
\end{equation}
\end{widetext}
The vison PSGs now can be easily obtained: 
\begin{equation}\label{eq:eps}
\begin{split}
&O_{\phi}(T_2)^{-1}O_{\phi}(T_1)O_{\phi}(T_2)O_{\phi}(T_1)^{-1}=-1,\\
&O_{\phi}(\sigma)^{-1}O_{\phi}(T_1)O_{\phi}(\sigma)O_{\phi}(T_1)^{-1}=-1,\\
&O_{\phi}(\sigma)^{-1}O_{\phi}(T_2)O_{\phi}(\sigma)O_{\phi}(T_2)=1,\\
&O_{\phi}(\sigma)^2=1,\\
&O_{\phi}(\sigma)^{-1}O_{\phi}(C_4)O_{\phi}(\sigma)O_{\phi}(C_4)=1,\\
&O_{\phi}(C_4)^4=-1,\\
&O_{\phi}(C_4)^{-1} O_{\phi}(T_1)  O_{\phi}(C_4)O_{\phi}(T_2)=-1,\\
&O_{\phi}(C_4)^{-1} O_{\phi}(T_2)  O_{\phi}(C_4)O_{\phi}(T_1)^{-1}=1.\\
\end{split}
\end{equation}

\section{Solutions of Abrikosov fermion PSG}
\label{app:fermionicPSG}
The Abrikosov fermion PSG on square lattice are studied in great detail in Ref.~\onlinecite{Wen-PRB02}. In this Appendix, we briefly summarize the algebraic solutions of fermionic PSGs and introduce some notations for convenience. 

The solutions to the fermionic PSGs are as follows:
\bea \label{F1}
G_{T_x}(\textbf{\emph{i}})&=&\eta_{xy}^{i_y}\tau_0, \\
G_{T_y}(\textbf{\emph{i}})&=&\tau_0, \\
G_{P_x}(\textbf{\emph{i}})&=&\eta_{xpx}^{i_x}\eta_{xpy}^{i_y}g_{P_x}, \\
G_{P_y}(\textbf{\emph{i}})&=&\eta_{xpy}^{i_x}\eta_{xpx}^{i_y}g_{P_y}, \\
G_{P_{xy}}(\textbf{\emph{i}})&=&\eta_{xy}^{i_xi_y}g_{P_{xy}}, \\
G_{\mathcal{T}}(\textbf{\emph{i}})&=&\eta_t^{i_x+i_y}g_{\mathcal{T}}. \label{F6}
\eea

In Wen's notation, $T_{x,y}$ stands for translation symmetry, and three parity operations $P_x$, $P_y$ and $P_{xy}$ acts as follows:
\bea\label{F7}
P_x: && (i_x,i_y)\rightarrow (-i_x,i_y),\\
P_y: && (i_x,i_y)\rightarrow (i_x,-i_y),\\
P_{xy}: && (i_x,i_y)\rightarrow (i_y,i_x).\label{F9}
\eea
stands for reflection along $y$ axis, reflection along $x$ axis and respectively.

Note that $C_4=P_{xy}P_y$, therefore we can obtain the PSG for $C_4$ from the product of $P_{xy}$ and $P_y$. And the $\sigma$ operation is just the operation $P_y$ in Wen's notation.

The fermionic PSGs are characterized by four numbers: $\eta_{xpx},\eta_{xpy},\eta_{xy},\eta_t$,
and numbers from the commutative relations of SU(2) matrices as defined in TABLE~\ref{table7}.

\begin{table}[h]
\begin{tabular}[t]{|c|c|}\hline
 Algebraic Identities  & $Z_2$ numbers from commutative relations\\
\hline
$\sigma^2=1$ & $[g_{P_y}]^2=\eta_{\sigma}$\\ \hline
$\sigma^{-1}C_4\sigma C_4=1$&$[g_{P_y}]^{-1}[g_{P_{xy}}g_{P_y}][g_{P_y}][g_{P_{xy}}g_{P_y}]=\eta_{\sigma C_4}$
\\ \hline
$C_4^4=1$&$[g_{P_{xy}}g_{P_y}]^4=\eta_{C_4}$\\ \hline
$C_4^{-1}\mathcal{T}^{-1}C_4\mathcal{T}=1$  & $[g_{P_{xy}}g_{P_y}]^{-1}g_{\mathcal{T}}^{-1}g_{P_{xy}}g_{P_y}g_{\mathcal{T}}=\eta_{C_4\mathcal{T}}$ \\ \hline
$\sigma^{-1}\mathcal{T}^{-1}\sigma\mathcal{T}=1$  & $g_{P_y}^{-1}g_{\mathcal{T}}^{-1}g_{P_y}g_{\mathcal{T}}=\eta_{\sigma\mathcal{T}}$ \\ \hline
\end{tabular}
\caption{
$Z_2$ numbers from the commutative relations between $SU(2)$ matrices. Here $\eta_i$ are numbers that can be $\pm 1$, and $g_i$ are $SU(2)$ matrices as defined in Eq.~\eqref{F1}-\eqref{F6}.
}\label{table7}
\end{table}

\section{Physical realization of Abrikosov fermion spin liquid states}\label{appendix:G}
In this Appendix we will analyze the physical realization of the fermionic PSGs as shown in TABLE~\ref{table:correspondenceofansatz}.
As discussed in Ref. \onlinecite{Wen-PRB02}, the mean-field ansatz $U_{ij}$ should take the form of \bea
U_{ij}=i\rho_{ij}W_{ij},
\eea (here $\rho_{ij}$ is a non-negative real number, and $W_{ij}\in SU(2)$) in order to describe a spin-rotational symmetric fermionic spin liquid state, and the time-reversal symmetry $\mathcal{T}$ is implemented as \bea\mathcal{T}: U_{ij}\rightarrow -U_{ij}.\eea

The symmetry operations are defined in Eq.~\eqref{F7}-\eqref{F9}.
\subsection{Realization of the $Z_2B\tau_{-}^1\tau_{-}^1\tau^1\tau_{-}^3$ state}
PSG elements of this state are:
\bea
&G_x(\bm{i})=(-)^{i_y}\tau_0;\\
&G_y(\bm{i})=\tau_0;\\
&G_{P_x}(\bm{i})=(-)^{i_x+i_y}i \tau_1;\\
&G_{P_y}(\bm{i})=(-)^{i_x+i_y}i \tau_1;\\
&G_{P_{xy}}(\bm{i})=(-)^{i_xi_y}i \tau_1;\\
&G_{\mathcal{T}}(\bm{i})=(-)^{i_x+i_y}i\tau_3.
\eea

There are two sites in a unit cell of $\pi$-flux fermionic spin liquid states distinguished by parity $(-)^x=\pm 1$, which are labeled by $u$ and $v$ respectively.

First, let us consider the Lagrangian multiplier $a_0^{l}$. Since the term $a_0^{l}\tau_l$ should be invariant under $G_{\mathcal{T}}\mathcal{T}$, we have $\{a_0^{l}\tau_l, \tau_3\}=0$. And it is also invariant under $G_{P_x}P_x,G_{P_y}P_y$ and $G_{P_{xy}}P_{xy}$, so we have $[a_0^{l}\tau_l, \tau_1]=0$. Hence we have $a_0^{1}\neq 0$ and $a_0^{2,3}=0$.

\subsubsection{Nearest-neighbor bond}
With translation along $x$ and $y$ direction, we can write down the general form of the nearest bonds:
\bea
&U_{i,i+\hat{x}}=u_{\hat{x}},\\
&U_{i,i+\hat{y}}=(-)^{i_x}u_{\hat{y}},
\eea
where $u_{\hat{x}}$ and $u_{\hat{y}}$ are site-independent.

Since $u_{\hat{x}}$ is invariant under time-reversal operation, we have $G_{\mathcal{T}}(i)u_{\hat{x}}G^{\dagger}_{\mathcal{T}}(i+\hat{x})=-u_{\hat{x}}$, which leads to $[u_{\hat{x}},\tau_3]=0$.

$u_{\hat{x}}$ is also invariant under $P_y$: $G_{P_y}(i)u_{\hat{x}}G^{\dagger}_{P_y}(i+\hat{x})=u_{\hat{x}},$
therefore $\{ u_{\hat{x}},\tau_1\}=0. $
Hence we have \begin{equation}
u_{\hat{x}}=\eta\tau_3.
\end{equation}

Under $P_{xy}$, $G_{P_{xy}}(i)u_{\hat{x}}G^{\dagger}_{P_{xy}}(i+\hat{x})=(-)^{i_y}u_{\hat{y}}$, therefore \begin{equation}
u_{\hat{y}}=-\eta\tau_3.
\end{equation}

\subsubsection{Next-nearest-neighbor bond}
The general form of the bonds is:
\bea\label{G12}
&U_{i,i+\hat{x}+\hat{y}}=(-)^{i_x}u_{\hat{x}+\hat{y}},\\
&U_{i,i+\hat{x}-\hat{y}}=(-)^{i_x}u_{\hat{x}-\hat{y}}.
\eea

Under $P_xP_y$, we have 
\begin{equation}\label{G14}
\begin{split}
&G_{P_x}P_xG_{P_y}P_yU_{O,O+\hat{x}+\hat{y}}[G_{P_x}P_xG_{P_y}P_y]^{\dagger}\\
&=U_{O,O-\hat{x}-\hat{y}},
\end{split}
\end{equation}
where $O$ represents the original point.

Therefore we have \begin{equation}
\label{G15}
u_{\hat{x}+\hat{y}}=U_{O,O+\hat{x}+\hat{y}}=U_{O-\hat{x}-\hat{y},O}=-u_{\hat{x}+\hat{y}},
\end{equation}
where the first and the third equation are due to Eq.~\eqref{G12}, and the second is due to Eq.~\eqref{G14} and the symmetry of the ansatz $U_{ij}=U_{ji}^{\dagger}$.

Finally we have $u_{\hat{x}+\hat{y}}=u_{\hat{x}-\hat{y}}=0$.

\subsubsection{Third neighbor bond}
The general form of the third neighbor bonds is:
\bea
&U_{i,i+2\hat{x}}=u_{2\hat{x}},\\
&U_{i,i+2\hat{y}}=u_{2\hat{y}}.
\eea

Considering time-reversal transformation $\mathcal{T}$, we immediately obtain $\{ u_{2\hat{x}},g_{\mathcal{T}}\}=\{u_{2\hat{y}},g_{\mathcal{T}}\}=0$.
Considering reflection $P_y$, we have $[u_{2\hat{x}},g_{P_y}]=[u_{2\hat{y}},g_{P_y}]=0$.
And under reflection $P_{xy}$, we have $u_{2\hat{y}}=i\tau_1u_{2\hat{x}}(i\tau_1)^{\dagger}$.

Therefore we have $u_{2\hat{x}}=u_{2\hat{y}}=\chi \tau_1$.

\subsubsection{Fourth neighbor bond}
The fourth neighbor is required to obtain a $Z_2$ spin liquid. The general form of the fourth neighbor is
\bea
&U_{i,i+2\hat{x}+\hat{y}}=(-)^{i_x}u_{2\hat{x}+\hat{y}},\\
&U_{i,i+2\hat{x}-\hat{y}}=(-)^{i_x}u_{2\hat{x}-\hat{y}},\\
&U_{i,i+\hat{x}+2\hat{y}}=u_{\hat{x}+2\hat{y}},\\
&U_{i,i+\hat{x}-2\hat{y}}=u_{\hat{x}-2\hat{y}}.
\eea

Considering time reversal transformation, we have $[u_{2\hat{x}+\hat{y}},g_{\mathcal{T}}]=0$.

For convenience we choose $u_{2\hat{x}+\hat{y}}$ to be $i\gamma \tau_0$, which ensures this state to be a $Z_2$ spin liquid state. The other bonds can be obtained under reflection $P_x$, $P_y$, and $P_{xy}$.

The results are
\bea
&U_{i,i+2\hat{x}+\hat{y}}=(-)^{i_x}i\gamma \tau_0,\\
&U_{i,i+2\hat{x}-\hat{y}}=-(-)^{i_x}i\gamma \tau_0,\\
&U_{i,i+\hat{x}+2\hat{y}}=i\gamma \tau_0,\\
&U_{i,i+\hat{x}-2\hat{y}}=-i\gamma \tau_0.
\eea
In conclusion, the symmetry allowed ansatz for the $Z_2B\tau_{-}^1\tau_{-}^1\tau^1\tau_{-}^3$ state are
\bea
&U_{i,i+\hat{x}}=   \eta \tau_3   , \\
&U_{i,i+\hat{y}}=-(-)^{i_x}\eta \tau_3,\\
&U_{i,i+2\hat{x}}=U_{i,i+2\hat{y}}=\chi\tau_1 , \\
&U_{i,i+2\hat{x}+\hat{y}}=(-)^{i_x}i\gamma \tau_0,\\
&U_{i,i+2\hat{x}-\hat{y}}=-(-)^{i_x}i\gamma \tau_0,\\
&U_{i,i+\hat{x}+2\hat{y}}=i\gamma \tau_0,\\
&U_{i,i+\hat{x}-2\hat{y}}=-i\gamma \tau_0,\\
&a_0^1\neq 0.
\eea

After Fourier transformation, the Hamiltonian becomes $H=\Psi_{\textbf{k}}^{\dagger}M\Psi_{\textbf{k}}$, where 
\bea
&M=2\eta \text{cos}(k_x)\Gamma_1-2\eta \text{cos}(k_y)\Gamma_2\\
&+\{a_0^1+2\chi [\text{cos}(2k_x)+\text{cos}(2k_y)]\}\Gamma_3\\
&+2\gamma[\text{sin}(k_x+2k_y)-\text{sin}(k_x-2k_y)]\Gamma_4\\
&+2\gamma[\text{sin}(2k_x+k_y)-\text{sin}(2k_x-k_y)]\Gamma_5,
\eea
where $\eta,\ \chi,\ \gamma$ are all real numbers, and we have used the Nambu spinor representation
\bea
\Psi_{\textbf{k}}=(f_{u\textbf{k}\uparrow},f_{v\textbf{k}\uparrow},f_{u-\textbf{k}\uparrow}^{\dagger},f_{v-\textbf{k}\downarrow}^{\dagger})^{T},
\eea
and \bea
&\Gamma_1=  \tau_3\otimes \tau_1 ,\\
&\Gamma_2= \tau_3\otimes \tau_3  , \\
&\Gamma_3= \tau_1\otimes \tau_0  , \\
&\Gamma_4=  \tau_0\otimes \tau_1 , \\ 
&\Gamma_5=   \tau_0\otimes \tau_3 .
\eea

Note that $k_x\in(0,\pi)$, $k_y\in(-\pi,\pi)$.

When there is only nearest-neighbor bond present, the state will have two Dirac points at wave-vectors $(k_x,k_y)=(\pi/2,\pm\pi/2)$.  By including the Lagrangian multiplier $a_0^1$, the third neighbor bond and the fourth neighbor bond, we find that two gaps open at the two Dirac points.

\subsection{Realization of the $Z_2B\tau_{-}^2\tau_{-}^2\tau^1\tau_{-}^3$ state}
PSG elements for this state are:
\bea
&G_x(\bm{i})=(-)^{i_y}\tau_0;\\
&G_y(\bm{i})=\tau_0;\\
&G_{P_x}(\bm{i})=(-)^{i_x+i_y}i \tau_2;\\
&G_{P_y}(\bm{i})=(-)^{i_x+i_y}i \tau_2;\\
&G_{P_{xy}}(\bm{i})=(-)^{i_xi_y}i \tau_1;\\
&G_{\mathcal{T}}(\bm{i})=(-)^{i_x+i_y}i\tau_3.
\eea

First, since the term $a_0^{l}\tau_l$ should be invariant under $G_{\mathcal{T}}\mathcal{T}$, we have $\{a_0^{l}\tau_l, \tau_3\}=0$. And it is also invariant under $G_{P_x}P_x$ and $G_{P_{xy}}P_{xy}$, so we have $[a_0^{l}\tau_l, \tau_1]=[a_0^{l}\tau_l, \tau_2]=0$. Hence the Lagrangian multipliers $a_0^{1,2,3}=0$.

The analysis of the symmetry allowed ansatz is similar to the previous case. A possible ansatz for the realization of a gapped $Z_2$ spin liquid state is

\bea
&U_{i,i+\hat{x}}=\eta\tau_3,\\
&U_{i,i+\hat{y}}=-(-)^{i_x}\eta\tau_3,\\
&U_{i,i+2\hat{x}}=\chi \tau_2,\\
&U_{i,i+2\hat{y}}=-\chi\tau_2,\\
&U_{i,i+2\hat{x}+\hat{y}}=(-)^{i_x}i\gamma\tau_0,\\
&U_{i,i+2\hat{x}-\hat{y}}=-(-)^{i_x}i\gamma\tau_0,\\
&U_{i,i+\hat{x}+2\hat{y}}=i\gamma\tau_0,\\
&U_{i,i+\hat{x}-2\hat{y}}=-i\gamma\tau_0,\\
&a_0^{1,2,3}=0.
\eea

After Fourier transformation, the Hamiltonian becomes $H=\Psi_{\textbf{k}}^{\dagger}M\Psi_{\textbf{k}}$, where 
\begin{equation}
\begin{split}
&M=2\eta \text{cos}(k_x)\Gamma_1-2\eta \text{cos}(k_y)\Gamma_2\\
&+2\chi [\text{cos}(2k_x)-\text{cos}(2k_y)]\Gamma_3\\
&+2\gamma[\text{sin}(k_x+2k_y)-\text{sin}(k_x-2k_y)]\Gamma_4\\
&+2\gamma[\text{sin}(2k_x+k_y)-\text{sin}(2k_x-k_y)]\Gamma_5,
\end{split}
\end{equation}
where $\eta,\ \chi,\ \gamma$ are all real numbers, and we have used the Nambu spinor representation
\bea
\Psi_{\textbf{k}}=(f_{u\textbf{k}\uparrow},f_{v\textbf{k}\uparrow},f_{u-\textbf{k}\uparrow}^{\dagger},f_{v-\textbf{k}\downarrow}^{\dagger})^{T},
\eea
and \bea
&\Gamma_1=  \tau_3\otimes \tau_1 ,\\
&\Gamma_2= \tau_3\otimes \tau_3  , \\
&\Gamma_3= \tau_2\otimes \tau_0  , \\
&\Gamma_4=  \tau_0\otimes \tau_1 , \\ 
&\Gamma_5=   \tau_0\otimes \tau_3 .
\eea

Note that $k_x\in(0,\pi)$, $k_y\in(-\pi,\pi)$. The fourth neighbor bond opens up gaps at two wave-vectors $(k_x,k_y)=(\pi/2,\pm \pi/2)$.

\subsection{Realization of the $Z_2B\tau^1_{-}\tau^1_{-}\tau^0\tau^3_{-}$ state}
PSG elements are:
\bea
&G_x(\bm{i})=(-)^{i_y}\tau_0;\\
&G_y(\bm{i})=\tau_0;\\
&G_{P_x}(\bm{i})=(-)^{i_x+i_y}i \tau_1;\\
&G_{P_y}(\bm{i})=(-)^{i_x+i_y}i \tau_1;\\
&G_{P_xy}(\bm{i})=(-)^{i_xi_y} \tau_0;\\
&G_{\mathcal{T}}(\bm{i})=(-)^{i_x+i_y}i\tau_3.
\eea

The analysis of the symmetry allowed ansatz is similar to the previous case, hence we will directly show the results here. The next-nearest-neighbor bonds are again prohibited by the symmetry in this case.

The symmetry allowed ansatz for the $Z_2B\tau^1_{-}\tau^1_{-}\tau^0\tau^3_{-}$ state are
\bea
&U_{i,i+\hat{x}}=   \eta \tau_3  ,  \\
&U_{i,i+\hat{y}}=(-)^{i_y}\eta \tau_3,\\
&U_{i,i+2\hat{x}}=u_{i,i+2\hat{y}}=\chi\tau_1,\\
&U_{i,i+2\hat{x}+2\hat{y}}=\xi \tau_2,\\
&U_{i,i+2\hat{x}-2\hat{y}}=-\xi \tau_2,\\
&a_0^1\neq 0.
\eea

After Fourier transformation, the Hamiltonian becomes
\begin{equation}
\begin{split}
&H=2\eta cos(k_x)\Gamma_1+2\eta cos(k_y)\Gamma_2\\
&+\{a_0^1+2\chi [cos(2k_x)+cos(2k_y)]\}\Gamma_3\\
&+2\xi[cos(2k_x+2k_y)-cos(2k_x-2k_y)]\Gamma_4,
\end{split}
\end{equation}
where $k_x\in(0,\pi)$, $k_y\in(\pi,\pi)$, and \bea
&\Gamma_1=  \tau_3\otimes \tau_1,\\
&\Gamma_2= \tau_3\otimes \tau_3  , \\
&\Gamma_3= \tau_1\otimes \tau_0  , \\
&\Gamma_4=  \tau_2\otimes \tau_0 .
\eea

The Lagrangian multiplier $a_0^1$ and the third neighbor bond open up gaps at the two Dirac points $(k_x,k_y)=(\pi/2,\pm\pi/2)$.

\subsection{Realization of the $Z_2A\tau^1_-\tau^1_-\tau^0\tau^3_+$ state}
PSG elements are:
\bea
&G_x(\bm{i})=G_y(\bm{i})=\tau_0;\\
&G_{P_x}(\bm{i})=(-)^{i_x+i_y}i \tau_1;\\
&G_{P_y}(\bm{i})=(-)^{i_x+i_y}i \tau_1;\\
&G_{P_{xy}}(\bm{i})=\tau_0;\\
&G_{\mathcal{T}}(\bm{i})=i\tau_3.
\eea

First, since the term $a_0^{l}\tau_l$ should be invariant under $G_{\mathcal{T}}\mathcal{T}$, we have $\{a_0^{l}\tau_l, \tau_3\}=0$. And it is also invariant under $G_{P_x}P_x,G_{P_y}P_y$ and $G_{P_{xy}}P_{xy}$, so $[a_0^{l}\tau_l, \tau_1]=0$. Hence we have $a_0^{1}\neq 0$ and $a_0^{2,3}=0$.

\subsubsection{Nearest-neighbor bond}
In this state, the gauge transformation $G_x,G_y,G_{P_{xy}}$ are trivial, we can simply write down the general form of the nearest bond:
\bea
U_{i,i+\hat{x}}=U_{i,i+\hat{y}}=u_{\hat{x}}.
\eea
The bond $u_{\hat{x}}$ is invariant under time-reversal transformation, therefore we have $G_{\mathcal{T}}(i)u_{\hat{x}}G_{\mathcal{T}}(i+\hat{x})^{\dagger}=-u_{\hat{x}}$, which leads to $\{u_{\hat{x}},\tau_3\}=0$.

$u_{\hat{x}}$ is also invariant under $P_y$, therefore we have $G_{P_y}u_{\hat{x}}G_{P_y}^{\dagger}=u_{\hat{x}}$, which leads to $\{u_{\hat{x}},\tau_1\}=0$.

In conclusion, the nearest bonds are as follows:
\bea
U_{i,i+\hat{x}}=U_{i,i+\hat{y}}=\eta \tau_2.
\eea

\subsubsection{Next-nearest-neighbor bond}
The general form of the next-nearest bonds are 
\bea
&U_{i,i+\hat{x}+\hat{y}}=u_{\hat{x}+\hat{y}},\\
&U_{i,i+\hat{x}-\hat{y}}=u_{\hat{x}-\hat{y}}.
\eea

The bond is invariant under time-reversal operation, therefore we have $G_{\mathcal{T}}u_{\hat{x}+\hat{y}}G_{\mathcal{T}}^{\dagger}=u_{\hat{x}+\hat{y}}$, which leads to $\{u_{\hat{x}+\hat{y}},\tau_3\}=0$. Thus we can choose $u_{\hat{x}+\hat{y}}$ to be $\chi\tau_1+\xi \tau_2$. And from $G_{P_y}$, we can obtain $u_{\hat{x}-\hat{y}}=\chi\tau_1-\xi \tau_2$.

\subsubsection{Third neighbor bond}
Since the $G_x,G_y,G_{P_{xy}}$ are trivial, we have
\bea
U_{i,i+2\hat{x}}=U_{i,i+2\hat{y}}=u_{2\hat{x}}.
\eea

It is invariant under time-reversal transformation, therefore we have $\{ u_{2\hat{x}},\tau_3 \}=0$. It is also invariant under $P_y$, which leads to $[u_{2\hat{x}},\tau_1]=0$. Therefore we have $u_{2\hat{x}}=\zeta \tau_1$. Note that the third neighbor bond is necessary for the $IGG$ to be $Z_2$.

In conclusion, the ansatz are
\bea
&U_{i,i+\hat{x}}=U_{i,i+\hat{y}}=\eta \tau_2,\\
&U_{i,i+\hat{x}+\hat{y}}=\chi \tau_1+\xi \tau_2,\\
&U_{i,i+\hat{x}-\hat{y}}=\chi \tau_1-\xi \tau_2,\\
&U_{i,i+2\hat{x}}=U_{i,i+2\hat{y}}=\zeta \tau_1,\\
&a_0^1\neq 0.
\eea

After Fourier transformation, the Hamiltonian becomes $\epsilon_1\tau_1+\epsilon_2\tau_2$, where 
\begin{equation}
\begin{split}
&\epsilon_1=a_0^1+2\zeta[\text{cos}(2k_x)+\text{cos}(2k_y)]+4\chi \text{cos}(k_x)\text{cos}(k_y),
\end{split}
\end{equation}
\begin{equation}
\epsilon_2=2\eta[\text{cos}(k_x)+\text{cos}(k_y)]-4\xi \text{sin}(k_x)\text{sin}(k_y).
\end{equation}

And the dispersion relation is $E_{\pm}=\pm\sqrt{\epsilon_1^2+\epsilon_2^2}$, it is easy to see that when $a_0^1$ is large enough, the spin liquid is gapped.

\subsection{Realization of the $Z_2A\tau^1_-\tau^1_-\tau^0\tau^3_-$ state}
PSG elements of this state are:
\bea
&G_x(\bm{i})=G_y(\bm{i})=\tau_0;\\
&G_{P_x}(\bm{i})=(-)^{i_x+i_y}i \tau_1;\\
&G_{P_y}(\bm{i})=(-)^{i_x+i_y}i \tau_1;\\
&G_{P_{xy}}(\bm{i})=\tau_0;\\
&G_{\mathcal{T}}(\bm{i})=(-)^{i_x+i_y}i\tau_3.
\eea

The analyze of this state is in much the same way as that of the $Z_2A\tau^1_-\tau^1_-\tau^0\tau^3_+$ case, here we will only show the results:
\bea
&U_{i,i+\hat{x}}=U_{i,i+\hat{y}}=\eta \tau_3,\\
&U_{i,i+\hat{x}+\hat{y}}=\chi \tau_1+\xi \tau_2,\\
&U_{i,i+\hat{x}-\hat{y}}=\chi \tau_1-\xi \tau_2,\\
&U_{i,i+2\hat{x}}=U_{i,i+2\hat{y}}=\zeta \tau_1,\\
&a_0^1\neq 0.
\eea
The mean-field Hamiltonian is $\epsilon_1\tau_1+\epsilon_2\tau_2+\epsilon_3\tau_3$, where 
\begin{equation}
\epsilon_1=a_0^1+2\zeta[\text{cos}(2k_x)+\text{cos}(2k_y)]+4\chi \text{cos}(k_x)\text{cos}(k_y),
\end{equation}
\begin{equation}
\epsilon_2=-4\xi \text{sin}(k_x)\text{sin}(k_y),
\end{equation}
\begin{equation}
\epsilon_3=2\eta[\text{cos}(k_x)+\text{cos}(k_y)].
\end{equation}

The energy dispersion is $E_{\pm}=\pm\sqrt{\epsilon_1^2+\epsilon_2^2+\epsilon_3^2}$. When $a_0^1$ is sufficiently large, the dispersion is necessarily gapped.
\subsection{Realization of the $Z_2A\tau^0_+\tau^0_+\tau^0\tau^3_+$ state}
PSG elements are:
\bea
&G_x(i)=G_y(i)=\tau^0,\\
&G_{P_x}(i)=G_{P_y}(i)=G_{P_{xy}}(i)=\tau^0,\\
&G_{\mathcal{T}}(i)=i\tau^3.
\eea

The term $a_0^l\tau_l$ is invariant under $G_{\mathcal{T}}\mathcal{T}$, therefore we have $\{ a_0^l\tau_l,\tau_3\}=0$, hence $a^{1,2}_0\neq 0$ and $a^3_0=0$.

The gauge transformations $G_x,G_y,G_{P_x},G_{P_y},G_{P_{xy}}$ are all trivial, therefore the only constraint comes from time-reversal transformation. 

From $G_{\mathcal{T}}(i)u_{ij}G_{\mathcal{T}}(j)^{\dagger}=-u_{ij}$, we have$\{u_{ij},\tau^3\}=0$. We can therefore write down a symmetry-allowed ansatz which realizes a gapped $Z_2$ spin liquid
\bea
&U_{i,i+\hat{x}}=U_{i,i+\hat{y}}=\eta_1\tau_1+\eta_2\tau_2,\\
&U_{i,i+\hat{x}\pm\hat{y}}=\chi\tau_1,\\
&U_{i,i+2\hat{x}}=U_{i,i+2\hat{y}}=\xi\tau_2,\\
&a^{1,2}_0\neq 0.
\eea

The mean-field Hamiltonian is $\epsilon_1\tau_1+\epsilon_2\tau_2$, where
\bea
&\epsilon_1=a^1_0+2\eta_1[\text{cos}(k_x)+\text{cos}(k_y)]+4\chi\text{cos}(k_x)\text{cos}(k_y),\\
&\epsilon_2=a^2_0+2\eta_2[\text{cos}(k_x)+\text{cos}(k_y)]+2\xi[\text{cos}(2k_x)+\text{cos}(2k_y)].
\eea

The dispersion relation is $E_{\pm}=\pm\sqrt{\epsilon_1^2+\epsilon_2^2}$. When $a_0^1,a_0^2$ are sufficiently large, the dispersion is necessarily gapped.

\subsection{Realization of the $Z_2A\tau^0_+\tau^0_+\tau^0\tau^0_-$ state}
PSG elements are:
\bea
&G_x(i)=G_y(i)=\tau^0,\\
&G_{P_x}(i)=G_{P_y}(i)=G_{P_{xy}}(i)=\tau^0,\\
&G_{\mathcal{T}}(i)=(-)^{i_x+i_y}\tau^0.
\eea

We can prove that this state is necessarily gapless which is protected by the PSG symmetry~\cite{Wen-Zee-PRB02}. Since $G_x$ and $G_y$  are trivial, the mean-field ansatz is manifestly translational invariant, therefore we can write down the mean-field Hamiltonian as 
\bea
H(\textbf{k})=\epsilon^{\mu}(\textbf{k})\tau_{\mu},
\eea
where $\mu=1,2,3$.

The condition $G_{\mathcal{T}}(i)u_{ij}G_{\mathcal{T}}^{\dagger}(j)=-u_{ij}$, when translated into $\textbf{k}$ space, becomes $g_{\mathcal{T}}H(k_x,k_y)g_{\mathcal{T}}^{\dagger}=-H(k_x+\pi,k_y+\pi),g_{\mathcal{T}}=\tau_0$. Therefore we have
\bea\label{constraint1}
\epsilon^{\mu}(k_x,k_y)=-\epsilon^{\mu}(k_x+\pi,k_y+\pi).
\eea

And from $G_{P_{xy}}H(k_x,k_y)G^{\dagger}_{P_{xy}}=H(k_y,k_x)$ we have
\bea\label{constraint2}
\epsilon^{\mu}(k_x,k_y)=\epsilon^{\mu}(k_y,k_x),
\eea

Eq.~\eqref{constraint1} and Eq.~\eqref{constraint2} indicate that the Hamiltonian is gapless along the line $(k_x,k_x+\pi)$. On the one hand, we have $\epsilon^{\mu}(k_x,k_x+\pi)=-\epsilon^{\mu}(k_x+\pi,k_x)$ from Eq.~\eqref{constraint1}. On the other hand, we have $\epsilon^{\mu}(k_x,k_x+\pi)=\epsilon^{\mu}(k_x+\pi,k_x)$ from Eq.~\eqref{constraint2}. Therefore $\epsilon^{\mu}(k_x,k_x+\pi)$ must vanish for any $k_x$, indicating that the Hamiltonian is gapless along this line. Similarly, we can prove that the Hamiltonian is gapless along the line $(k_x,\pi-k_x)$. Thus, we have completed our proof that the $Z_2A\tau^0_+\tau^0_+\tau^0\tau^0_-$ state is necessarily gapless.

\subsection{Realization of the $Z_2A\tau^0_+\tau^0_+\tau^0\tau^3_-$ state}
PSG elements are:
\bea
&G_x(i)=G_y(i)=G_{P_x}(i)=G_{P_y}(i)=G_{P_{xy}}(i)=\tau^0,\\
&G_{\mathcal{T}}(i)=(-)^{i_x+i_y}i\tau^3.
\eea

The term $a_0^l\tau_l$ is invariant under $G_{\mathcal{T}}\mathcal{T}$, therefore we have $\{ a_0^l\tau_l,\tau_3\}=0$, hence $a^{1,2}_0\neq 0$ and $a^3_0=0$.

We can label site $i$ with its parity $(-1)^{i_x+i_y}$. Then for a bond $u_{ij}$ connecting even site $i$ with odd site $j$, we have $[u_{ij},\tau^3]=0$. And for a bond $u_{ij}$ connecting two even sites or two odd sites, we have $\{u_{ij},\tau^3\}=0$. Since $u_{ij}$ becomes $u_{ji}$ under $C_4^2=(P_{xy}P_y)^2$ and translation, we have an additional constraint $u_{ij}=u_{ji}=u_{ij}^{\dagger}$ for any bond $u_{ij}$.

Therefore we could write down a mean-field ansatz which realizes a gapped $Z_2$ spin liquid
\bea
&U_{i,i+\hat{x}}=U_{i,i+\hat{y}}=\eta\tau_3,\\
&U_{i,i+\hat{x}\pm\hat{y}}=\chi\tau_1,\\
&U_{i,i+2\hat{x}}=U_{i,i+2\hat{y}}=\xi\tau_2,\\
&a_0^{1,2}\neq 0.
\eea

The mean-field Hamiltonian is therefore $\epsilon_1\tau_1+\epsilon_2\tau_2+\epsilon_3\tau_3$, where
\bea
&\epsilon_1=a_0^1+4\chi\text{cos}(k_x)\text{cos}(k_y),\\
&\epsilon_2=a_0^2+2\xi[\text{cos}(2k_x)+\text{cos}(2k_y)],\\
&\epsilon_3=2\eta[\text{cos}(k_x)+\text{cos}(k_y)].
\eea

The energy dispersion is $E_{\pm}=\pm\sqrt{\epsilon_1^2+\epsilon_2^2+\epsilon_3^2}$. When $a_0^1$ and $a_0^2$ are sufficiently large, the dispersion is necessarily gapped.

\end{document}